\documentclass[traditabstract]{aa} % for the abstract without structuration
\usepackage[pdftex]{graphicx}
\usepackage{breqn}
\usepackage[flushleft]{threeparttable}

\def\la{\lower.5ex\hbox{$\; \buildrel < \over \sim \;$}}
\def\ga{\lower.5ex\hbox{$\; \buildrel > \over \sim \;$}}
\def\SFR{\dot{\Sigma}_{*}}
\begin{document}
   \title{Deciphering the radio-star formation correlation on kpc-scales}
   \subtitle{I. Adaptive kernel smoothing experiments}

   \author{B.~Vollmer\inst{1}, M.~Soida\inst{2}, R.~Beck\inst{3}, \and M.~Powalka\inst{1}}

   \institute{Universit\'e de Strasbourg, CNRS, Observatoire astronomique de Strasbourg, UMR 7550, F-67000 Strasbourg, France \and
          Astronomical Observatory, Jagiellonian University, Krak\'ow, Poland \and
	  Max-Planck-Institut f\"{u}r Radioastronomie, Auf dem H\"{u}gel 69, 53121 Bonn, Germany}

   \date{Received ; accepted }

% \abstract{}{}{}{}{}
% 5 {} token are mandatory

  \abstract
{
One of the tightest correlations in astronomy is the relation between the integrated radio continuum and the far-infrared emission.
Within nearby galaxies, variations in the radio-FIR correlation have been observed, mainly because the cosmic ray electrons migrate before they 
lose their energy via synchrotron emission or escape. The major cosmic ray electron transport mechanisms within the plane of galactic
disks are diffusion and streaming. 
A predicted radio continuum map can be obtained by convolving the map of comic ray electron sources, represented by that of the star formation, with adaptive
Gaussian and exponential kernels. The ratio between the smoothing lengthscales at $6$~cm and
$20$~cm can be used to distinguish between diffusion and streaming as the dominant transport mechanism.
The dependence of the smoothing lengthscale on the star formation rate bares information on dependence of the magnetic field strength or the
ratio between the ordered and turbulent magnetic field strengths on star formation.
Star formation maps of eight rather face-on local and Virgo cluster spiral galaxies were constructed from Spitzer and Herschel infrared and 
GALEX UV observations.
These maps were convolved with adaptive Gaussian and exponential smoothing kernels to obtain model radio continuum emission maps. 
It is found that in asymmetric ridges of polarized radio continuum emission the total power emission is enhanced with respect to the star 
formation rate. At a characteristic star formation rate of $\dot{\Sigma}_*=8 \times 10^{-3}$~M$_{\odot}$yr$^{-1}$kpc$^{-2}$ the typical 
lengthscale for the transport of cosmic ray electrons is $l=0.9 \pm 0.3$~kpc at $6$~cm and  $l=1.8 \pm 0.5$~kpc at $20$~cm.
Perturbed spiral galaxies tend to have smaller lengthscales. This is a natural consequence of the enhancement of the magnetic 
field caused by the interaction. 
The discrimination between the two cosmic ray electron transport mechanisms, diffusion and streaming, is based on 
(i) the convolution kernel (Gaussian or exponential),
(ii) the dependence of the smoothing kernel on the local magnetic field and hence on the local star formation rate,
(iii) the ratio between the two smoothing lengthscales via the frequency-dependence of the smoothing kernel, and 
(iv) the dependence of the smoothing kernel on the ratio between the ordered and the turbulent magnetic field.
Based on our empirical results, methods (i) and (ii) cannot be used to determine the cosmic ray transport mechanism. 
Important asymmetric large-scale residuals and a local dependence of the smoothing length on $B_{\rm ord}/B_{\rm turb}$ are most probably responsible 
for the failure of method (i) and (ii), respectively.
On the other hand, the classifications based on $l_{\rm 6cm}/l_{\rm 20cm}$ (method iii) and $B_{\rm ord}/B_{\rm turb}$ (method iv) are well consistent and complementary.
%The ratio between the ordered and turbulent magnetic field strengths is found to be $(B_{\rm ord}/B_{\rm turb}) \propto \dot{\Sigma}_*^{\ -0.24}$.
%From this we obtained the relation between the total magnetic field strength and star formation
%$B \propto \dot{\Sigma}_*^{\ 0.30 \pm 0.24}$, which is lower, but still consistent with energy equipartition between the energy density of the magnetic field and the 
%kinetic energy density of the interstellar medium. 
We argue that in the six Virgo spiral galaxies the turbulent magnetic field is globally enhanced in the disk. Therefore, the regions where the magnetic field 
is independent of the star formation rate are more common. In addition, $B_{\rm ord}/B_{\rm turb}$ decreases leading to a diffusion
lengthscale which is smaller than the streaming lengthscale. Therefore, cosmic ray electron streaming dominates in most of the Virgo
spiral galaxies.
}


   \authorrunning{Vollmer et al.}
   %\titlerunning{Large-scale radio continuum properties of 19 Virgo galaxies}

   \maketitle
%
%________________________________________________________________

\section{Introduction\label{sec:introduction}}

One of the tightest correlations in astronomy is the relation between the integrated radio continuum (synchrotron) and the far-infrared emission
(Helou et al. 1985; Condon  1992; Mauch \& Sadler  2007; Yun et al. 2001; Bell 2003; Appleton et al. 2004; Kovacs et al. 2006;
Murphy et al. 2009; Sargent et al. 2010, Li et al. 2016). It holds over five orders of magnitude in various types of galaxies including starbursts.
The common interpretation of the correlation is that both emission types are proportional to star formation:
the radio emission via (i) the cosmic ray source term which is due to supernova explosions and the turbulent amplification of
the small-scale magnetic field (small-scale dynamo, see, e.g., Schleicher \& Beck 2013) and (ii) the far-infrared emission via
the dust heating mainly through massive stars. Within nearby galaxies, variations in the radio-FIR correlation have been observed
by Gordon et al. (2004), Murphy et al. (2006, 2008), Dumas et al. (2011), Hughes et al. (2006), Tabatabaei et al. (2013a).
The latter authors found that the  slope  of  the  radio-FIR  correlation  across  the  galaxy varies as a function of the star
formation rate and the magnetic field strength. For the resolved correlation between the radio
continuum emission and the star formation rate Heesen et al. (2014, 2019) found a non-linear slope of $0.79 \pm 0.25$ and $0.75 \pm 0.10$, respectively.
When the star formation distribution was smoothed owing to the transport of cosmic ray electrons away
from star formation site, the slope became about unity (Heesen et al. 2019).
Thus, the radio continuum emission is smoothed with respect to the star formation rate owing to the transport of cosmic ray electrons.

Radio continuum emission observed at frequencies below a few GHz is most frequently dominated by synchrotron emission, which
is emitted by cosmic ray electrons with relativistic velocities that spiral around galactic magnetic fields.
The magnetic field can be regular, i.e. structured on large-scales (kpc), or tangled on small-scales via turbulent motions.
The turbulent magnetic field has an isotropic and an anisotropic component.
The total magnetic field $B$ is the total magnetic field, which is the quadratic sum of the ordered and turbulent magnetic
field components. The ordered magnetic field includes the large-scale regular and the anisotropic small-scale magnetic fields.
The turbulent magnetic field dominates in spiral arms with total magnetic field strengths of $20$-$30$~$\mu$G,
the ordered field dominates in the interarm regions with total magnetic field strengths of $5$-$15$~$\mu$G (Beck 2005, Beck \& Wielebinski 2013).
The evolution of the magnetic field is determined by the induction equation.

The energy loss caused by synchrotron emission depends on the magnetic field strength and the electron energy or Lorentz factor $\gamma$:
\begin{equation}
\frac{d\,E}{d\,t}=b(E)=3.4 \times 10^{-17} {\rm GeV\,sec^{-1}} \times \big(\frac{E}{\rm 1~GeV} \big)^2 \big( \frac{B}{3~\mu{\rm G}} \big)^2\ .
\end{equation}
A cosmic ray electron with energy $E$ will emit most of its energy at a critical frequency $\nu_{\rm c}$ where
\begin{equation}
\nu_{\rm c}=1.3 \times 10^{-2} \big(\frac{B}{\mu {\rm G}} \big) \big(\frac{E}{\rm GeV}  \big)^2~{\rm GHz} \ .
\end{equation}
The timescale for synchrotron emission is
\begin{equation}
\label{eq:synch}
t_{\rm syn}=\frac{E}{b(E)} \simeq 4.5 \times 10^7 \big( \frac{B}{10~\mu{\rm G}} \big)^{-3/2} \nu_{\rm GHz}^{-1/2}~{\rm yr}\ .
\end{equation}

Cosmic ray particles are mainly produced in supernova shocks via Fermi acceleration. 
However, the relativistic electrons do not stay at the location of their creation. They propagate either via diffusion,
or by streaming with the Alfv\'en velocity. In addition, cosmic ray electrons can be transported into the halo
by advection, i.e. a galactic wind. During the transport process, the cosmic ray electron lose energy via synchrotron emission.
The associated diffusion--advection--loss equation for the cosmic ray electron density $N$ reads
\begin{dmath}
\label{eq:diffeq}
\frac{\partial N}{\partial t} = D {\nabla}^2 N + \frac{\partial}{\partial E} \big( b(E) N(E) \big) - 
(\vec{u}+\vec{v}) \nabla N + \frac{p}{3} \frac{\partial N}{\partial p} \nabla \vec{u} + Q(E)\ ,
\end{dmath}
where $D$ is the diffusion coefficient, $E$ the cosmic ray electron energy, $\vec{u}$ the advective flow velocity, 
$\vec{v}$ the streaming velocity, $p$ the cosmic ray electron pressure,
$Q$ the source term, and $b(E)$ the energy loss rate.
The first part of the right-hand side is the diffusion term, followed by the synchrotron loss, advection, streaming, adiabatic energy gain or loss,
and the source terms. The advection and adiabatic terms are only important for the transport in vertical direction. 
For the cosmic ray electron transport within the disk, these terms play a minor role.
Energy can be lost via inverse Compton radiation, Bremsstrahlung, pion or ionization energy loss, and most importantly synchrotron emission
(see, e.g., Murphy 2009, Lacki et al. 2010). In non-starburst galaxies, the synchrotron loss term is expected to dominate.

To predict the distribution of non-thermal radio continuum emission, where the thermal and AGN emission has been removed, one has to
solve the diffusion equation (Eq.~\ref{eq:diffeq}). If we assume that advection, i.e. galactic winds, are negligible and
adiabatic gains and losses are unimportant, the equation becomes a diffusion-advection equation with a source term.
In a steady state this gives: $D {\nabla}^2 N + \vec{v} {\nabla} N + Q(E) = 0$. 
The assumption of steady state at a spatial resolution of $\sim 1$~kpc is justified by the tightness of the correlation between the resolved star formation 
rate and non-thermal radio continuum emission, with a dispersion of $0.3$~dex at 20~cm (Heesen et al. 2019)
and $0.2$~dex at 6~cm (Fig~\ref{fig:sfrrad_allc}). The predicted radio continuum map can thus be obtained by
convolving the source map, represented by that of the star formation rate, with Gaussian (diffusion) and exponential (streaming) kernels (cf. middle panels of Figs.~5 and 6 
in Heesen et al. 2016)\footnote{A leaky-box model, i.e., with a diffusion, source, and loss terms leads to an exponential kernel (Eq.~12 of Atoyan et al. 1995).}. 

The transport mechanism of cosmic ray electrons within galactic disks can be diffusion, as a result of random motions across
tangled magnetic field lines, or streaming, as a result of ordered motion along magnetic field lines down a cosmic ray
pressure gradient (e.g., Yan et al. 2012).
The two transport mechanisms give rise to different dependencies of the kernel lengthscale with respect to
the magnetic field and the frequencies of radio continuum observations.
It should thus be possible to discriminate between the two transport mechanism by the study of the lengthscale of
the Gaussian/exponential convolution kernels and their dependence on frequency and magnetic field strength in face-on galaxies. 

Diffusion is found to be  the dominant transport process in the Galaxy (Strong et al. 2011).
Murphy et al. (2006, 2008, 2009) created modeled radio continuum distributions by appropriately smoothed far infrared images
of a sample of local spiral galaxies.
For unperturbed spiral galaxies, Murphy et al. (2008) showed that  the  dispersion in the far infrared-radio ratios on
subkiloparsec scales within galaxies can be reduced by a factor of $\sim 2$, on average. 
These authors used a spatially constant lengthscale for the smoothing kernel within a galactic disk. 
However, for cosmic ray diffusion or streaming the lengthscale  
should depend on the synchrotron timescale which depends on the magnetic field strength.
Since the magnetic field strength is related to the star formation rate (e.g. Tabatabaei et al. 2013a, Heesen et al. 2014), we expect that the
smoothing kernel is proportional to the local star formation rate density. 
This behaviour was found globally by Murphy et al. (2006, 2008) using one single smoothing lengthscale for the whole galaxy. 
We want to see if we can find evidence for such a local behaviour.
In addition, the cosmic ray electron source term 
$Q(E)$ should be directly linked to the star formation distribution, rather than the far-infrared emission distribution (see, e.g., Heesen et al. 2014, 2019).

In this paper we test our hypothesis by applying a Gaussian and exponential convolution with an adaptive Kernel to the star formation maps 
of eight local rather face-on spiral galaxies and compare the results with synchrotron emission maps. 
For the comparison we use the same minimization as Murphy et al. to find the ``best fit'' model and thus the smoothing lengthscale.
Heesen et al. (2019) used a different criterion: the relation between the star formation rate $\dot{\Sigma}_*$ and the radio surface brightness
becomes linear for the appropriate smoothing lengthscale. In addition to the work of Murphy et al. and Heesen et al. we use
an adaptive smoothing kernel which depends on the local star formation rate $\dot{\Sigma}_*$. 
This is physically motivated, because the smoothing lengthscale depends on
the synchrotron timescale which depends on the magnetic field. The latter is supposed to depend on $\dot{\Sigma}_*$.
The possible advection of cosmic electrons via a galactic outflow is also taken into account in our models.
Compared to Murphy et al. (2006, 2008, 2009) we derive the smoothing lengthscales at two frequencies (as Heesen et al. 2019).

Our aim is to discriminate between the two cosmic ray transport mechanisms, diffusion and streaming, based on (i) the convolution kernel (Gaussian or exponential),
(ii) the dependence of the smoothing kernel on the local magnetic field and hence on the local star formation rate,
(iii) the ratio between the two smoothing lengthscales via the frequency-dependence of the smoothing kernel, and 
(iv) the dependence of the smoothing kernel on the ratio between the ordered and the turbulent magnetic field.
 
The article is structured in the following way:
the observations, on which the input star formation and radio continuum maps are based, are described in Sect.~\ref{sec:observations}.
The correlations between the integrated star formation, radio continuum emission, and average magnetic field are
presented in Sect.~\ref{sec:integ}. The adaptive kernel smoothing method is described in Sect.~\ref{sec:method} and tested
on toy model images in Sect.~\ref{sec:tests}. We present the results of the adaptive kernel smoothing in Sect.~\ref{sec:results}.
Sect.~\ref{sec:diffstream} is dedicated to the question which transport mechanism of cosmic ray electrons dominates in
the disks of our sample galaxies. We discuss our results in Sect.~\ref{sec:discussion} and give our conclusions in Sect.~\ref{sec:conclusions}.

\section{Data\label{sec:observations}}

The studied eight local face-on galaxies are presented in Table~\ref{tab:gals}.
For the resolved radio-SFR correlation we use published radio continuum, UV, and infrared data.
\begin{table*}[!ht]
      \caption{Galaxy sample.}
         \label{tab:gals}
      \[
         \begin{array}{lrcccccll}
           \hline
           \noalign{\smallskip}
           \hline
           {\rm galaxy\ name } &  {\rm m_{\rm B}} & i^{\rm (a)} & {\rm Dist^{\rm (b)}} & {\rm type} & v_{\rm rot}  & \Theta & B_{\rm ord}^{\rm (c)} & B_{\rm turb} \\
            & & {\rm (degrees)} & {\rm (Mpc)} & & {\rm (km\,s}^{-1}) & ($''$) & (\mu {\rm G}) & (\mu {\rm G}) \\
           \hline
           {\rm NGC}~6946 & 9.76 & 18 & 6.0  &  {\rm  SAB(rs)cd} & 200 & 18 & 9.9^{\rm (d)} & 12.6^{\rm (d)} \\ 
           {\rm M}~51 & 9.26 & 33 & 8.4 & {\rm SAbc} & 200 & 22  & 11.5^{\rm (e)} & 16.0^{\rm (e)} \\
           {\rm NGC}~4321 & 10.02 & 27 & 17 & {\rm SAB(s)bc} & 200 & 18  & 4.3 & 10.5 \\
           {\rm NGC}~4303 & 9.97 & 25 & 17 & {\rm SAB(rs)bc} & 200 & 18  & 6.5 & 15.4 \\
           {\rm NGC}~4535 & 10.73 & 43 & 17 & {\rm SAB(s)c} & 180 & 18  & 3.6 & 7.2 \\
           {\rm NGC}~4254 & 10.44 & 20 & 17 &  {\rm SA(s)c} & 180 & 18  & 6.5^{\rm (f)} & 20.3^{\rm (f)} \\
           {\rm NGC}~4501 & 10.50 & 57 & 17 & {\rm SA(rs)b} & 260 & 18  & 7.1 & 15.3 \\
           {\rm NGC}~4654 & 11.31 & 51 & 17 & {\rm SAB(rs)cd} & 150 & 18  & 2.5 & 9.2 \\ 
       \noalign{\smallskip}
       \hline
       \noalign{\smallskip}
 \hline
        \end{array}
      \]
\begin{list}{}{}
\item[$^{\rm (a)}$ Inclination angle]
\item[$^{\rm (b)}$ This research has made use of the GOLD Mine Database (Gavazzi et al. 2003)]
\item[$^{\rm (c)}$ The ordered magnetic field $B_{\rm ord}$ includes the large-scale ordered and the anisotropic small-scale magnetic fields]
\item[$^{\rm (d)}$ from Tabatabaei et al. (2013b)]
\item[$^{\rm (e)}$ from Fletcher et al. (2011)]
\item[$^{\rm (f)}$ mean value based on Chy{\.z}y (2008)]
\end{list}
\end{table*}

\subsection{Radio continuum}

Four Virgo spiral galaxies (NGC~4321, NGC~4535, NGC~4501, NGC~4654) were observed at 4.85 and 1.4~GHz between
November 8, 2005 and January 10, 2006 with the Very Large
Array (VLA) of the National Radio Astronomy Observatory
(NRAO) in the D array configuration. In addition, we observed the 8 galaxies at $1.4$~GHz on August 15, 2005 in the C array configuration.
The band passes were $2 \times ×50$~MHz. 
The final cleaned maps were convolved to a symmetric Gaussian beam size of $18'' \times 18''$ ($1.5 \times 1.5$~kpc).
The total power images at 6~cm and 20~cm are presented in Vollmer et al. (2010).
NGC~4303 was observed with the VLA
at 4.86~GHz between October 12, 2009 and December 23,
2009 with the VLA in the D array configuration. In addition, we observed all galaxies at $1.4$~GHz on March 21, 2008 in the C array configuration.
The band widths were $2 \times 50$~MHz.
The final cleaned maps were convolved to a symmetric Gaussian beam size of $22'' \times 22''$ ($1.8 \times 1.8$~kpc).
The total power images at 6~cm and 20~cm are presented in Vollmer et al. (2013).
In addition, we used VLA 4.85 and 1.4~GHz images of NGC~4254 (Chy\.zy et al. 2007), NGC~6946 (Beck \& Hoernes 1996), 
and M~51 (Fletcher et al. 2011) with spatial resolutions of $15'' \times 15''$.
We subtracted strong point sources and the thermal free-free radio emission according to the recipe of Murphy et al. (2008)
\begin{equation}
\big(\frac{S_{\rm therm}}{\rm Jy}\big) = 7.9 \times 10^{-3} \big(\frac{T}{10^4~{\rm K}}\big)^{0.45} \big(\frac{\nu}{\rm GHz}\big)^{-0.1}
\big(\frac{I_\nu(24~\mu{\rm m})}{\rm Jy}\big)\ ,
\end{equation}
where $T$ is the electron temperature and $I_\nu(24~\mu{\rm m})$ is the flux density at a wavelength of $24~\mu$m.
We note that there are other alternative methods to
account for the thermal free-free emission using, e.g., the extinction-corrected H$\alpha$ emission (Tabatabaei et al. 2007, Heesen et al. 2014).

In the inner regions of many galaxies the ratio between the radio continuum emission and star formation is significantly different from that of the disk.
Tests with radio continuum emission maps including the nuclei showed that these regions can behave significantly different
than the disks in terms of cosmic ray electron transport.
To prevent a strong influence of the nucleus on our results, we therefore removed the emission of the inner kpc in many cases.

\subsection{Star formation rate}

The star formation rate was calculated from the FUV luminosities corrected by the total infrared to FUV luminosity ratio (Hao et al. 2011).
This method takes into account the UV photons from young massive stars which escape the galaxy and those which
are absorbed by dust and re-radiated in the far infrared:
\begin{equation}
\dot{\Sigma}_{*} = 8.1 \times 10^{-2}\ (I({\rm FUV}) + 0.46\ I({\rm TIR}))\ ,
\end{equation}
where $I({\rm FUV})$ is the GALEX far ultraviolet and $I({\rm TIR})$ the total infrared intensity based on Spitzer IRAC and MIPS data
in units of MJy\,sr$^{-1}$. $\dot{\Sigma}_{*}$ has the units of M$_{\odot}$kpc$^{-2}$yr$^{-1}$.
This prescription only holds for a constant star formation rate over the last few $100$~Myr.

The full width at half-maximum (FWHM) of the point spread functions (PSFs), as stated in the Spitzer Observer's Manual 
(Spitzer Science Centre 2006), are 1.7, 2.0, 6, and 18~arcsec at 3.6, 8.0, 24, and 70~$\mu$m, respectively.
In addition, we used Herschel Virgo Cluster Survey (HeViCS) 100 and 160~$\mu$m images (Davies et al. 2010) which have spatial resolutions 
of 7 and 12~arcsec, respectively.
First, the data are convolved with Gaussian kernels that match the PSFs of the images in the 3.6, 8, 24 70, 100, and 160~$\mu$m bands to the PSF of 
the radio continuum data. 
Next, the data were re-binned to the common pixel size of the radio continuum maps. 
We exclude from the analysis regions not detected at the $3 \sigma$ level in one or more wave bands.
This resulted in a loss of surface of $10$\,\% to $20$\,\% compared to the regions with $24~\mu$m $3\sigma$ detections
with a tendency of less loss for truncated galaxies.
Following Helou et al. (2004), we subtracted the stellar continuum from the 8 and 24~$\mu$m surface brightnesses (in MJy\,sr$^{-1}$) using 
\begin{equation}
I_{\nu}({\rm PAH}\ 8\mu {\rm m})=I_{\nu}(8\mu {\rm m})-0.232\ I_{\nu}(3.6\mu {\rm m})\ {\rm and}
\end{equation}
\begin{equation}
I_{\nu}(24\mu {\rm m})=I_{\nu}(24\mu {\rm m})-0.032\ I_{\nu}(3.6\mu {\rm m})\ .
\end{equation}
We calculated the TIR surface brightnesses using
\begin{eqnarray}
I({\rm TIR})=2.064\, \nu I_{\nu}(24\mu {\rm m}) + 0.539\, \nu I_{\nu}(70\mu {\rm m}) + \nonumber \\
0.277\, \nu I_{\nu}(100\mu {\rm m}) + 0.938\, \nu I_{\nu}(160\mu {\rm m})
\label{eq:tir}
\end{eqnarray}
based on Table~3 from Galametz et al. (2013).
In most galactic environments the error introduced to $I({\rm TIR})$ is about 50\,\% based on the Draine \& Li (2007) models.

One should keep in mind that these star formation rates are not perfect: e.g., the SF timescales probed by FIR and GALEX UV might not match — 
the GALEX UV probes star formation timescales of $\sim 50$-$100$~Myr, whereas the FIR probes timescales that are probably shorter, although this depends 
on the star formation history. Leroy et al. (2008) estimated the typical uncertainty on their 24~$\mu$m--FUV star formation rate to be of the
order of $50$\,\% or $0.2$~dex. The uncertainty of our star formation rates is of the same order. For a comparison between different star formation
indicators based on mixed processes (direct stellar light (FUV), dust-processed stellar light (FIR, TIR), ionised gas emission(H$\alpha$)) 
we refer to Calzetti (2013).

\section{Integrated and resolved correlation \label{sec:integ}}

\begin{figure}[!ht]
  \centering
  \resizebox{\hsize}{!}{\includegraphics{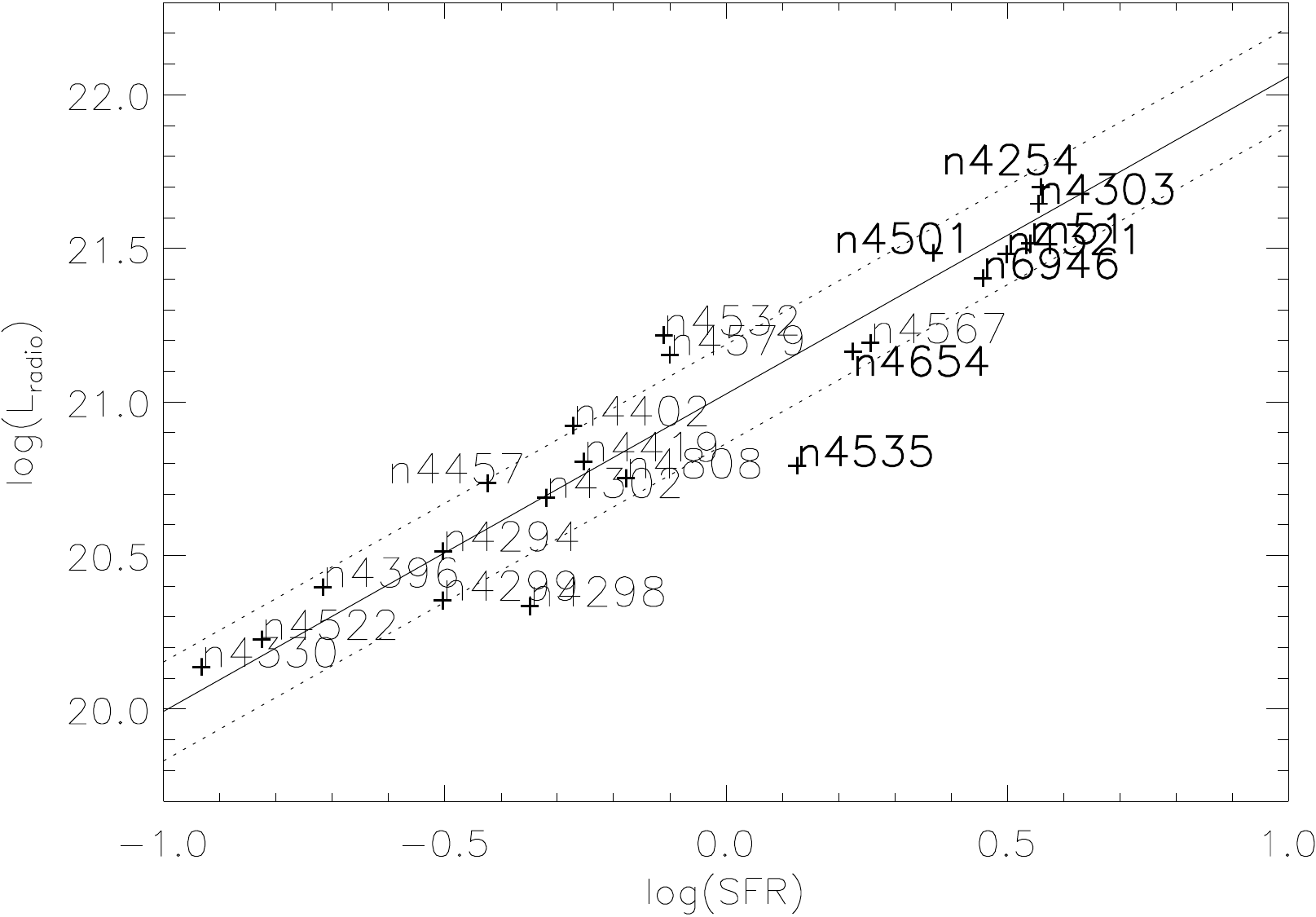}}
  \caption{Integrated SFR -- radio continuum correlation at $6$~cm. The star formation rate is in $M_{\odot}\,{\rm pc}^{-2}$, the radio luminosity 
    in W\,Hz$^{-1}$.
  \label{fig:sfrradc_int_small}}
\end{figure}

The correlation between the integrated star formation rate (SFR) and the non--thermal radio continuum emission of 
the sample of Vollmer et al. (2012, 2013) is shown in Fig.~\ref{fig:sfrradc_int_small}.
The integrated SFR -- radio continuum correlation is approximately linear ($\log(SFR) = 1.03\, \log(L_{\rm 6cm}) + 21.03$) with a scatter of $0.16$~dex.
Only two galaxies lie significantly below the SFR -- radio continuum correlation: 
NGC~4298, which is part of a galaxy pair together with NGC~4299, and NGC~4535.
On the other hand, all perturbed Virgo spiral galaxies lie somewhat above the correlation: NGC~4522 (ram pressure stripping),
NGC~4330 (ram pressure stripping), NGC~4396 (H{\sc i} tail), NGC~4457 (probable minor merger), NGC~4532 (accreting H{\sc i} filament), 
NGC~4501 (ram pressure stripping), and NGC~4254 (tidal interaction and probably ram pressure stripping).

\begin{figure}[!ht]
  \centering
  \resizebox{\hsize}{!}{\includegraphics{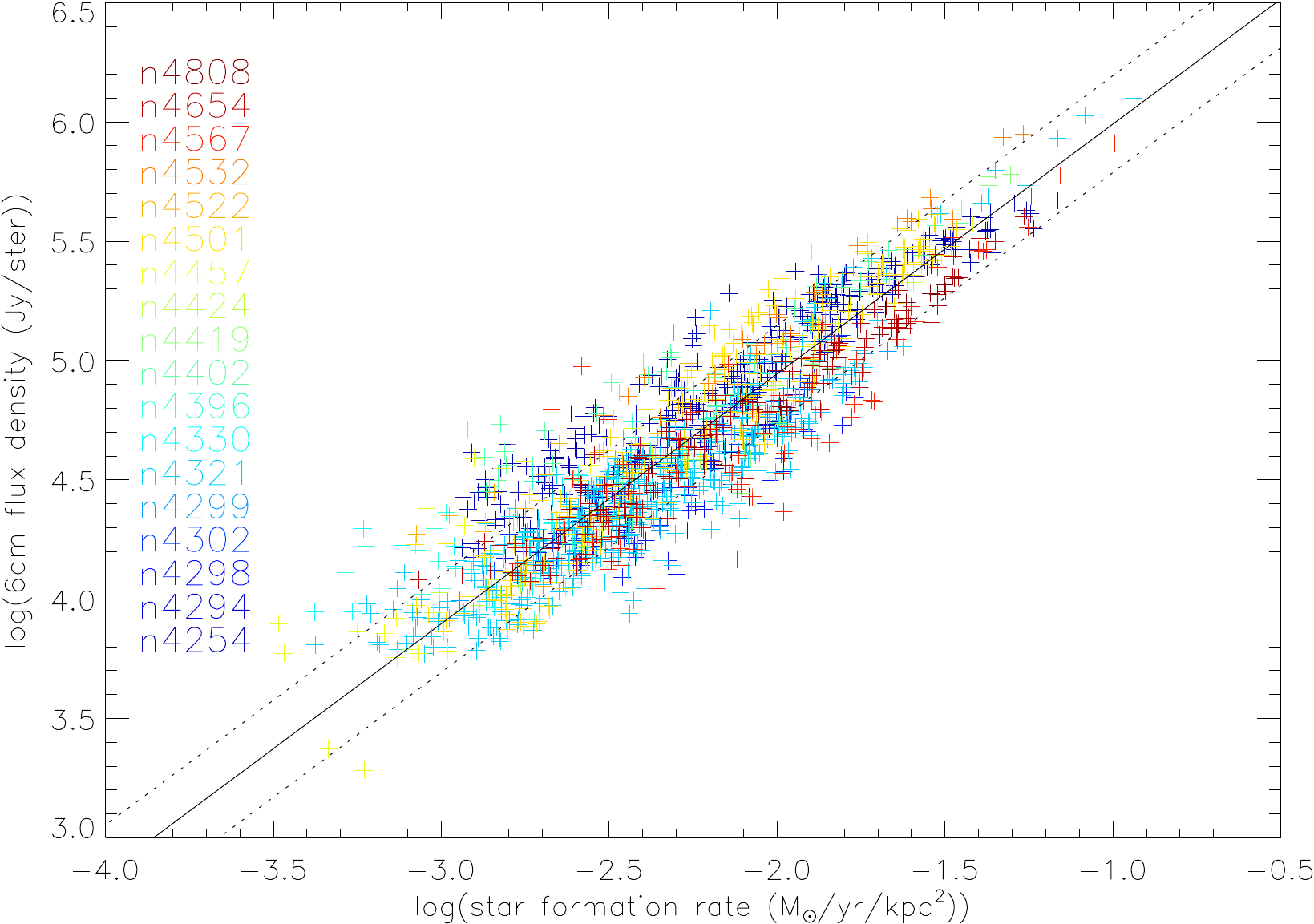}}
  \caption{Resolved $\dot{\Sigma}_*$ -- radio continuum correlation at $6$~cm.
  \label{fig:sfrrad_allc}}
\end{figure}

Fig.~\ref{fig:sfrrad_allc} shows the resolved star formation -- non-thermal radio continuum relation at 6~cm.
With an outlier-resistant two-variable linear regression (IDL robust\_linefit) of 18 Virgo galaxies (Vollmer et al. 2012, 2013)
we find an exponent of $1.05$ and a scatter of $0.2$~dex.
For comparison, Heesen et al. (2014) found a slope of $0.8$ for their sample of SINGS galaxies observed at 20~cm and Tabatabaei et al. (2013a)
determined a slope of $0.9$ for NGC~6946 observed at 20~cm.

\begin{figure}[!ht]
  \centering
  \resizebox{\hsize}{!}{\includegraphics{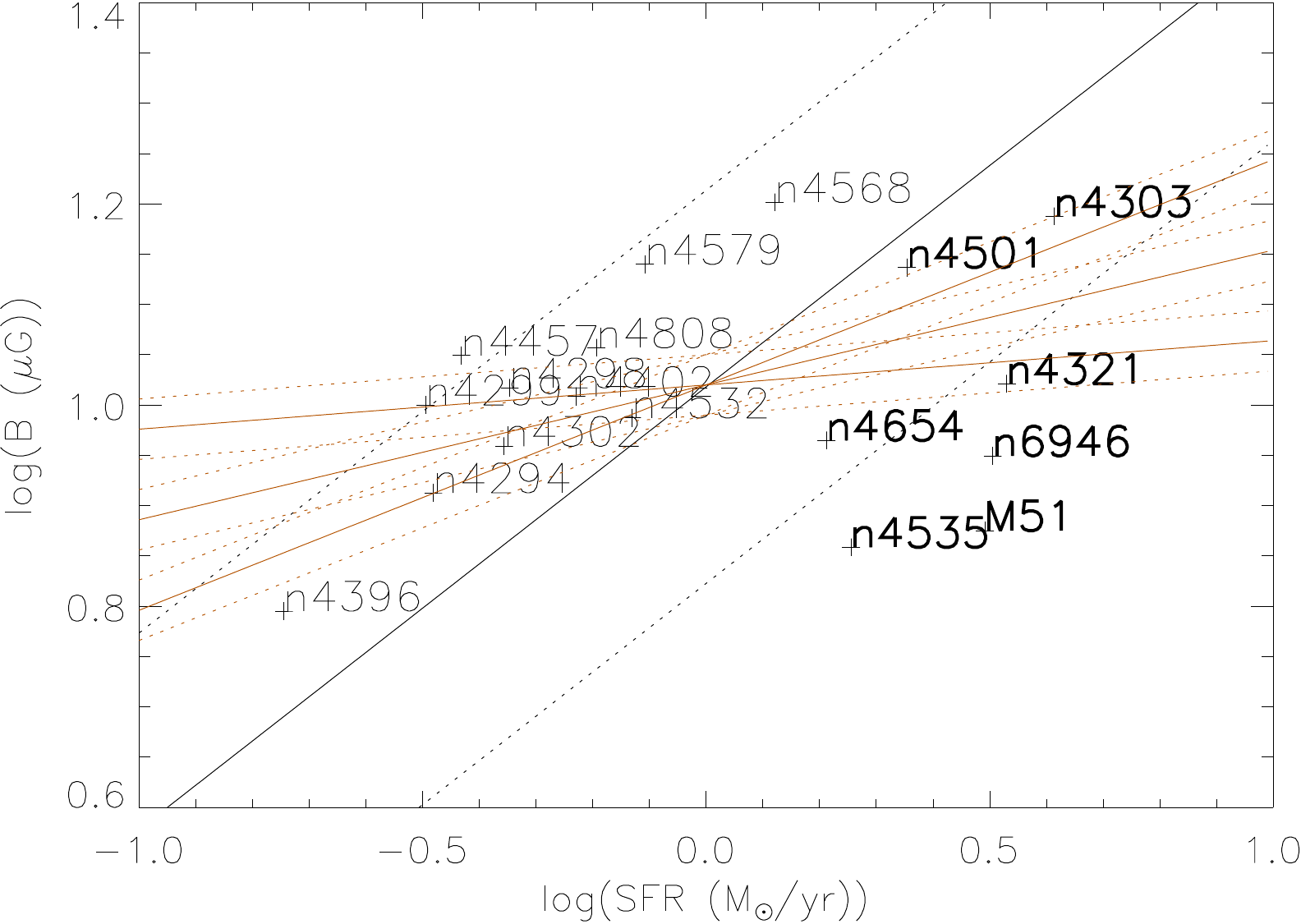}}
  \caption{Average magnetic field strength as a function of the total star formation rate.
    The black line correspond to an outlier-resistant two-variable linear regression,
    the red lines to a Bayesian approach to linear regression.
  \label{fig:sfr_Bfield}}
\end{figure}

The average magnetic field strengths based on the integrated properties of the galaxies are presented in Table~\ref{tab:gals} and Fig.~\ref{fig:sfr_Bfield}.
They were calculated using the revised formula of Beck \& Krause (2005) implying the spectral index.
We assumed a proton-to-electron number density ratio of $100$ and a pathlength of $1$~kpc$/\cos(i)$, where $i$ is the galaxy inclination, with a maximum of 5~kpc.
In addition, we assumed a constant thermal fraction of $20$\,\%. This approximation is sufficient, because a variation by a factor of two leads to
a much smaller uncertainty on the magnetic field strength than a comparable variation of the pathlength or non-thermal spectral index.
Overall, the magnetic field strengths of Virgo galaxies are high, exceeding $8$~$\mu$G except in NGC~4396 and NGC~4535.
There is no clear correlation between the star formation rate and the magnetic field strength:
for our sample, we determine $\log B = 0.44 \, \log (SFR) + 1.02$ with an outlier-resistant two-variable linear regression
(IDL robust\_linefit). With a Bayesian approach to linear regression (IDL linmix\_err) 
we find $\log B = (0.13 \pm 0.09) \, \log (SFR) + (1.02 \pm 0.03)$ assuming errors of $0.18$~dex and $0.08$~dex on the star formation rate and magnetic field, respectively. For comparison,  Heesen et al. (2014) found $\log B \propto 0.3 \, \log (SFR)$
for integrated values of local spiral galaxies, whereas Tabatabaei et al. (2013a) found $\log B \propto 0.14 \, \log (SFR)$ 
within NGC~6946.
There is no clear dependence of the average magnetic field on the integrated star formation rate in our galaxy sample. 
One explanation for the absence of a clear correlation is a possible enhancement of the total magnetic field and of
the ratio between the ordered and turbulent magnetic fields $(B_{\rm ord}/B_{\rm turb})$ of Virgo cluster galaxies due to environmental interactions. 
In the Sect.~\ref{sec:discussion}, we will show that there is evidence for an increase of $B_{\rm turb})$ on the local star formation rate ($\dot{\Sigma}_*$).

\section{Method \label{sec:method}}

The lengthscale of cosmic ray electron diffusion is given by $l_{\rm diff}=\sqrt{4 D \, t_{\rm syn}}$ (see Sect.~\ref{sec:diffcoeff}). 
The synchrotron timescale $t_{\rm syn}$ is given by
Eq.~\ref{eq:synch}. The diffusion coefficient might be constant or proportional to the ratio between the ordered and turbulent magnetic 
field (Chuvilgin \& Ptuskin 1993, Breitschwerdt et al. 2002) $D \propto (B_{\rm ord}/B_{\rm turb})^2$. 
Tabatabaei et al. (2013b) indeed found that the diffusion length scales with the ratio $(B_{\rm ord}/B_{\rm turb})$ (their Fig.~12).
For a power law dependence of the magnetic field on the star formation rate $B \propto \SFR^j$ the diffusion lengthscale is
\begin{equation}
\label{eq:ldiff1}
l_{\rm diff}   \propto (B_{\rm ord}/B_{\rm turb}) \, \SFR^{-0.75\,j} \nu^{-0.25}\ .
\end{equation}
If the ratio between the ordered and the turbulent magnetic fields depends on the star formation rate via a power law
$(B_{\rm ord}/B_{\rm turb}) \propto \SFR^{-m}$ (as observed by Stil et al. 2009) the expression for the diffusion lengthscale becomes
\begin{equation}
\label{eq:pdegree}
l_{\rm diff} \propto  \SFR^n \nu^{-0.25} \propto \SFR^{-(0.75\,j+m)} \nu^{-0.25}\ .
\end{equation}

If streaming dominates the transport of cosmic ray electrons, the lengthscale is $l_{\rm stream}=v_{\rm A} \, t_{\rm syn}$,
where $v_{\rm A}=B/\sqrt{\rho_i}$ is the Alfv\'en velocity and $\rho_i$ the ionized gas density.
Assuming energy equipartition between the magnetic field and the turbulent kinetic energy $\rho v_{\rm turb}^2=B^2/(8 \pi)$
and a constant gas velocity dispersion $v_{\rm turb}$, the streaming length scale becomes
\begin{equation}
l_{\rm stream} \propto  \SFR^n \nu^{-0.5} \propto \SFR^{-1.5 \, j} \nu^{-0.5}\ .
\end{equation}
Streaming occurs preferentially along the anisotropic component of the small-scale or the ordered large-scale magnetic field lines 
which are traced by the polarized radio continuum emission.

If diffusion is the dominating transport mechanism, a Gaussian convolution of the source term, i.e. the star formation map, is
expected. If streaming dominated the cosmic ray electron transport, the convolution kernel is expected to be an exponential.
This is justified by the results of Heesen et al. (2016), who found that within their 1D cosmic ray transport model advection leads 
to approximately exponential radio continuum intensity profiles, whereas diffusion leads to profiles that can be better approximated by Gaussian functions.
Since advection and streaming are formally similar (the outflow velocity has to be replaced by the Alfv\'en velocity),
the effect of streaming can be approximated by a convolution with an exponential kernel.

The exponent $j$ might vary between $0.3$ (Heesen et al. 2014) and $0.5$. The latter value is based on the model presented in 
Vollmer \& Leroy (2011) and assuming energy equipartition between the magnetic field and the turbulent kinetic energy.
It is close to the value found in Sect.~\ref{sec:integ} (Fig.~\ref{fig:sfr_Bfield}).
For the simple case $m=0$, we thus obtain
\begin{eqnarray} \nonumber
l_{\rm diff} & \propto & \SFR^{-(0.23-0.38)} \nu^{-0.25}\\ 
l_{\rm stream} & \propto & \SFR^{-(0.45-0.75)} \nu^{-0.5}\ . 
\label{eq:akernel}
\end{eqnarray}
If $m$ is non-zero, the dependence of the diffusion lengthscale on the star formation becomes steeper.
The exponents of the star formation rate and the frequency are thus well separated between the cases of cosmic ray electron
diffusion and streaming if $m$ is not too large. The absolute values of both exponents are always smaller in the case of diffusion.
If the magnetic field were independent of the star formation rate, i.e. it is generated by other mechanisms than the
small-scale turbulence induced by supernova explosions, an exponent $n$ close to zero would be expected for both cases. 
If the smoothing lengthscales are known at two frequencies, the different frequency-dependences in Eq.~\ref{eq:akernel} lead to 
different ratios between the smoothing lengthscales: $l_{\rm 6cm}/l_{\rm 20cm}=1.34$ for diffusion and $l_{\rm 6cm}/l_{\rm 20cm}=1.81$ for streaming.
It is expected that the ratio between the smoothing lengthscales is consistent with the smoothing kernel (Gaussian or exponential). 
We can thus hope to separate diffusion as the dominant transport mechanism from streaming in our galaxies based on:
(i) the smoothing kernel (Sects.~\ref{sec:bestfit} and \ref{sec:goodness}), (ii) the ratio $l_{\rm 6cm}/l_{\rm 20cm}$ (Sect.~\ref{sec:lrat}), and 
(iii) the ratio ($B_{\rm ord}/B_{\rm turb}$) (Eq.~\ref{eq:ldiff1}; Sect.~\ref{sec:ratiord}).

To produce the predicted radio continuum maps, we convolved the star formation maps with Gaussian and exponential functions 
\begin{equation}
f(r) \propto \exp(-(r/l)^2);\ \ f(r) \propto \exp(-(r/l))
\label{eq:kernel}
\end{equation}
using an adaptive kernel with the star formation dependence $l=l_0 \, (\dot{\Sigma}_{*}/\dot{\Sigma}_{*,\ 0})^n$ (Eq.~\ref{eq:akernel}) with 
$\dot{\Sigma}_{*,\ 0}=8 \times 10^{-3}$~M$_{\odot}$yr$^{-1}$kpc$^{-2}$.
This star formation rate is characteristic for the outer edges of spiral arms in the galaxies of our sample.
Spiral arm generally show $\dot{\Sigma}_{*} > \dot{\Sigma}_{*,\ 0}$, whereas interarm regions have $\dot{\Sigma}_{*} < \dot{\Sigma}_{*,\ 0}$.

To determine the smoothing lengthscale at a given star formation rate, we convolved the ``observed'' star formation rate
with kernels of different $l$ and calculated the goodness $\phi$ of the fit\footnote{A smaller goodness corresponds to a better fit.} 
following Murphy et al. (2006):
\begin{equation}
\label{eq:phi}
\phi=\frac{\sum (radio_{obs}-radio_{\rm model})^2}{\sum radio_{\rm obs}^2} \ . 
\end{equation}
To obtain a minimum $\phi$, the model radio map has to be normalized with the factor
\begin{equation}
\label{eq:qq}
Q=\frac{\sum \dot{\Sigma}_{*,\ {\rm convolved}}^2}{\sum (radio_{\rm obs} \times \dot{\Sigma}_{*,\ {\rm convolved}})}\ ,
\end{equation}
so that $radio_{\rm model}=\dot{\Sigma}_{*,\ {\rm convolved}}/Q$.

We considered 6 different circular and elliptical smoothing kernels with major and minor half-axis $a$ and $b$.
The kernel lengthscale depends on the local star formation rate according to Eq.~\ref{eq:akernel}.
\begin{enumerate}
\item
circular constant smoothing kernel: \\ $a=b=l_0$;
\item
circular variable smoothing kernel:  \\ $a=b=l_0 (\SFR/\dot{\Sigma}_{*,\ 0})^{0.225}$;
\item
circular variable smoothing kernel:  \\ $a=b=l_0 (\SFR/\dot{\Sigma}_{*,\ 0})^{0.5}$;
\item
elliptical constant smoothing kernel. This is motivated by the expectation that the cosmic ray electrons are transported
preferentially along the field lines of the ordered magnetic field:  \\ $b=l_0$, $a=b/(1-p/0.75)$;
the ellipse is rotated so that its major axis has the same orientation as the ordered magnetic field.
The field orientation is determined from the polarized radio continuum emission.
\item
elliptical variable smoothing kernel:  \\ $b=l_0 (\SFR/\dot{\Sigma}_{*,\ 0})^{0.225}$, $a=b/(1-p/0.75)$;
\item
elliptical variable smoothing kernel:  \\ $b=l_0 (\SFR/\dot{\Sigma}_{*,\ 0})^{0.5}$, $a=b/(1-p/0.75)$.
\end{enumerate}
For NGC~6946 and M~51 we also calculated models with exponents $0.7$ to $1.2$.

Within our method we neglect the mean free path of the UV photons, i.e. between the young stars and the dust which absorbs and reradiates the 
UV/optical photons. We thus assume that the average UV path length is generally less than the diffusion length of relativistic electrons.

In the presence of a galactic wind, launched from sites of high local star formation rates, cosmic ray electrons 
can be lost into the halo by advection. In the absence of a detailed treatment of the cosmic ray electron distribution via
the diffusion energy-loss equation (e.g., Eq.~1 of Mulcahy et al. 2016), we model this effect by a loss term which is directly subtracted from the source term,
i.e. the star formation map. The model radio map becomes thus $radio_{\rm model}=(\SFR-loss)/Q$.
Without a physical model of the galactic wind, the loss rate is assumed to have the form of a power law with respect to
star formation: 
\begin{equation}
\label{eq:loss}
loss=10^{-3} \times c \times (10^3 \times \SFR)^k \ . 
\end{equation}
This is an ad hoc prescription which mainly removes the peaks of the emission distribution, e.g. spiral arms, for high $k > 1$.
The dependence of the loss rate on the synchrotron lifetime is contained within the constant $c$.
The normalization factor and the exponent are varied in a systematic way: $k=1.0$-$2.0$ and $c=0.01$-$0.08$.
The constant $c$ is adapted to $k$ to produce losses between a few and $\sim 50$\,\% of the total emission.

Heesen et al. (2018) and Davies et al. (2019) found the following relation between the wind velocity and the
local star formation rate: $v_{\rm wind} \propto \dot{\Sigma}_*^{0.4}$. The mass outflow rate per area is the
$\dot{\Sigma} \sim \Sigma v_{\rm wind}/H$, where $H$ is the height of the gas disk. With a Schmidt-Kennicutt law
of the form $\dot{\Sigma}_* \propto \Sigma^{1.0-1.5}$ and a constant disk height, one obtains $\dot{\Sigma} \sim \dot{\Sigma}_*^{1.4-1.9}$.
We thus expect $1.4 \le k \le 1.9$ and this is what we found (Tables~\ref{tab:table_6cm} and \ref{tab:table_20cm}).

In total we carried out $12$ convolution series within which $l_0$, $k$, and $c$ were varied.
The derived smoothing lengthscales can well be smaller than the beam size (FWHM). This is possible, because
the rms error of a source position is given by $\Delta \alpha = \Delta \delta \sim 0.6\,({\rm S/N})^{-1}$\,FWHM\footnote{This is valid for
a symmetric beam shape and no sidelobes, which is the case for our data.} (Eq.~{B2} of Ivison et al. 2007).
As a consequence, beam shifts of the same order can also be detected and this is what is done by our method. 
For (S/N)$ \sim 5$ the source position can be determined with an accuracy better than $0.12$\,FWHM. For a Virgo cluster
distance of $17$~Mpc and FWHM$=18''$ this leads to an accuracy better than $\sim 180$~pc compared to FWHM$=1.5$~kpc.

In reality, the convolution of the star formation distribution should be performed in three dimensions. However, the vertical
dimension is not accessible in our sample of face-on galaxies. We therefore have to assume that diffusive losses in the
vertical direction are not significant (except in the radio-deficient galaxy NGC~4535). This assumption is motivated by
the ubiquitous existence of radio continuum halos in edge-on spiral galaxies (Krause et al. 2018). The cosmic ray electrons thus are 
expected to diffuse into the halo, where they radiate away most of their energy. 
However, Mulcahy et al. (2016) found a diffusive vertical loss in M~51 by fitting the radial profile of the spectral index between $151$~MHz and $1.4$~GHz.
We argue that the loss is only significant at a frequency of $151$~MHz, is small at $1.4$~GHz, and negligible at $5$~GHz.

\section{Tests \label{sec:tests}}

Before the application of the adaptive smoothing kernel to observations, we tested our method on simple model data.
We generated the star formation map of a toy model galaxy with two symmetric spiral arms and orderedly spaced point sources of different flux densities.
The ``perfect'' model radio continuum map was obtained by convolving the model star formation map with an adaptive kernel whose
lengths scale $l$ is proportional to the star formation rate $l=a \, \SFR^n$.
Finally, we convolved the model star formation and ``perfect'' radio continuum maps with a Gaussian kernel of constant
lengthscale to obtain the ``observed'' maps. 

\begin{figure*}[!ht]
  \centering
  \resizebox{18cm}{!}{\includegraphics{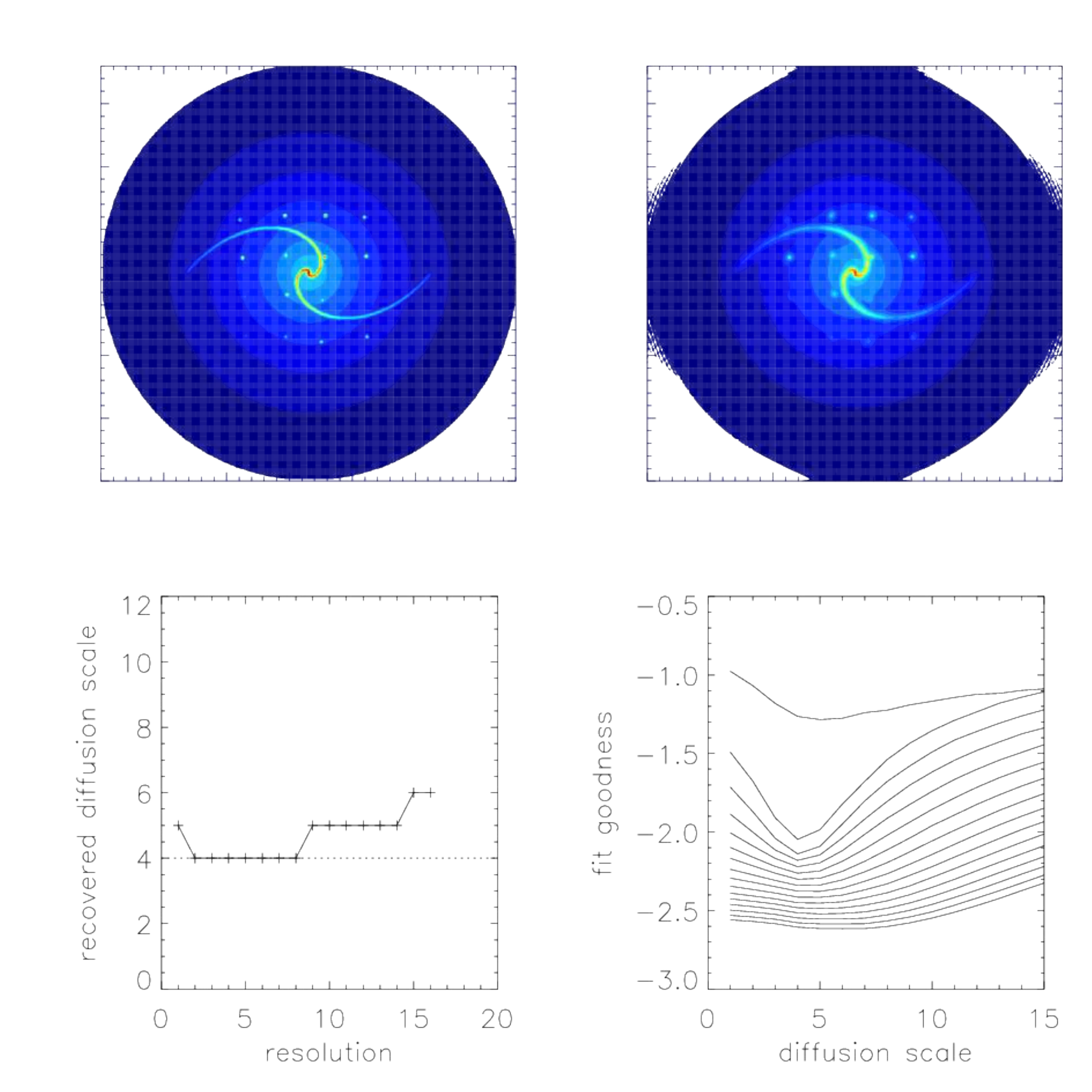}\includegraphics{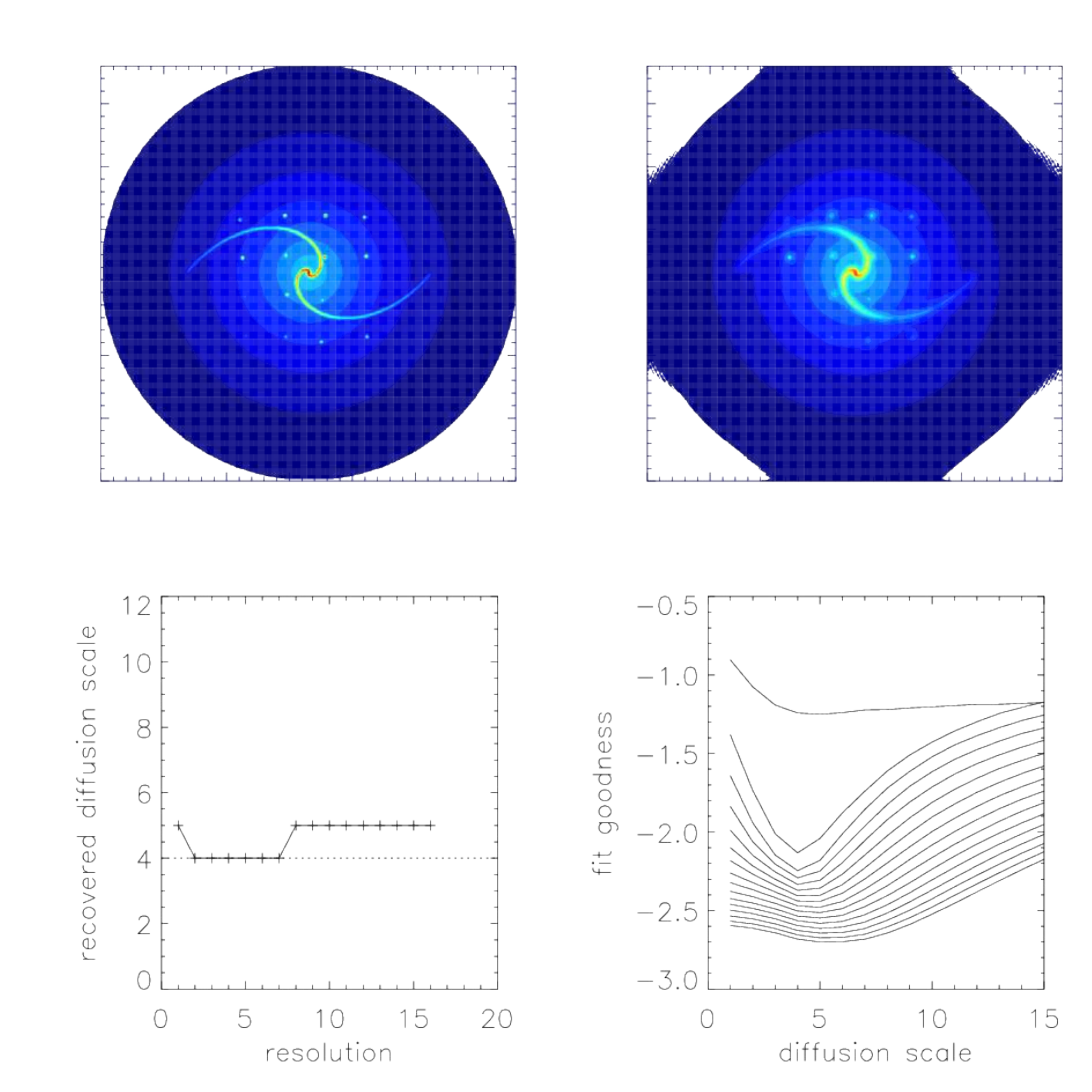}}
  \resizebox{18cm}{!}{\includegraphics{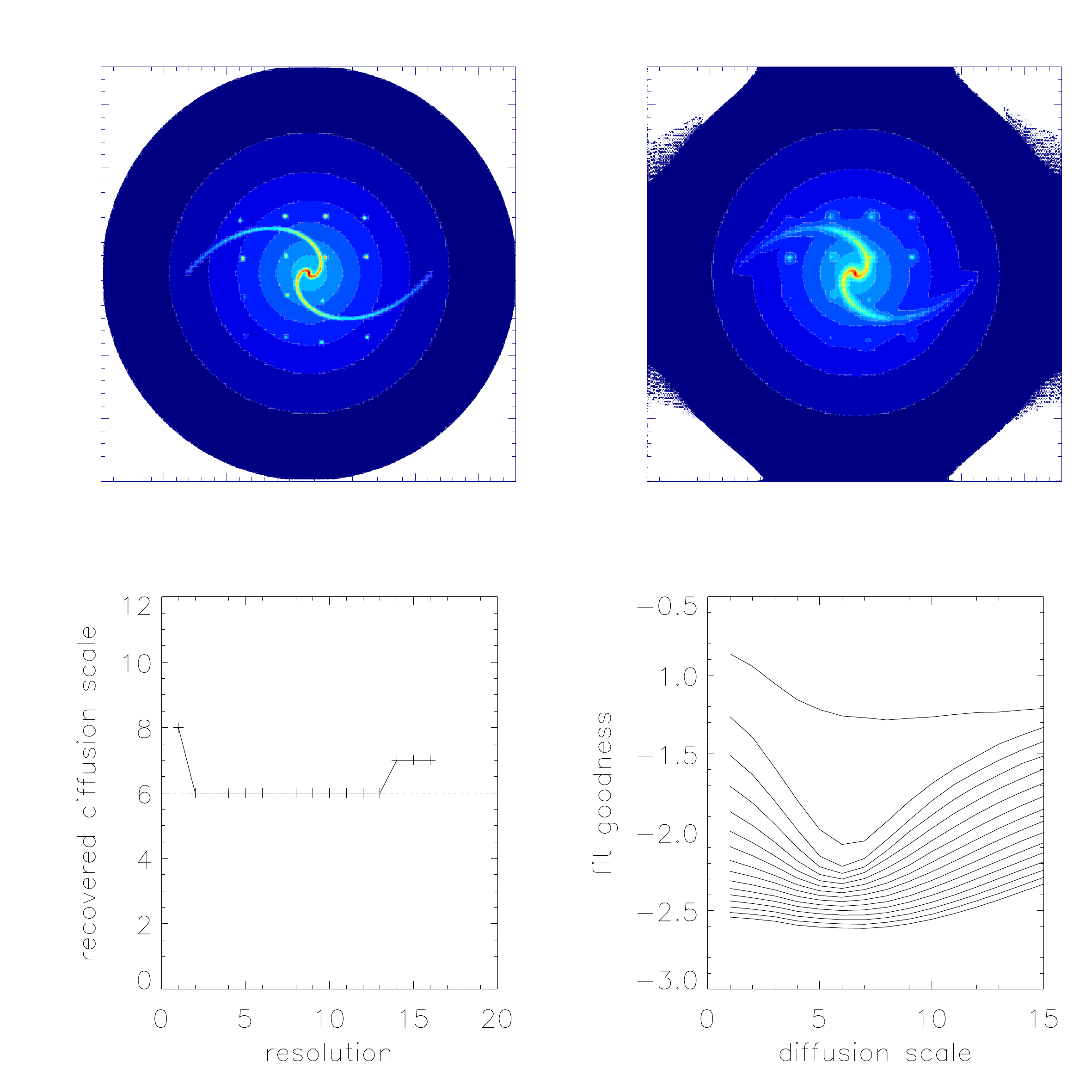}\includegraphics{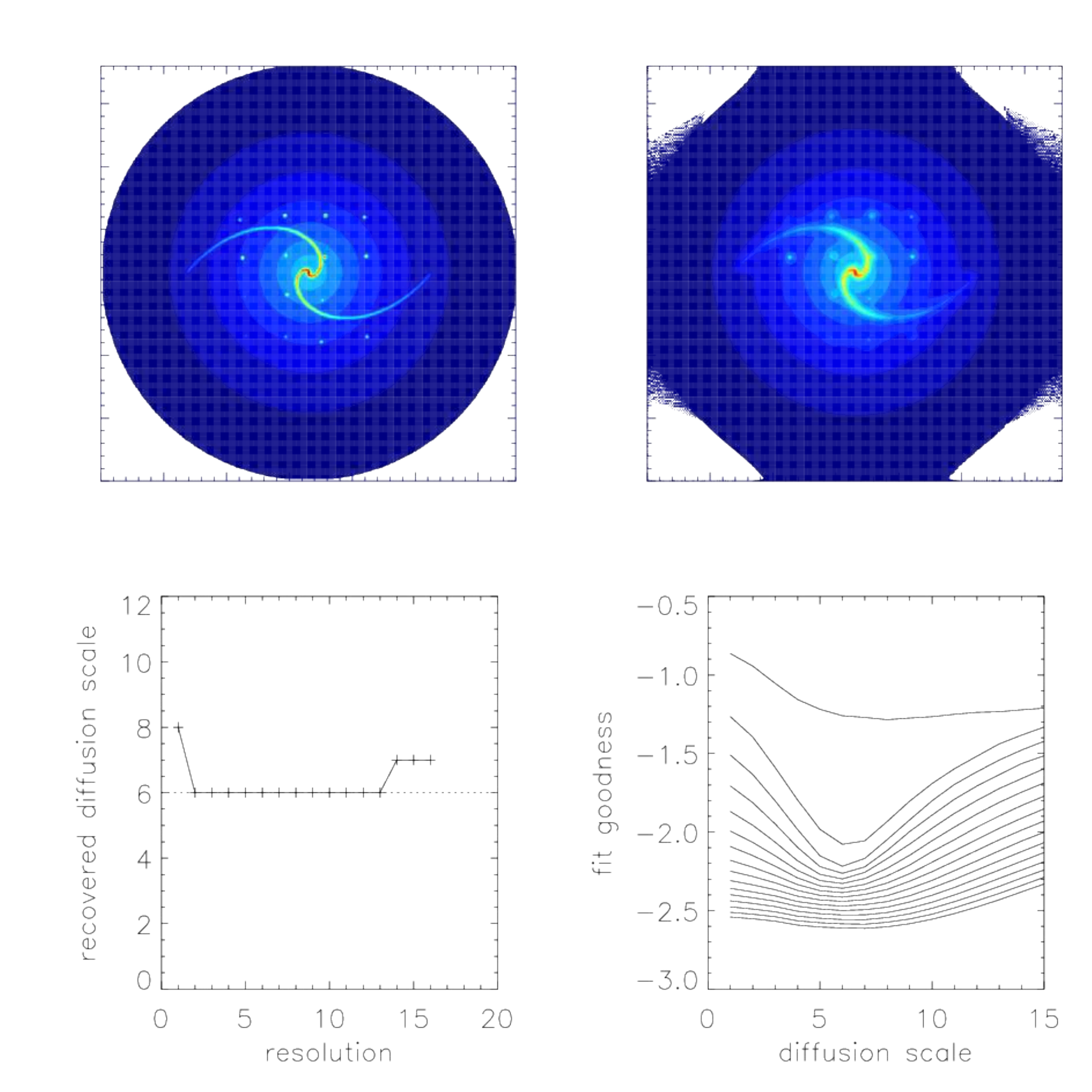}}
  \caption{Adaptive Gaussian kernel (left row) and exponential (right row) convolution of model star formation maps.
    Smoothing lengthscale at a given local star formation rate equals 4 pixels (upper four panels) and 6 pixels 
    (lower four panels). Upper left: ``perfect'' star formation distribution; upper right: ``observed'' star formation distribution;
    lower left: recovered (solid line) and injected (dotted line) lengthscale of the adaptive Gaussian kernel.
    The resolution and diffusion scale correspond to $1.67 \, FWHM$.
  \label{fig:testgal_adapt_4b}}
\end{figure*}

The ``perfect'' star formation maps were convolved with Gaussian kernels of lengths of 4, 6, 8, and 10 pixels to obtain
the ``perfect'' radio maps.
These ``perfect'' maps were then convolved with Gaussian and exponential kernels of lengths between 2 and 16 pixels.
Fig.~\ref{fig:testgal_adapt_4b} and \ref{fig:testgal_adapt_8b} show the results of our tests.

For both kinds of kernels we observe that the curve of goodness shows less pronounces minima for maps of lower resolution.
Between 2 and 10 pixels we can recover the lengthscale of the Gaussian/exponential smoothing kernel at a given
local star formation rate within $10$-$20$~\%. 

As an additional test, we used the ``perfect'' model radio continuum map obtained with a Gaussian convolution and determined the
best-fitting exponential kernel with which the star formation map is convolved. Conversely, a  ``perfect'' model radio continuum map 
obtained with an exponential convolution was used to determine the best-fitting Gaussian kernel with which the star formation map is convolved.
It turned out the lengthscale of the best-fitting exponential is half of that of the Gaussian kernel, because the half-light
radius of an exponential is half that of a Gaussian light profile.

\begin{figure*}[!ht]
  \centering
  \resizebox{14cm}{!}{\includegraphics{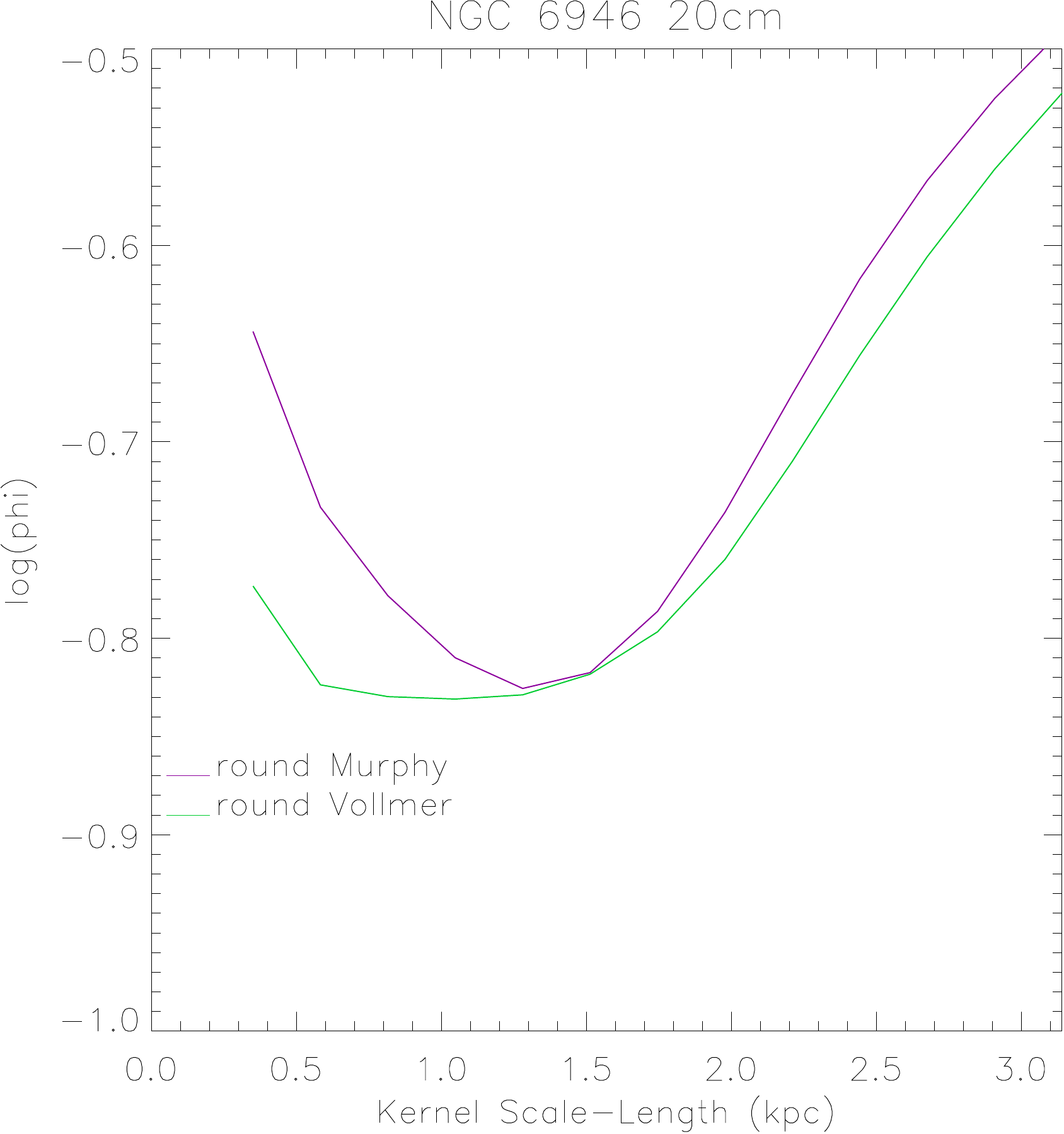}\includegraphics{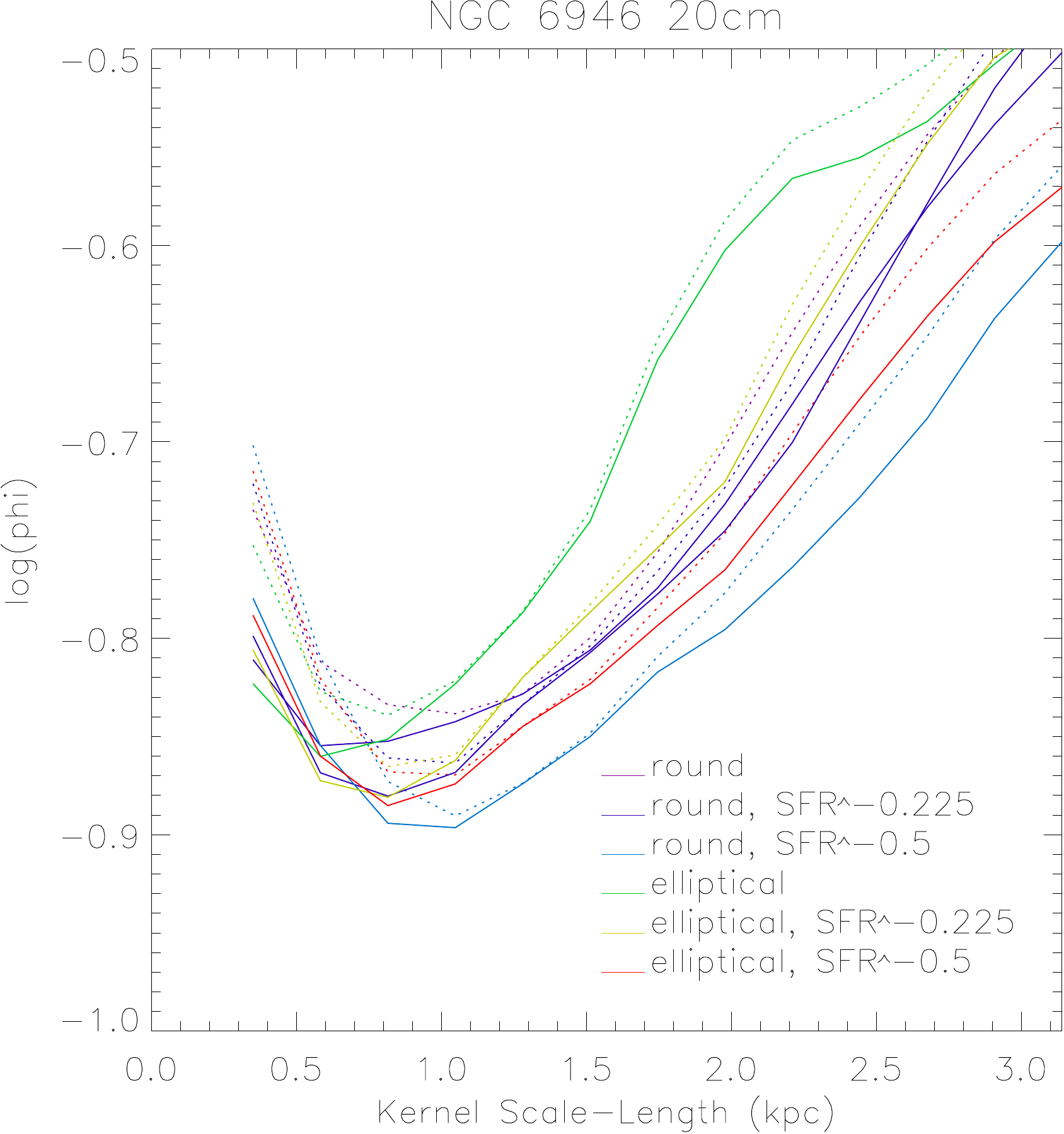}}
  \resizebox{14cm}{!}{\includegraphics{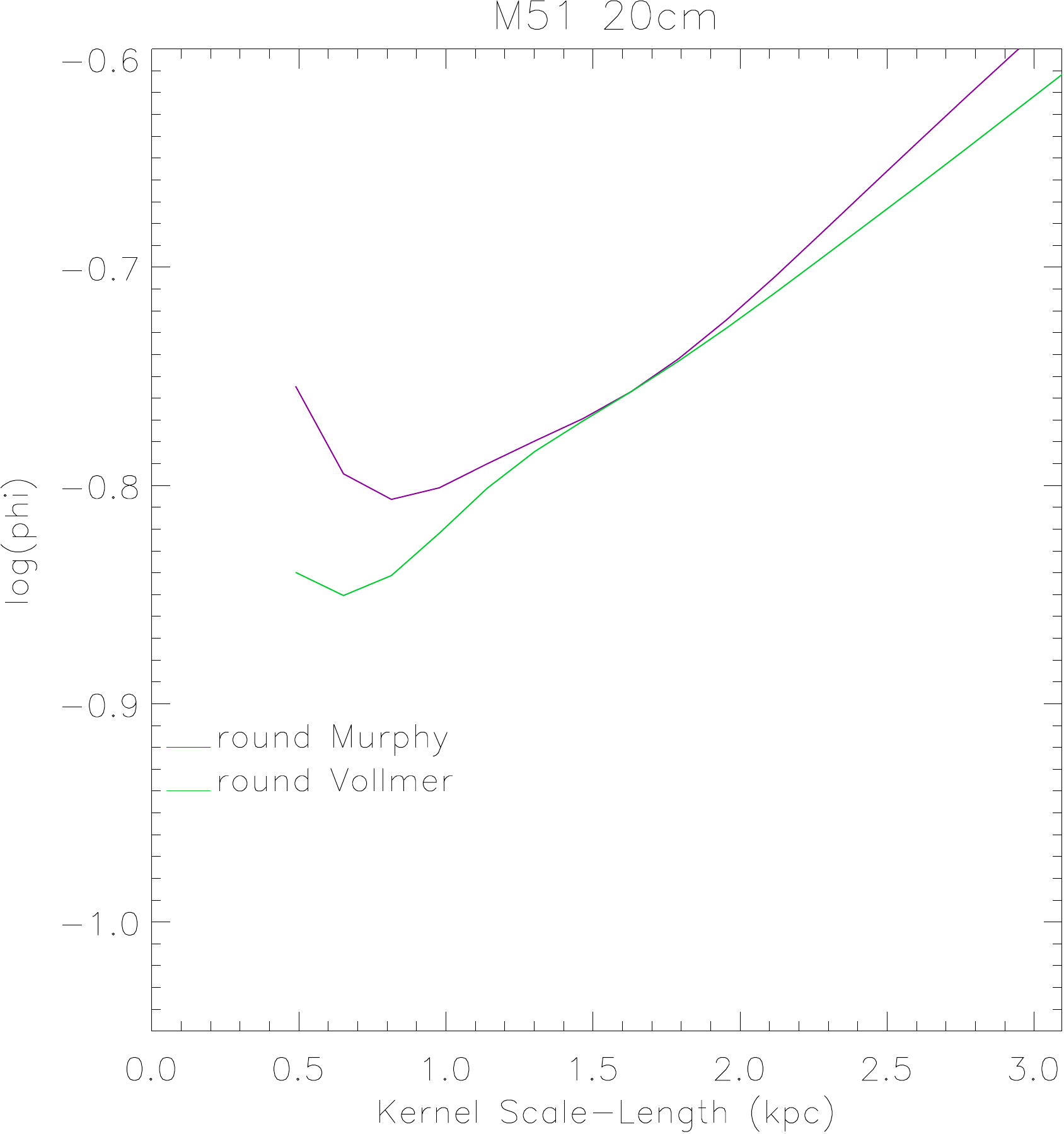}\includegraphics{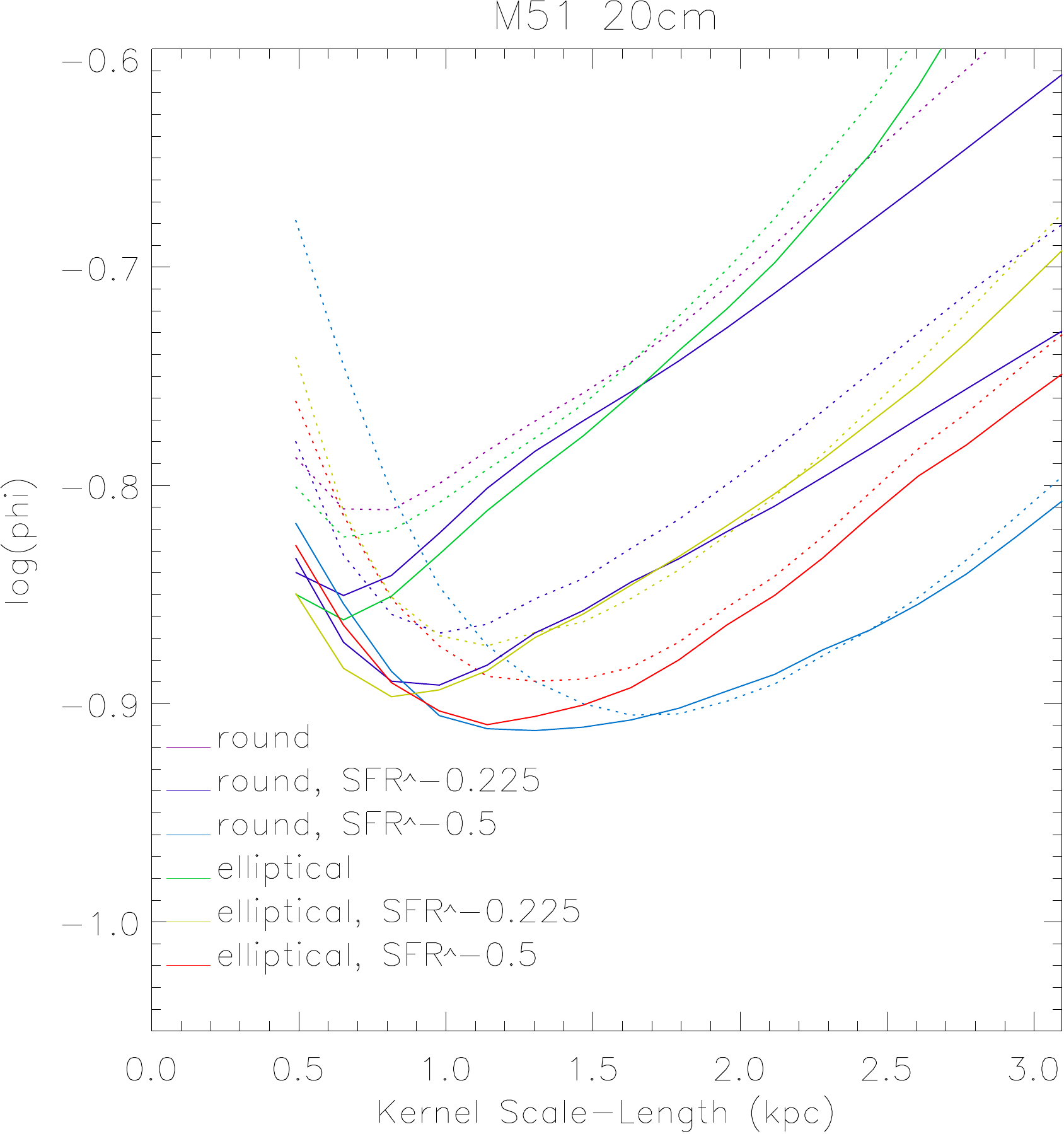}}
    \caption{Left-hand panels: exponential convolutions of star formation maps using the Murphy et al. (2006) and our normalization $Q$ (Eq.~\ref{eq:qq}).
      Right-hand panels: exponential convolution using adaptive round and elliptical kernels.
  \label{fig:graphn6946_20_murphy1}}
\end{figure*}

In a second step we compare our constant exponential smoothing lengthscales with those of Murphy et al. (2006) for the two closest galaxies
that we have in common: NGC~6946 and M~51 (Fig.~\ref{fig:graphn6946_20_murphy1}). 
One has to keep in mind that Murphy et al. (2006) used the Spitzer $70~\mu$m
maps a proxy for the star formation map. An additional difference between our method and that of Murphy et al. (2006)
is the normalization of the model radio map. Whereas Murphy et al. (2006) used $Q=\frac{\sum \dot{\Sigma}_{*,\ {\rm convolved}}}{\sum radio_{\rm obs}}$,
we used Eq.~\ref{eq:qq} which minimizes the goodness ($\phi$, Eq.~\ref{eq:phi}).
The two graphs on the left-hand side of Fig.~\ref{fig:graphn6946_20_murphy1} show that the lengthscales using the two normalizations $Q$
are well comparable, but the lengthscales with the $\phi$-minimizing normalization are smaller than those with the normalization of
Murphy et al. (2006). A comparison between our smoothing lengthscales and those of Murphy et al. (2006, 2009) can be found in Sect.~\ref{sec:discussion}.

The goodnesses for the convolutions with SFR-dependent smoothing kernels are shown on the right-hand side of Fig.~\ref{fig:graphn6946_20_murphy1}.
Using these adaptive smoothing kernels increases the ``best-fit'' lengthscale and decreases the minimum goodness up to $20$\,\% for NGC~6946 and M~51. 
The decrease of $\phi$ is consistent with the findings of Murphy et al. (2008) who convolved the small- and large-scale structures
of the FIR maps separately.

The model data are noiseless and have a perfect point spread function (PSF). Though our radio continuum data have a circular Gaussian PSF,
they contain noise and areas of stochastically increased and decreased radio continuum emission most probably due to time-delay effects
between star formation and the subsequent radio continuum emission.
Thanks to the proximity of NGC~6946 and M~51, we could degrade the spatial resolution of the radio continuum emission to study its
influence on the derived smoothing length (Sect.~\ref{sec:testsres}).

\section{Results \label{sec:results}}

\begin{figure*}[!ht]
  \centering
  \resizebox{12cm}{!}{\includegraphics{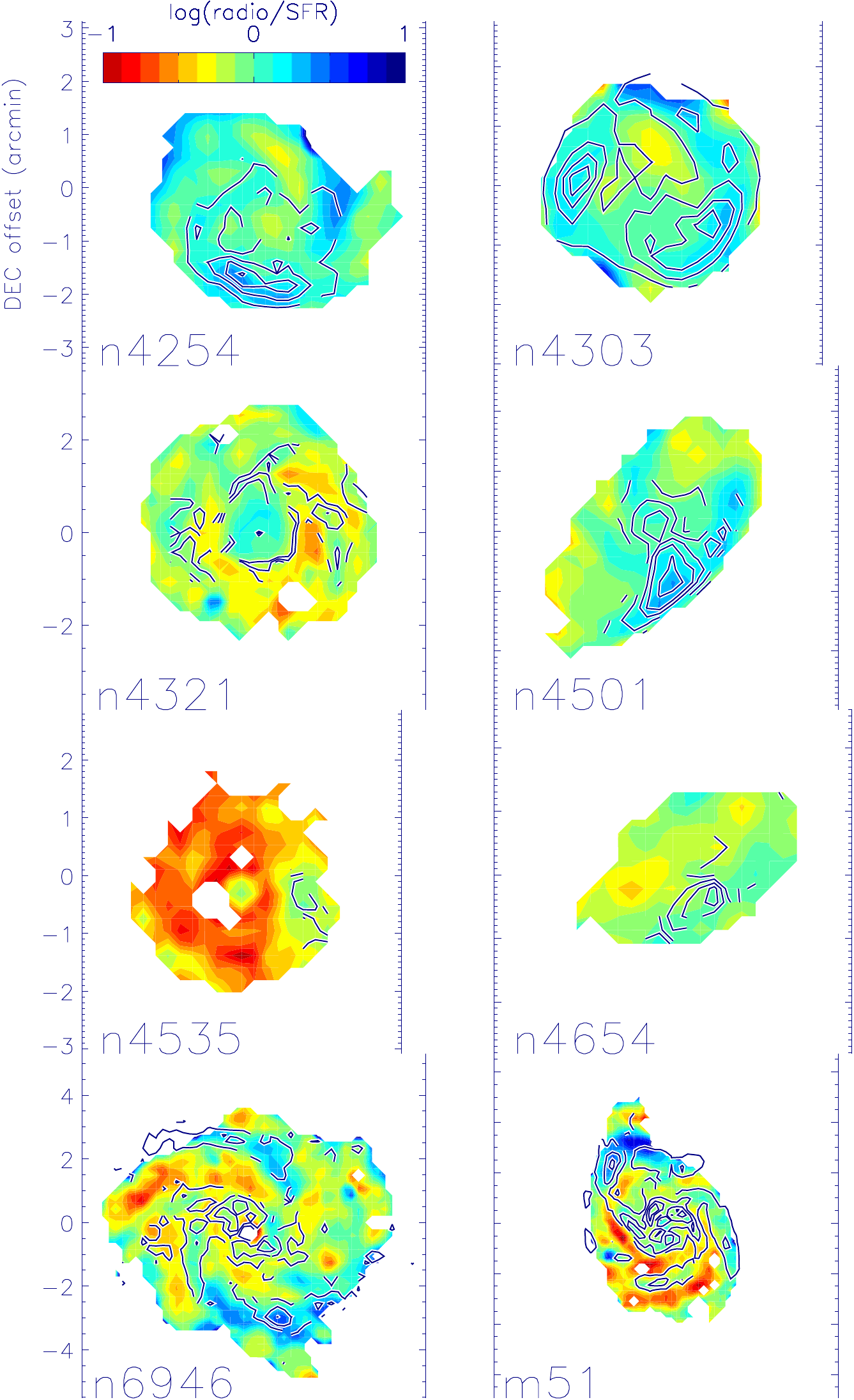}}
    \caption{Color: radio/SFR maps (in arbitrary units). Blue is radio-bright, red is radio-dim. Blue contours: polarized radio continuum emission (Vollmer et al.
      2010, 2013). The contour levels were chosen such that the asymmetric ridges of polarized emission are well visible.
  \label{fig:rcfir_spixx1c1_nice_pol_smoothing}}
\end{figure*}

The adaptive smoothing kernel convolution is applied to the observed star formation maps. Gaussian and exponential kernels are used.
In a second step, losses were applied to the star formation maps.
The obtained model radio continuum maps were then compared to the observed $6$~cm and $20$~cm radio data.
Since we realized that regions of high degree of polarization at $6$~cm are radio-bright with respect to star formation
(Fig.~\ref{fig:rcfir_spixx1c1_nice_pol_smoothing} and Vollmer et al. in prep.),
we decided to enhance the source term (the star formation map) using the degree of polarization $p$:
\begin{equation}
\label{eq:penhance}
\SFR'=\frac{\SFR}{1.0-(p/0.75)}\ .
\end{equation}
Possible physical explanations are (i) ordered magnetic fields generated by a large-scale dynamo (Tabatabaei et al. 2013a), 
(ii) magnetic field compression by large-scale shocks in strong density waves or in interaction regions, or (iii) less cosmic ray losses due to a slower wind.
It turned out that this modification of the source term significantly improved the goodness of the fits 
for NGC~4303, NGC~4535, NGC~4254, NGC~4501, and NGC~4654.

In the following we first present the results for the ``best fit'' models yielding the lowest goodness $\phi$ (Eq.~\ref{eq:phi}).
The ``best fit'' analysis yields unique results for the different model facilitating their visual inspection (Sect.~\ref{sec:bestfit}).
In a statistical approach, the goodness distribution were taken into account (Sect.~\ref{sec:goodness}). 
This allowed us to calculate uncertainties for the parameter determined by our model.

\subsection{The influence of resolution, sensitivity, and the galaxy central region \label{sec:testsres}}

In this section we study the influence of resolution, sensitivity, and the galaxy central region on the results
of our smoothing experiments.

\subsubsection{Resolution}

NGC~6946 and M~51 are located at much smaller distances than the Virgo cluster galaxies ($D=17$~Mpc).
The spatial resolution of $18''$ corresponds to physical distances of $0.5$~kpc, $0.7$~kpc, and $1.5$~kpc for
the distances of NGC~6946, M~51, and the Virgo cluster, respectively.
We thus convolved the radio and SFR images of NGC~6946 and M~51 to obtain the same physical resolution as for the
Virgo cluster galaxies. In principle, our method should be able to recover the same smoothing lengthscale
as those derived from the higher resolution images (Sect.~\ref{sec:tests}).
The results are presented in Tables~\ref{tab:table_bestfitcc} to \ref{tab:table_20cm}.
Since the results for the ``best fit'' models are similar to those taking into account the goodness distribution (Sect.~\ref{sec:results}),
we only discuss the results presented in Tables~\ref{tab:table_6cm_ccn} to \ref{tab:table_20cm}.

Using the low resolution input images does not change the normalization $Q$ in all cases.
However, it changes the smoothing lengthscale $l$. At $20$~cm the smoothing lengthscales of NGC~6946 decrease by $\sim 20$\,\%,
those of M~51 stay constant, with and without advective losses. At $6$~cm without losses the smoothing lengthscales decrease by
$28$\,\% and $17$\,\% for NGC~6946 and M~51, respectively. At $6$~cm with losses the decrease is 
$52$\,\% for NGC~6946 and $9$\,\% for M~51. 

For NGC~4535 we decreased the resolution to the VIVA (Chung et al. 2009) resolution of $27''$.
Surprisingly, the smoothing lengthscale increased by a factor of two at $6$~cm 
with respect to the analysis based on the high resolution images (Table~\ref{tab:table_6cm_ccn}). 
This is due to the changing depth of the input images (see below): the lower resolution results in
the detection of low surface brightness emission in the northern part of the galaxy where a large
smoothing lengthscale is needed. At $20$~cm less diffuse emission is recovered by the VIVA observations, leading to a more modest
increase ($\sim 20$\,\%) of the smoothing lengthscale.

We conclude that an increase of resolution (better resolution) of the input images by a factor $1.5$-$2$ can lead to an increase of
the smoothing lengthscale by $\sim 20$-$30$\,\%. If losses are included, a decrease in resolution can lead 
in extreme cases to a decrease of the smoothing lengthscale by $\sim 50$\,\%.
The ratio between the smoothing lengthscales at $6$~cm and $20$~cm is only significantly modified in the cases where losses
were included (Table~\ref{tab:tables_SI}). Without losses the ratio increases by $15$\,\% for NGC~6946 and M~51.

\subsubsection{Sensitivity}

The ``best fit'' models of NGC~6946 and M~51 at $6$~cm were those models with $l \propto \SFR^n$ and high exponents 
$n \sim 1.5$ (Table~\ref{tab:table_6cm_ccn}).
This means that regions of low local star formation rate have a huge smoothing lengthscale. These regions are located within the interarm regions 
and, most importantly, at the border of the galactic disk. Indeed, there is faint diffuse $6$~cm radio continuum emission observed in NGC~6946 
(symmetric) and M~51 (mainly to the south) that extends well beyond the star formation distribution. 
To test the influence of this faint emission, we clipped the
$6$~cm radio continuum image at the surface brightness of the faintest levels detected in the star formation map.
As expected, the clipping decreased the smoothing lengthscale by $20$-$40$\,\%, but it did not change the normalization $Q$.
NGC~6946 and M~51 have strong ordered magnetic fields (Tabatabaei et al. 2013a, Fletcher et al. 2011) giving
rise to larger diffusion lengthscales (Eq.~\ref{eq:ldiff1}). The radio continuum emission from the interarm regions is quite strong, 
which explains that the clipping did not change the exponent $n$ significantly.

We thus conclude that deep radio continuum images with emission that extends well beyond the optical disk can
lead to an overestimation of the smoothing lengthscale by $20$-$40$\,\%.

\subsubsection{Galaxy central region}

We realized that the radio properties of the central few kpc of NGC~4321 behave differently than those of the galactic disk.
This effect was already noted by Murphy et al. (2008) who decomposed the galaxies into star-forming structures and diffuse disks, 
and looked at the propagation physics for those two components independently. These authors noted that the smoothing lengthscales
were generally shorter for the star-forming structures compared to those of the diffuse disks.
We thus decided to increase the central region which was removed from the input maps.
At $20$~cm this procedure did not change the derived parameters.
However, at $6$~cm the normalization $Q$ increases to the mean value found for the other galaxies (Table~\ref{tab:table_6cm_ccn})
and the smoothing lengthlength decreases by $\sim 40$\,\%. This means that the derived parameters of the model with a 
significant part of the central region included in the analysis was dominated by the central emission.
This seems to be the limit of our rather simple model convolution and seems to indicate that the magnetic field strength
at a star formation rate $\Sigma_{*\,0}$ is lower in the central region than in the disk. In this case the synchrotron emission
timescale and the smoothing lengthscale increase.
We thus conclude that one has to be cautious about including the central region of a galaxy into the smoothing experiments.

\subsubsection{Summary}

The variation of the smoothing lengthscale with resolution is expected to be of the order of $\sim 30$\,\%. If losses are included in the model,
the variation can reach $\sim 50$\,\% in extreme cases. The presence of an extended low surface brightness disk emission can lead to
an overestimation of the smoothing lengthscale by up to $\sim 40$\,\%. The ratio between the smoothing lengthscales at $6$~cm and $20$~cm is robust
with respect to a variation of the smoothing lengthscale (within less than $20$\,\%). It is often preferable to exclude the galaxy central region 
from the input radio continuum image, because the dominant cosmic ray transport mechanism can be different from that of the disk.

\subsection{``Best fit models'' \label{sec:bestfit}}

We applied the convolution series described in Sect.~\ref{sec:method} to the $8$ local spiral galaxies of Table~\ref{tab:gals}.

\subsubsection{Model radio continuum maps}

\begin{figure*}[!ht]
  \centering
  \resizebox{16cm}{!}{\includegraphics{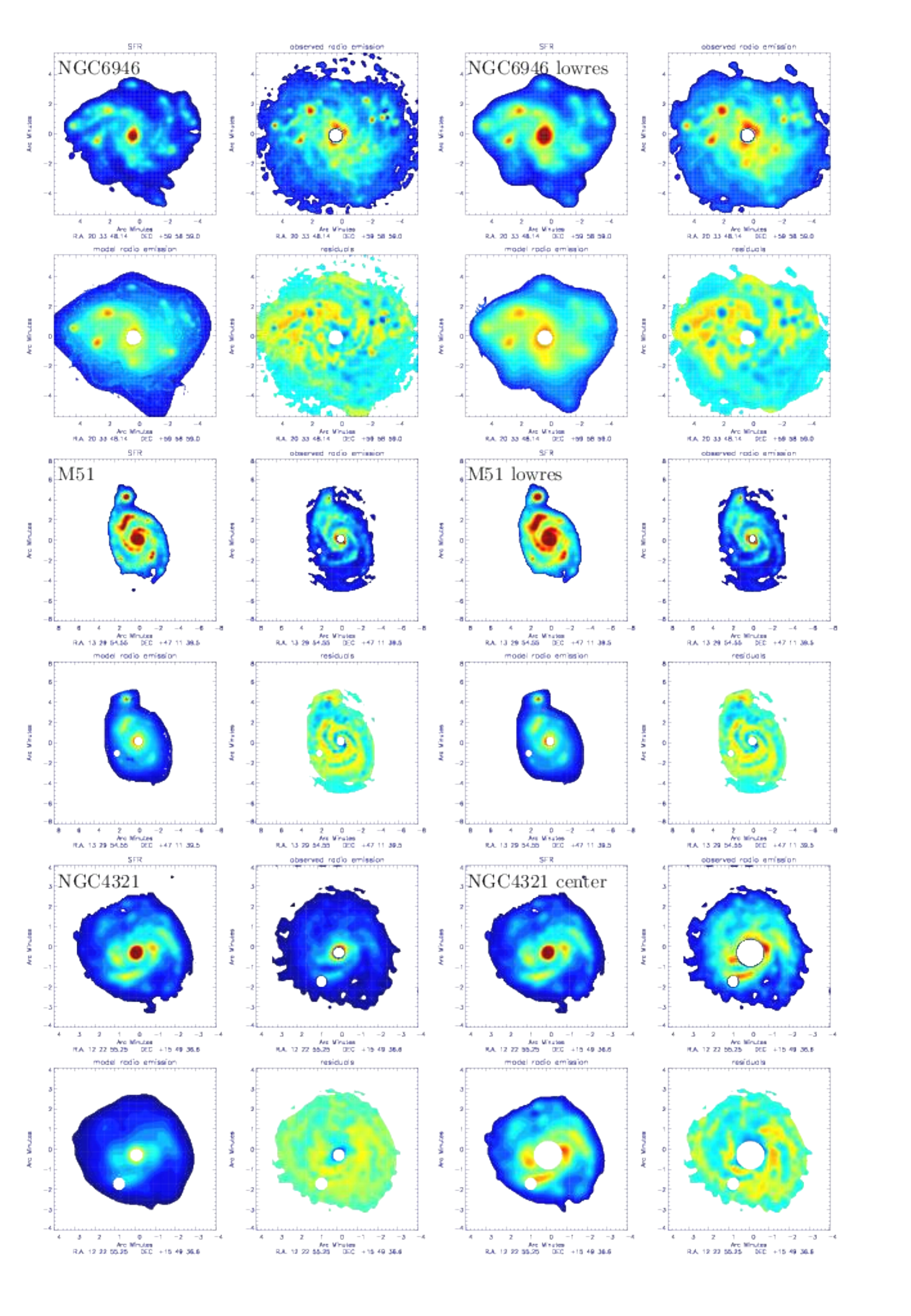}}
  \caption{Gaussian convolution: ``best fit'' model radio continuum maps at $6$~cm. Four maps are shown for each galaxy.
    Upper left: observed star formation; upper right: observed radio continuum emission; lower left: model radio continuum;
    lower right: residuals; blue is radio-bright, red radio-dim.
  \label{fig:zusammenplots1}}
\end{figure*}

The ``best fit'' model radio continuum maps are presented in Fig~\ref{fig:zusammenplots1},  \ref{fig:zusammenplots1exp} 
and Figs.~\ref{fig:zusammenplots2} to \ref{fig:zusammenplots5} for the convolution with Gaussian and exponential kernels, respectively.
The residual maps of the Gaussian and exponential methods are very similar.
The inspection of the residual maps shows important large-scale asymmetries. These asymmetries are similar
for the $6$~cm and $20$~cm data. NGC~6946 has a north-south asymmetry with a radio-bright southern and a radio-dim
northern half of the galactic disk. The disk of M~51 is rather radio-dim, whereas the region between M~51 and its
companion is radio-bright. The northeastern half of the galactic disk of NGC~4321 is radio-bright, whereas the southwestern
half is radio-dim. The southern spiral arm of NGC~4303 is strongly radio-bright, whereas a part of the northern spiral arm is radio-dim.
The disk of NGC~4535 is radio-dim, except in the region of the strongly polarized radio continuum arm (Vollmer et al. 2007), which
is radio-bright. The southeastern half of the galactic disk of NGC~4254 is radio-bright, whereas the northwestern half is radio-dim.
The southwestern side of the galactic disk of NGC~4501 where the asymmetric ridge of polarized emission is located is radio-bright,
the opposite side being radio-dim. The galactic disk of NGC~4654 behaves in a similar way as that of NGC~4501.

We conclude that large-scale asymmetries are observed in the residual emission of all galaxies. If the galaxy has an asymmetric
ridge of polarized emission, the ridge region is always radio-bright. The ring of strong polarized radio continuum emission
in NGC~4303 is also radio-bright. It is surprising to us that in most of the galaxies (NGC~6946, NGC~4321, NGC~4535, NGC~4254, NGC~4501, NGC~4654)
these asymmetries follow the disk's kinematic minor axis, i.e. they are separated by the kinematic major axis.
The distribution of radio-bright and radio-dim regions is not linked to disk orientation, i.e. to whether a disk half is located
in front or behind the galaxy center.

These large-scale asymmetries of radio-bright and radio-dim regions dominate the goodness of the fit
and thus explain the rather small decrease of the goodness ($\sim 10$-$20$\,\%) when the different model convolutions are applied.

\subsubsection{Smoothing lengthscales}

\begin{figure*}[!ht]
  \centering
  \resizebox{\hsize}{!}{\includegraphics{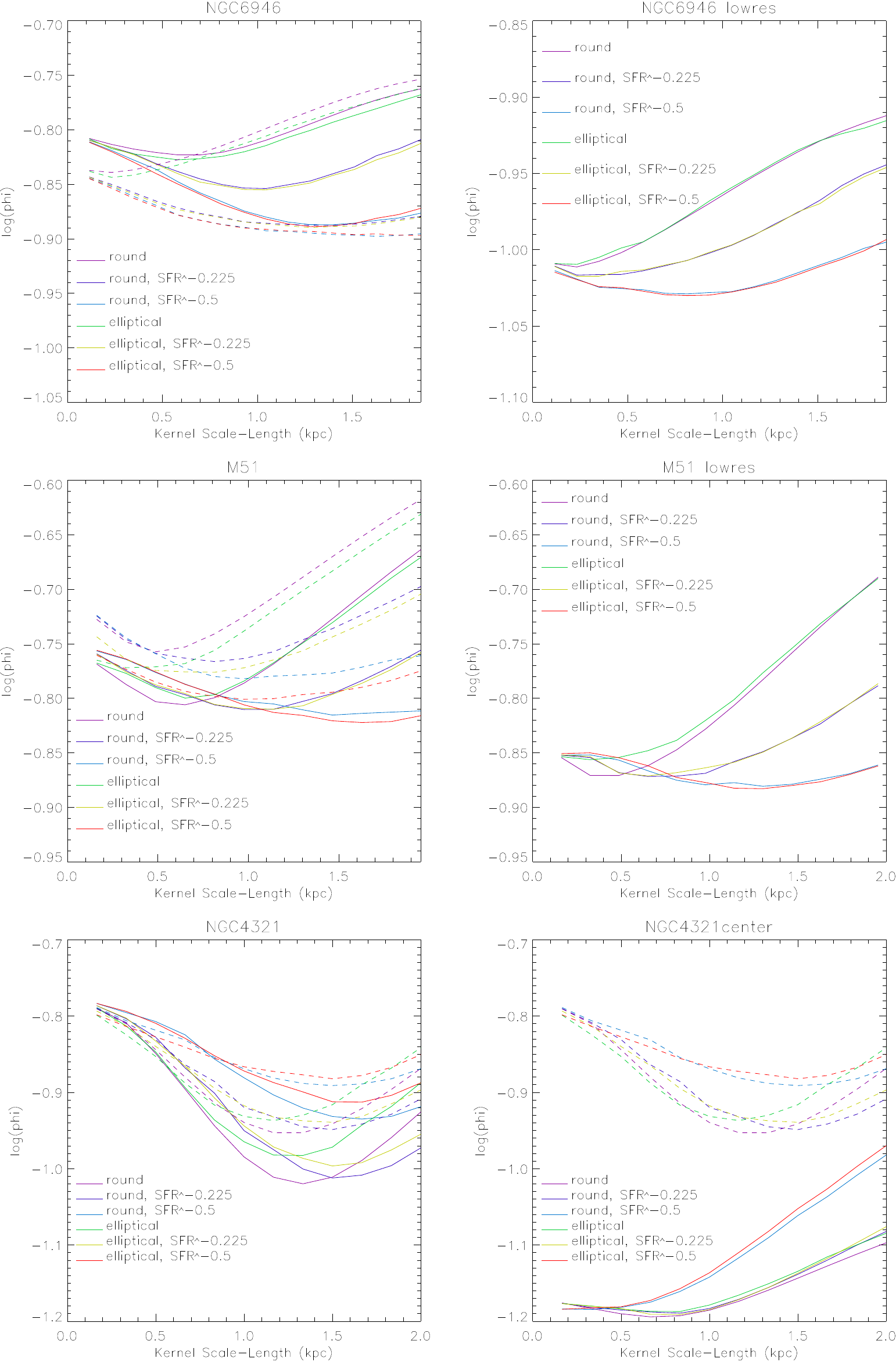}\includegraphics{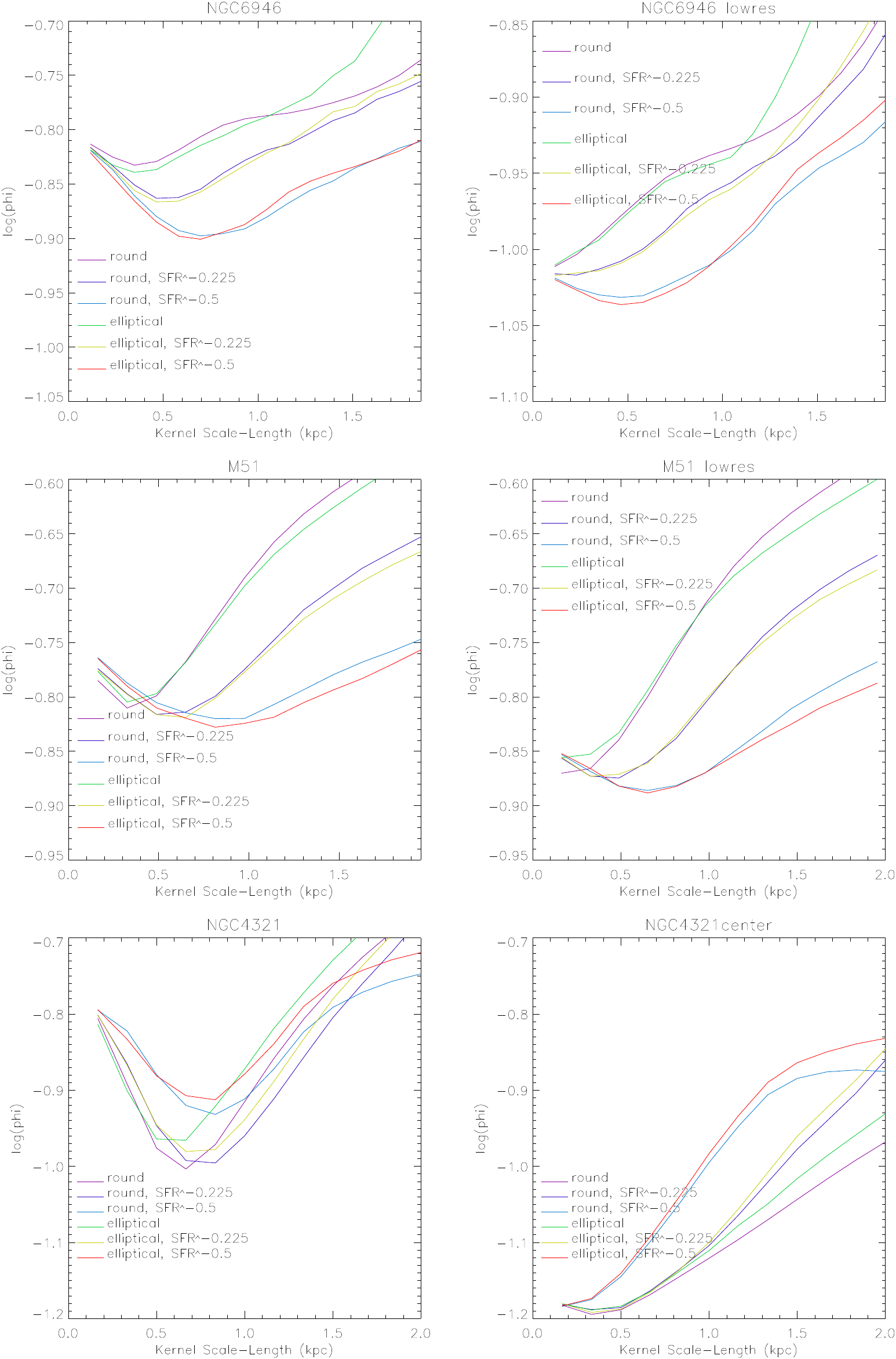}}
  \put(-315,290){Gaussian}
  \put(-445,290){Gaussian}
   \put(-60,290){exponential}
  \put(-190,290){exponential}
  \put(-315,155){Gaussian}
  \put(-445,155){Gaussian}
   \put(-60,155){exponential}
  \put(-190,155){exponential}
  \put(-315,20){Gaussian}
  \put(-445,20){Gaussian}
   \put(-60,20){exponential}
  \put(-190,20){exponential}
  \caption{Curves of goodness $\phi$ associated to the minimum $\phi$ for the $12$ models.
    The $6$~cm radio continuum observations are used. 
    The model convolutions with different kernels are shown in different colors. The models with an enhanced source
    term based on the degree of polarization are shown as dashed lines.
  \label{fig:zusammen1}}
\end{figure*}

\begin{figure}[!ht]
  \centering
  \resizebox{\hsize}{!}{\includegraphics{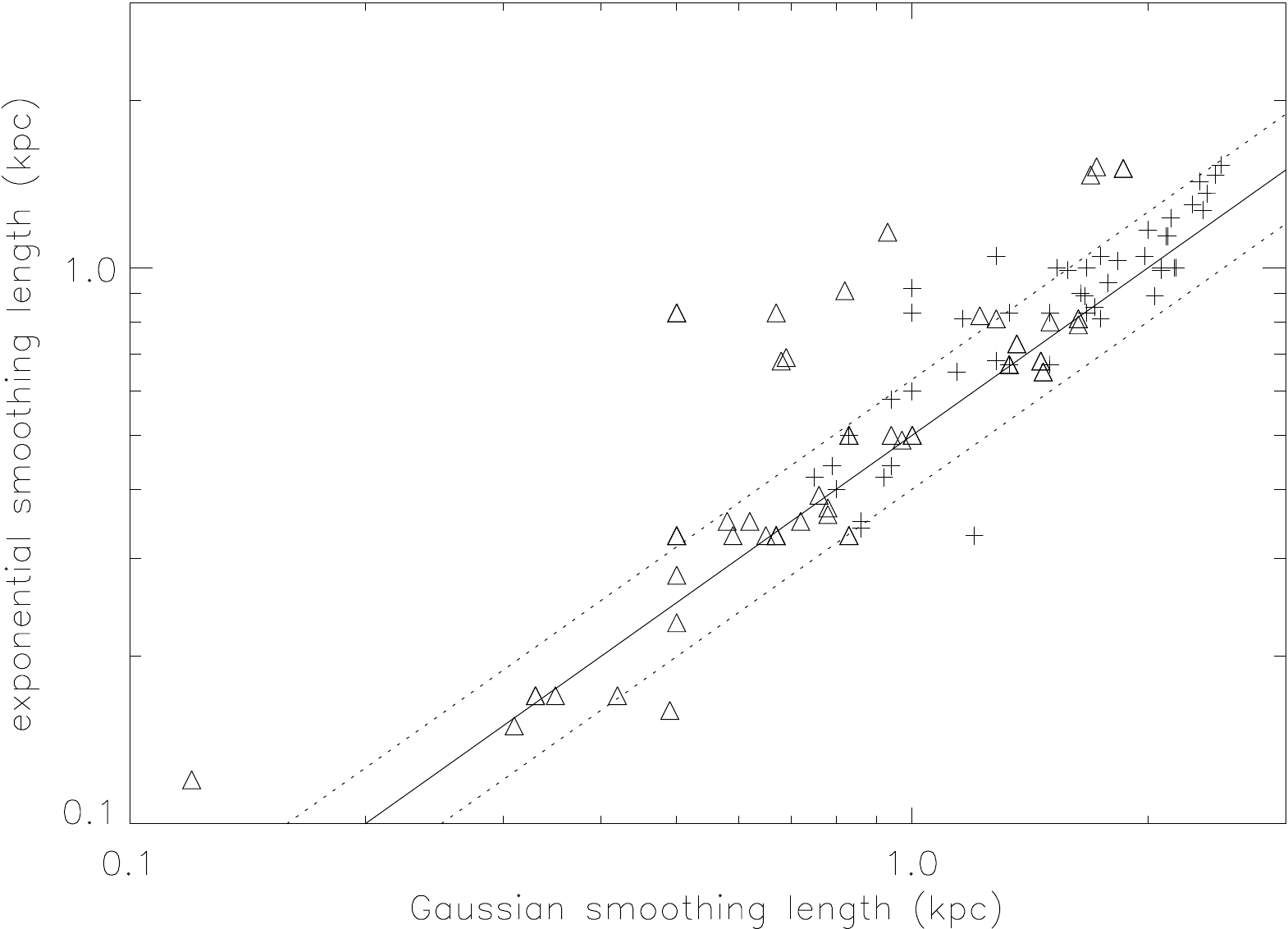}}
  \caption{Comparison between the Gaussian and exponential smoothing lengthscales. Crosses: 6~cm. Triangles: 20~cm.
  \label{fig:smoothinglength206}}
\end{figure}

The resulting curves of goodness $\phi$ associated to the minimum $\phi$ are presented in Figs.~\ref{fig:zusammen1} to \ref{fig:zusammen4}.
Only models without an advective loss term are shown. 
The model convolutions with different kernels are shown in different colors. The models with an enhanced source
term based on the degree of polarization are shown as dashed lines.

The resulting Gaussian and exponential smoothing lengthscales are compared to each other in Fig.~\ref{fig:smoothinglength206}.
Overall, the exponential lengthscale equals half the Gaussian lengthscale with a scatter of $0.1$~dex or $25$\,\%,
because the half-light radius of an exponential is half that of a Gaussian light profile (see Sect.~\ref{sec:tests}).
At $6$~cm the outliers are NGC6946, NGC6946 lowres, M51 lowres, and NGC4654. At $20$~cm the outliers are 
NGC6946, NGC6946 lowres, NGC4303, and NGC4535 VIVA. In all the cases, except NGC4303, the exponential lengthscale
is larger than half of the Gaussian lengthscale.
In addition, the minimum goodnesses of the ``best-fit'' models for the Gaussian and exponential kernels are in most of the
cases well comparable (Fig.~\ref{fig:zusammen1} to \ref{fig:zusammen4}).
These figures also show that the enhancement of the source term (the star formation map) 
using the degree of polarization $p$ leads to the same decrease of the goodness $\phi$ for the Gaussian and exponential kernels.
This is consistent with the statement of Murphy et al. (2006) that exponential kernel ``works as well as
or better than the other kernel types''.
Thus, the minimum goodness of the ``best-fit'' models of the Gaussian and exponential convolutions cannot used
to discriminate between the two transport mechanisms (diffusion or streaming).
In the following we will only comment on the Gaussian lengthscales given that the exponential lengthscales are about two times smaller.

\subsubsection{Source term enhancement based on polarization}

At $6$~cm and $20$~cm the models with an enhanced source term based on the degrees of polarization lead to lower $\phi$
for NGC~4303, NGC~4535, NGC~4254, and NGC~4501. The last three galaxies show indeed asymmetric ridges of polarized
radio continuum emission which are caused by the compression of the magnetic fields by ram pressure (Vollmer et al. 2007). 
NGC~4303 shows a strongly polarized asymmetric ring structure (Vollmer et al. 2013).
For NGC~4654, only for the $20$~cm model an enhanced source term is preferred. In general, the convolution of the star formation maps
leads to a decrease of the fit goodness $\phi$ by $\sim 0.1$~dex or $25$\,\%. The improvement of the goodness due to the enhancement 
of the source term based on polarization is more important than that due to the variation of the smoothing lengthscale.

\subsubsection{Typical smoothing lengthscales}

The minimum values of $\phi$, the associated smoothing lengthscales, and the ratio between the model and observed radio
continuum maps at $6$~cm $Q=radio_{\rm model}/radio_{\rm obs}$ are presented in Tables~\ref{tab:table_bestfitcc} and \ref{tab:table_bestfitc}.
In the case of a constant smoothing kernel (Table~\ref{tab:table_bestfitcc}) the minimum goodness varies between 
$\phi_{\rm min}=0.028$ (NGC~4254) and $\phi_{\rm min}=0.16$ (NGC~6946).
The Gaussian smoothing lengthscales range between $0.42$~kpc (NGC~4535) and $1.33$~kpc (NGC~4321). However, the latter large 
smoothing lengthscale seems to be mostly needed for the galaxy center. Once the center excluded, the smoothing lengthscale is $1.00$~kpc.
We thus find smoothing lengthscales between $0.4$ and $1$~kpc. Berkhuijsen et al. (2013) found a comparable smoothing lengthscale for M~31.

The normalization constant $Q$ (Eq.~\ref{eq:qq})\footnote{The unit of $Q$ is $167$\,M$_{\odot}$kpc$^{-2}$/(MJy/sr).}
is different from SFR/L$_{\rm radio}=\sum \dot{\Sigma}_{*,\ {\rm convolved}} / \sum radio_{\rm obs}$, because it
is weighted towards and thus sensitive to regions of high $\dot{\Sigma}_{*,\ {\rm convolved}}$. The comparison between both
values is thus a sanity check for our model. The values of $Q$ 
lie in the range between $17$ and $19$ for half of the galaxies. Such a narrow range is expected for galaxies which follow the radio-SFR correlation. 
However, NGC~4303, NGC~4254, and NGC~4501 show a higher radio continuum emission than expected and NGC~4535 shows
a much lower radio continuum emission than expected. This is consistent with their location in the $L_{\rm radio}$--SFR plot
(Fig.~\ref{fig:sfrradc_int_small}).

The typical Gaussian smoothing lengthscale at $20$~cm (Tables~\ref{tab:table_bestfitcc_20} and \ref{tab:table_bestfitc_20}) is $l=0.8$--$2.4$~kpc.

\subsubsection{Asymmetric smoothing kernels with star formation dependence}

The addition of models with asymmetric smoothing kernels with a dependence on star formation leads to only a small decrease ($3$-$5$\,\%) 
of $\phi_{\rm min}$ for most of the galaxies except for NGC~6946 and M~51 where we observe a decrease of $25$\,\%.
The smoothing lengthscales stay the same within $20$\,\% for all galaxies except NGC~6946 and M~51 where they almost triple.
Elliptical smoothing kernels improve the fits for NGC~6946, M~51, NGC~4535, NGC~4501, and NGC~4654.

The addition of a star formation dependent loss term (Eq.~\ref{eq:loss}, Table~\ref{tab:table_bestfit}) leads to significant losses only for 
NGC~6946, NGC~4321 (excluding the galaxy central region), 
NGC~4303, NGC~4254, and NGC~4654. The most prominent effect is observed in NGC~6946 where the losses lead to a two times smaller $Q$ 
without changing the smoothing lengthscale. For NGC~4303, NGC~4254, and NGC~4654 the inclusion of a loss term leads to 
$\sim 25$\,\% smaller smoothing lengthscales. The decrease of $\phi$ caused by the inclusion of a loss term is small ($3$-$5$\,\%),
comparable to that for the inclusion of an asymmetric smoothing kernel.

At $20$~cm we see similar tendencies for the smoothing goodness $\phi_{\rm min}$, the smoothing lengthscale $l$, and the normalization $Q$:
$\phi_{\rm min}$ at $20$~cm and $6$~cm are comparable (within $0.1$~dex) except for NGC~4321 (excluding the center) and NGC~4501
where the $\phi_{\rm min}$ at $20$~cm are $\sim 0.2$~dex lower than the $\phi_{\rm min}$ at $6$~cm (Table~\ref{tab:table_bestfitcc_20}).
We observe a significantly modified $Q$ for the same galaxies at $6$~cm. The greatest difference lies in the
smoothing lengthscales which are about two times higher at $20$~cm than the values at $6$~cm. 

As for the $6$~cm data, the addition of models with asymmetric smoothing kernels (Table~\ref{tab:table_bestfitc_20}) 
with a dependence on star formation leads
to a significantly increased smoothing lengthscale only for NGC~6946 and M~51 at $20$~cm. NGC~6946, M~51, NGC~4535, and NGC~4501, for which an
asymmetric kernel is preferred at $6$~cm, seem to prefer a symmetric kernel at $20$~cm.
The addition of a star formation dependent loss term (Table~\ref{tab:table_bestfit_20}) leads to significant losses only for 
NGC~6946, M51, NGC~4303, and NGC~4254.

\subsection{Taking into account the goodness distribution \label{sec:goodness}}

In a second step we want to quantify the uncertainty on our model choice as the most relevant.
In contrast to a classical $\chi^2$ value, the goodness $\phi$ (Eq.~\ref{eq:phi}) is a relative merit, because it is normalized
with respect to the content of the radio continuum image.
In addition, the existence of important large-scale asymmetries in the residual maps well above the detection limit ($> 5 \sigma$),
make $\phi$ and its minimum vary from galaxy to galaxy. The only way to select ``best fit'' models is to apply a relative
cut to $\phi$.
To exploit the full wealth of goodness values $\phi$ of our models, we apply such a cut to the goodness distribution
to determine the family of model convolutions that reproduce the radio continuum observations best.
We decided to define the limiting $\phi_{\rm lim}$ as the goodness for which $1$\,\% of the models have $\phi < \phi_{\rm lim}$.
The distribution of the goodness and its cut are presented in Fig.~\ref{fig:zusammenphi1} and Figs.~\ref{fig:zusammenphi2} to 
\ref{fig:zusammenphi4}. The results with and without losses are shown separately.
As in the previous section, we only discuss the results for the Gaussian smoothing. 

In most of the cases the goodness distribution rises steeply from its minimum value to higher values.
A notable exception to this rule is NGC~4321, which shows a maximum of the $\phi$ distribution that is
much farther away from $\phi_{\rm min}$ than that of the other galaxies.
With the cut in $\phi$ we can give uncertainties on the derived parameters and thus determine the
most probable models with the most probable parameters. It has to be kept in mind that these uncertainties
are based on relative cuts on relative figures of merit. They are only valid in a loose qualitative sense and should not be over-interpreted. 

The resulting parameters are with the associated uncertainties are listed in Tables~\ref{tab:table_6cm_ccn} 
for the $6$~cm data and \ref{tab:table_20cm_ccn} for the $20$~cm data. The results are well comparable to
those for the ``best fit'' models (Tables~\ref{tab:table_bestfitcc} and \ref{tab:table_bestfitcc_20}).

The results for models with losses are presented in Tables~\ref{tab:table_6cm} and \ref{tab:table_20cm}.
Again, the derived parameters are close to those for the ``best-fit'' models.

At a characteristic star formation rate of $\dot{\Sigma}_*=8 \times 10^{-3}$~M$_{\odot}$yr$^{-1}$kpc$^{-2}$ the typical lengthscale for the
diffusive transport of cosmic ray electrons is $l=0.9 \pm 0.3$~kpc at $6$~cm (Table~\ref{tab:table_6cm_ccn}) and  
$l=1.8 \pm 0.5$~kpc at $20$~cm (Table~\ref{tab:table_20cm_ccn}). Perturbed spiral galaxies tend to have smaller Gaussian lengthscales.
This is a natural consequence of the enhancement of the magnetic 
field caused by the interaction (see also Otmianowska-Mazur \& Vollmer 2003; Drzazga et al. 2011),

\subsection{Summary}

Large-scale asymmetries are observed in the residual emission of all galaxies. If the galaxy has an asymmetric
ridge of polarized emission, the ridge region is always radio-bright. 
The decrease of the fit goodness caused by the inclusion an asymmetric smoothing kernel and a loss term is generally small ($3$-$5$\,\%).
The improvement of the goodness due to the enhancement 
of the source term based on polarization is more important than that due to the variation of the smoothing lengthscale.
The large-scale asymmetries of radio-bright and radio-dim regions dominate the goodness of the fit
and thus explain the rather small decrease of the goodness ($\sim 10$-$20$\,\%) when the different model convolutions are applied.
Taking into account the goodness distribution leads to results well comparable to those for the ``best fit'' models.
Overall, the exponential lengthscale equals half the Gaussian lengthscale with a scatter of $0.1$~dex or $25$\,\%.
Typical Gaussian smoothing lengthscales are $l_{\rm 6cm}=0.9 \pm 0.3$~kpc and $l_{\rm 20cm}=1.8 \pm 0.5$~kpc.

The minimum goodness of the ``best-fit'' models of the Gaussian and exponential convolutions cannot used
to discriminate between the two transport mechanisms (diffusion or streaming).

\section{Diffusion or Streaming? \label{sec:diffstream}}

As mentioned in Sect.~\ref{sec:method}, Eq.~\ref{eq:akernel} can be used to distinguish between cosmic ray electron
diffusion or streaming as the dominant transport mechanism in a face-on spiral galaxy:
the different frequency dependence of the lengthscales leads to an expected ratio between the smoothing lengthscales at $6$~cm and $20$~cm of 
$l_{\rm 6cm}/l_{\rm 20cm}=1.34$ for diffusion and $l_{\rm 6cm}/l_{\rm 20cm}=1.81$ for streaming.
In Sect.~\ref{sec:testsres} we showed that the lengthscale ratio $l_{\rm 6cm}/l_{\rm 20cm}$ is more robust with respect to
the image resolution than the derived lengthscales. With an uncertainty of $\Delta (l_{\rm 6cm}/l_{\rm 20cm}) \sim 15$\,\% it is
possible to discriminate between diffusion and streaming.
We decided to fix the border for the ratio of lengthscales to $l_{\rm 6cm}/l_{\rm 20cm}=1.57$. 
The ratios between the lengthscales at the two different frequencies and the associated cosmic ray electron transport mechanism
are presented in Table~\ref{tab:tables_SI} for models with and without losses.
The more distant Virgo galaxies might have a $15$\,\% lower ratio than observed due to a resolution effect (Sect.~\ref{sec:testsres}).
The inclusion of losses in the analysis changed the dominant transport mechanism only in the NGC~4303 model which has the strongest losses.
The analysis of $l_{\rm 6cm}/l_{\rm 20cm}$ is thus robust in the presence of moderate losses.

\subsection{Lengthscale ratio $l_{\rm 6cm}/l_{\rm 20cm}$ \label{sec:lrat}}

In principle, the smoothing kernels should be Gaussians if diffusion is the dominant transport mechanism and exponential if
streaming is the dominant transport mechanism of cosmic ray electrons. However, as we saw in Sects.~\ref{sec:bestfit} and \ref{sec:goodness}
the two convolution methods did not lead to significantly different minimum goodnesses $\phi$. 

The convolution kernel (Gaussian or exponential) should be consistent with the ratio of lengthscales $l_{\rm 6cm}/l_{\rm 20cm}$ 
(bold-face galaxy names in Table~\ref{tab:tables_SI}):
diffusion is the dominant transport mechanism in NGC~6946, M~51, and the center of NGC~4321. 
Streaming is the dominant transport mechanism in the disk of NGC~4321 and NGC~4254.
In NGC~4303 both mechanism seem to work, whereas we find contradicting results for NGC~4501 and NGC~4654.
NGC~4535 is a special case, because of its low surface brightness radio continuum disk which is only partly detected
at $6$~cm with a $18''$. We therefore rely on analysis based on the $27''$ resolution maps. We thus suggest that the dominant transport 
mechanism for NGC~4535 is diffusion.

Diffusion or streaming along the large-scale ordered magnetic field line only plays a major role in NGC~4535 ($e=1.0$ parameter in
Table~\ref{tab:table_6cm_ccn}) and a minor role in NGC~4501 and NGC~4654. These galaxies show asymmetric ridges on polarized
radio continuum emission. In the Virgo galaxies where streaming is the dominant transport mechanism the cosmic ray electrons
travel most probably along the anisotropic component of the turbulent magnetic field.

\subsection{Ratio between the ordered and the turbulent magnetic field \label{sec:ratiord}}

\begin{figure*}[!ht]
  \centering
  \resizebox{\hsize}{!}{\includegraphics{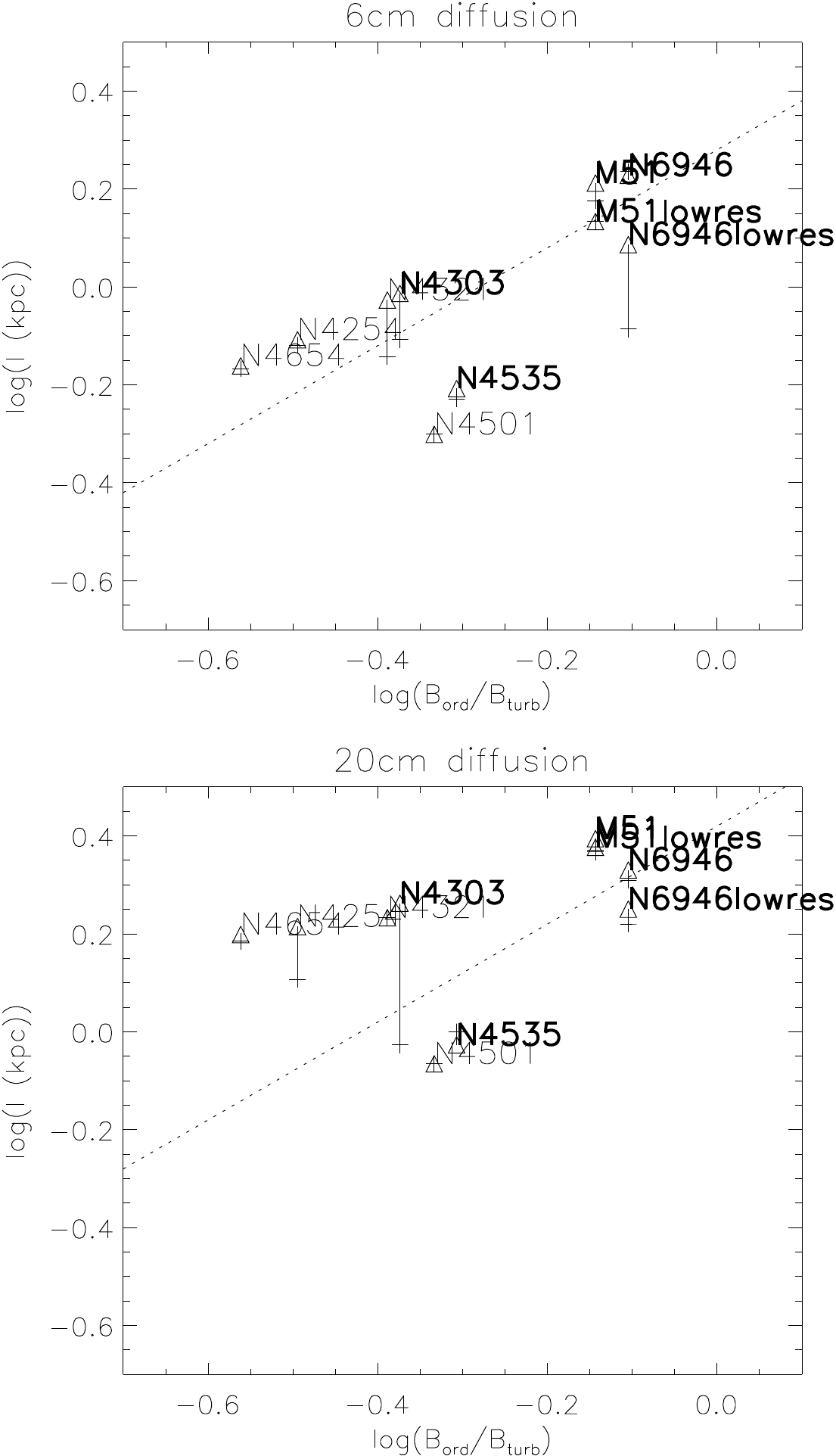}\includegraphics{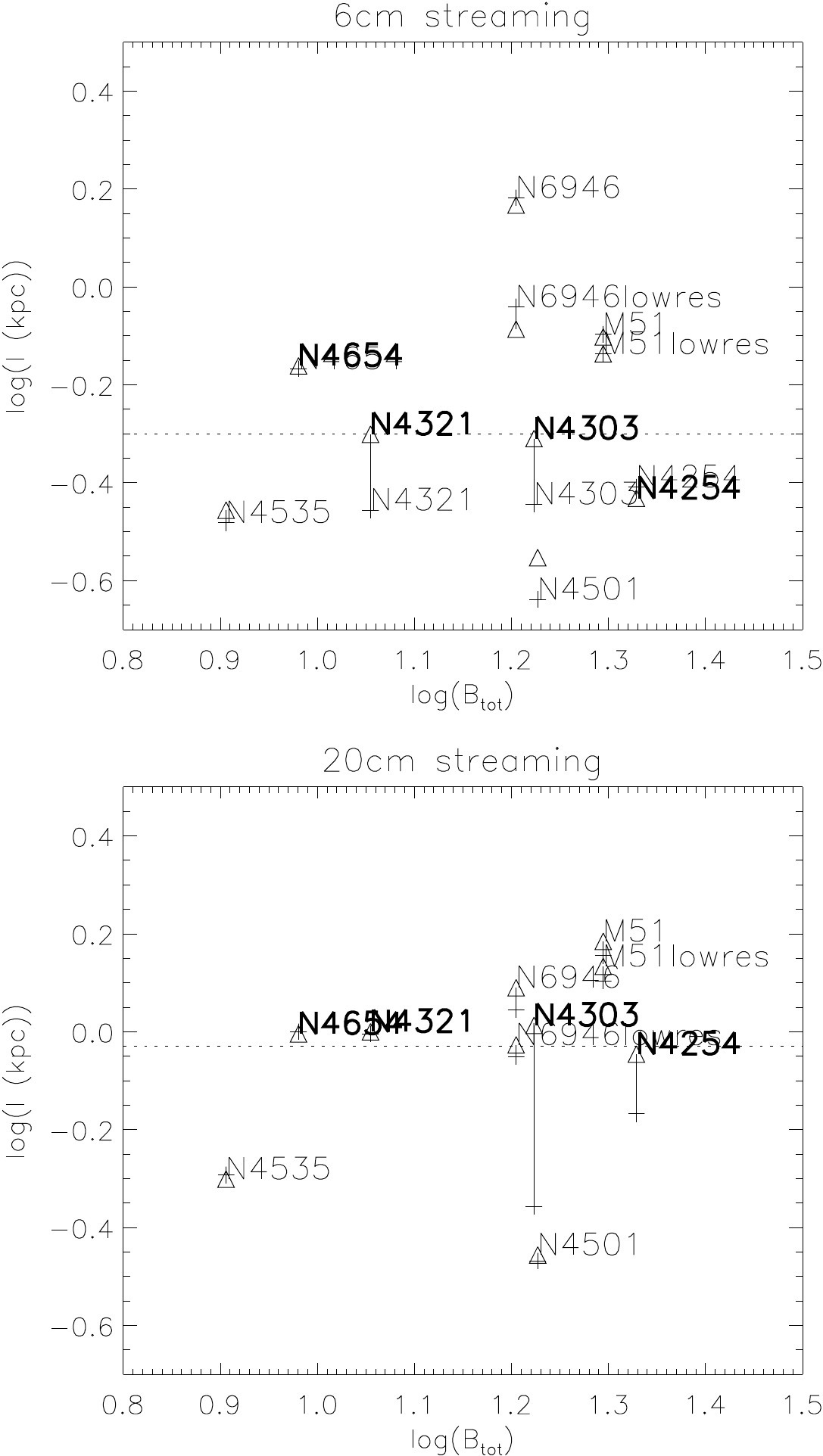}}
  \caption{Left-hand part: diffusion lengthscale (Gaussian kernels) as a function of the ratio between the ordered and turbulent
    magnetic field. Right-hand part: streaming lengthscale (exponential kernels) as a function of the total magnetic field.
    Pluses correspond to smoothing experiments with losses, triangles to experiments without losses.
    The expected relations are shown with dotted lines. The galaxies where we think that diffusion/streaming
    is the dominant transport mechanism are marked in boldface.
  \label{fig:tabadiff}}
\end{figure*}

As stated in Sect.~\ref{sec:method}, the Gaussian smoothing lengthscale is expected to be proportional to the ratio between
the ordered and the turbulent magnetic field if diffusion dominates the cosmic ray electron transport.
In the left-hand part of  Fig.~\ref{fig:tabadiff} the Gaussian smoothing lengthscale of all galaxies are shown for all galaxies.
Galaxies where we believe that diffusion is the dominant transport mechanism are marked in boldface. 
Whereas all galaxies except NGC~4501 and NGC~4535 lie close to this relation at 6~cm,
most of the galaxies deviated from this relation at 20~cm except NGC~6946 and M~51.
These plots illustrate why diffusion is the preferred transport mechanism in the latter two galaxies.
NGC~4535 and NGC~4501 are special cases: NGC~4535 is radio-deficient (Fig.~\ref{fig:sfrradc_int_small}). We think that
the cosmic ray electron diffuse out of a thin magnetic disk with a low magnetic field strength causing the radio-deficiency.
The diffusion lengthscale is thus not set by the synchrotron timescale, but by the disk height.
NGC~4501 seems to have a higher magnetic field strength at $\Sigma_{*\,0}$ which leads to a smaller lengthscale.
The steep spectral index observed in this galaxy (Vollmer et al. 2010) corroborates this hypothesis.

The right-hand part of  Fig.~\ref{fig:tabadiff} shows the exponential smoothing lengthscale for all galaxies.
Galaxies where we believe that streaming is the dominant transport mechanism are marked in boldface. 
The lengthscale is expected to be independent of the magnetic field. This is approximately the case for
NGC~4321, NGC~4303, NGC~4254, and NGC~4654.

We therefore conclude that diffusion is the main transport mechanism in NGC~6946, M~51, NGC~4535, and the central region of NGC~4321.
Cosmic ray electron streaming dominates in NGC~4303, NGC~4254, NGC~4501, NGC~4654, and the disk of NGC~4321, which show the highest 
degrees of polarization in the Virgo cluster sample exceeding $12$\,\% (Vollmer et al. 2013).

\begin{figure}[!ht]
  \centering
  \resizebox{\hsize}{!}{\includegraphics{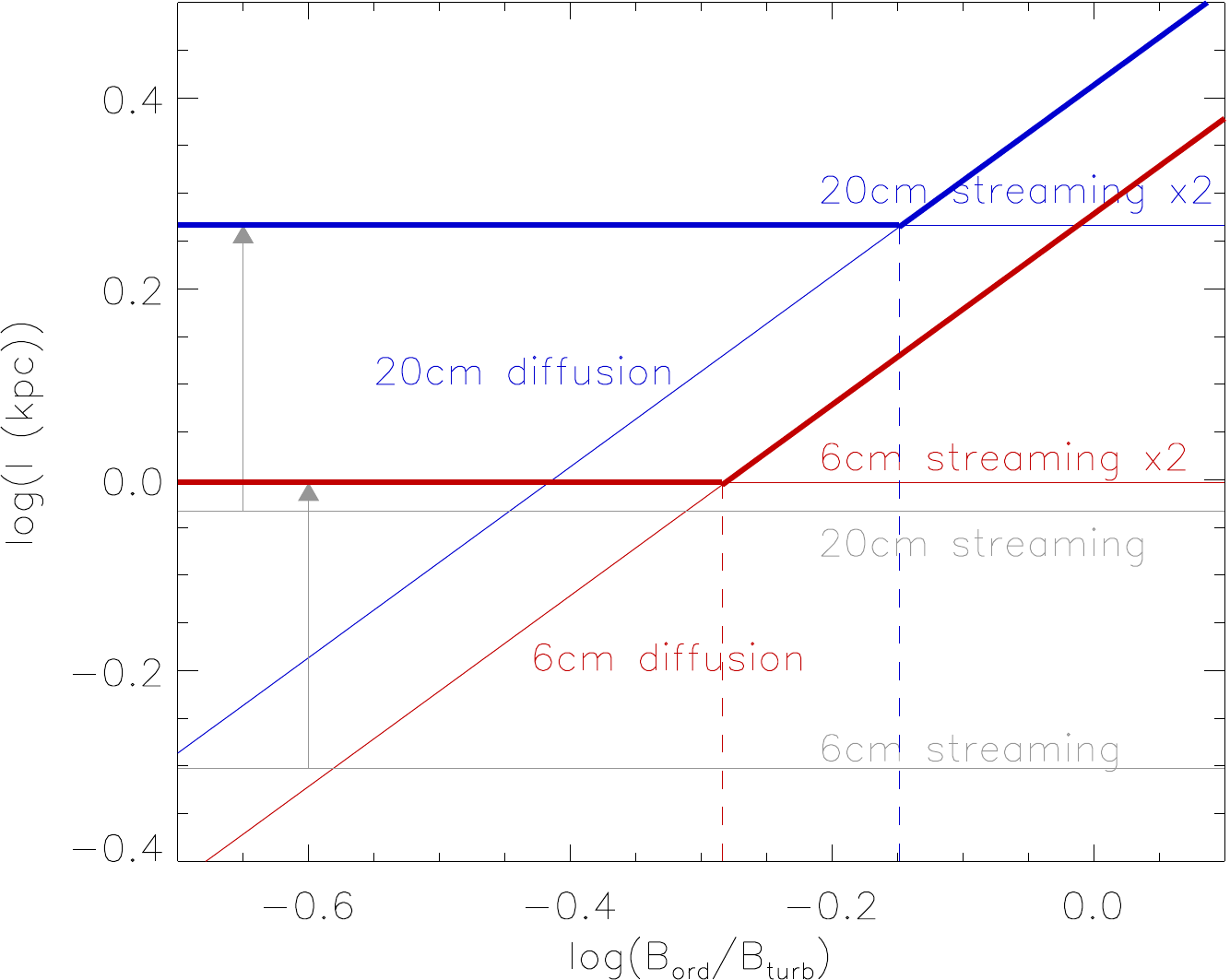}}
  \caption{Schematic sketch to explain the behaviour of the convolution lengthscales as a function of the ratio between the ordered and turbulent 
    magnetic field strengths (Fig.~\ref{fig:tabadiff}). Red: Gaussian convolution lengthscales at 6cm. Blue: Gaussian convolution lengthscales at 20cm.
    Grey: exponential convolution lengthscales at 6cm and 20cm. The grey arrow indicate that the Gaussian convolution lengthscales
    are about two times larger than the exponential convolution lengthscales (see Sect.~\ref{sec:tests}).
    The dashed line indicate the transition from diffusion (Gaussian convolution) to streaming (exponential convolution).
    Thick lines mark the region populated by observed galaxies (cf. Fig.~\ref{fig:tabadiff}).
  \label{fig:difflengthsketch}}
\end{figure}

Taken together, a consistent picture emerges from Fig.~\ref{fig:sfrradc_int_small}:
we measure the lengthscale at a constant star formation rate $\dot{\Sigma}_{*\,0}$ and thus a constant magnetic field strength.
Whereas the streaming lengthscale is constant, the diffusion lengthscale increases with increasing $B_{\rm ord}/B_{\rm turb}$ (Eq.~\ref{eq:ldiff1}).
As long as the diffusion lengthscale is larger than the streaming lengthscale, diffusion is the dominant
cosmic ray electron transport mechanism. Once the diffusion lengthscale is smaller than the streaming lengthscale, the transport is
dominated by streaming. To be able to compare the two lengthscales, we compare the Gaussian lengthscales for diffusion and 
streaming, i.e. we multiply the streaming lengthscales by a factor two (see Sect.~\ref{sec:tests}).
The transition between streaming and diffusion occurs around $\log(B_{\rm ord}/B_{\rm turb}) \sim -0.3$ at 6cm and 
$\log(B_{\rm ord}/B_{\rm turb}) \sim -0.155$ at 20cm,
which corresponds to a lengthscale of $l \sim 1$~kpc at 6~cm and $l \sim 1.9$~kpc at 20~cm (Fig.~\ref{fig:difflengthsketch}).
At this point $l_{\rm diff} = l_{\rm stream} = v_{\rm stream}\,t_{\rm syn}$, where  $v_{\rm stream}$ is the cosmic ray electron streaming velocity.
With a typical magnetic field strength of $B=10$~$\mu$G leading to synchrotron timescales of $t_{\rm syn}^{\rm 20cm}=3.8\,10^7$~yr
and $t_{\rm syn}^{\rm 6cm}=2.0\,10^7$~yr,
we obtain an effective streaming velocity of $v_{\rm stream}=l_{\rm diff}/t_{\rm syn} \sim 50$~km\,s$^{-1}$ at 6 and 20~cm.

In a magnetized plasma, the streaming velocity is expected to be of the order of the Alfv\'en speed, because cosmic ray electrons scatter
on self-generated Alfv\'en waves. In the case of energy density equipartition between the magnetic field and the ISM turbulence, the Alfv\'en speed 
is $v_{\rm A} \sim v_{\rm turb}/\sqrt{X_i}$, where $X_i$ is the degree of ionization and $v_{\rm turb}$ the turbulent velocity.
For the warm ionized medium $v_{\rm A} \sim v_{\rm turb} \sim 20$~km\,s$^{-1}$. This is about a factor of two too low compared to the value we found.
In the warm neutral medium with a turbulent velocity dispersion of $v_{\rm turb} \sim 10$~km\,s$^{-1}$ and a typical degree of ionization of 
$X_i \sim 0.01$ (Draine 2011), the Alfv\'en speed is $v_{\rm A} \sim 100$~km\,s$^{-1}$, 
about a factor of two higher than our estimated value. However, in such a weakly ionized medium the Alfv\'en waves are expected to
be strongly damped due to frequent ion-neutral collisions. This damping should lead to an even further increase of the streaming velocity with respect 
to the Alfv\'en velocity. We therefore argue that the we have measured the effective streaming velocity of the cosmic ray electrons which is a mixture of the
streaming velocities in the neutral and ionized warm ISM.

\subsection{Star formation dependence of the smoothing lengthscale}

Is the discrimination between diffusion and streaming based on $l_{\rm 6cm}/l_{\rm 20cm}$ and $B_{\rm ord}/B_{\rm turb}$
consistent with the derived exponents of the smoothing lengthscale $l \propto \dot{\Sigma}_{*}^n$, where
$n=-0.23$ to $-0.38$ for diffusion and $n=-0.45$ to $-0.75$ for streaming are expected (Eq.~\ref{eq:akernel})?
 
For all galaxies, except NGC~4303 at 6~cm and NGC~4535 at 6 and 20~cm, the results are not altered in the presence of advection losses (Eq.~\ref{eq:loss}).
Based on the exponent $n$ (Tables~\ref{tab:table_6cm_ccn} and \ref{tab:table_6cm}) streaming seems to be relevant  
(NGC~6946, M~51, NGC~4535, and NGC~4654), whereas diffusion is the
dominant transport mechanism in NGC~4321, NGC~4303, NGC~4254, and NGC~4501.
If losses are included in the model, NGC~4303 and NGC~4535 become dominated by streaming.

The determination of the transport mechanism based on the ratio of (i) the smoothing lengthscales and (ii) the ordered to the turbulent magnetic field
leads to consistent results for NGC~6946, M~51, NGC~4321, NGC~4535, and NGC~4254. Both methods are thus complementary.
The determination of the transport mechanism based on the star formation dependence of the smoothing lengthscale leads to results which
contradict the results based on $\log(B_{\rm ord}/B_{\rm turb})$ for all galaxies, except the central region of NGC~4321 and NGC~4654.
Therefore, the exponent $n$ cannot be used to discriminate between diffusion and streaming for the following reason:

by assuming $l \propto \dot{\Sigma}_*^n$ we assumed that there is no dependence of the smoothing length on $\log(B_{\rm ord}/B_{\rm turb})$ ($m=0$).
The apparent contradiction between the determination of the transport mechanism based on $n$ can be resolved if we assume $m > 0$ locally as 
suggested in Sect.~\ref{sec:ratiord}. This hypothesis will be discussed in Sect.~\ref{sec:discussion}.
\begin{table*}[!ht]
\begin{center}
\caption{Ratio between the smoothing lengthscales at 20~cm and 6~cm.\label{tab:tables_SI}}
\begin{tabular}{lcccccc}
\hline
         Gaussian convolution & & & & & & \\
\hline
            name & polarization$^{\rm a}$  & $l_{\rm 20cm}/l_{\rm 6cm}$ & $l_{\rm 20cm}/l_{\rm 6cm}$ & CR transport & $l_{\rm 20cm}/l_{\rm 6cm}$ &  CR transport   \\
             & & (``best fit'') & & (with losses) & & (without losses)  \\
\hline
             {\bf NGC6946} & & 2.2 & 1.19 $\pm$ 0.21 & diffusion & 1.26 $\pm$ 0.26 & diffusion  \\
      {\bf NGC6946 low resolution} & & 10.7 & 2.02 $\pm$ 0.64 & streaming &  1.46 $\pm$ 0.55 & diffusion  \\
                 {\bf M51} & & 1.8 & 1.55 $\pm$ 0.33 & diffusion & 1.52 $\pm$ 0.33 & diffusion  \\
          {\bf M51 low resolution} & & 2.3 & 1.73 $\pm$ 0.29 & streaming & 1.75 $\pm$ 0.30 & streaming   \\
             {\bf NGC4321} & & 1.1 & 1.17 $\pm$ 0.24 & diffusion & 1.17 $\pm$ 0.24 & diffusion   \\
      NGC4321 disk & & 2.2 & 2.89 $\pm$ 0.50 & streaming & 2.21 $\pm$ 0.42 & streaming  \\
             {\bf NGC4303} & POL & 2.4 & 1.21 $\pm$ 0.33 & diffusion & 1.89 $\pm$ 0.38 & streaming   \\
             NGC4535 & POL & 2.2 & 2.58 $\pm$ 0.59 & streaming & 2.29 $\pm$ 0.55 & streaming   \\
        {\bf NGC4535 VIVA} & POL & 1.2 & 1.56 $\pm$ 0.34 & diffusion & 1.44 $\pm$ 0.36 & diffusion   \\
             NGC4254 & POL & 2.0 & 1.69 $\pm$ 0.27 & streaming & 2.11 $\pm$ 0.36 & streaming  \\
             NGC4501 & POL & 1.7 & 1.72 $\pm$ 0.60 & streaming & 1.72 $\pm$ 0.60 & streaming  \\
            NGC4654  & POL & 2.7 & 2.26 $\pm$ 0.61 & streaming & 2.28 $\pm$ 0.64 & streaming  \\
\hline
         exponential convolution & & & & & & \\
\hline
      NGC6946 & & 3.0 & 0.73 $\pm$ 0.18 & diffusion & 0.83 $\pm$ 0.26 & diffusion  \\
      NGC6946 low resolution & & 8.8 & 0.98 $\pm$ 0.29 & streaming &  1.14 $\pm$ 0.34 & diffusion  \\
                 M51 & & 2.0 & 1.79 $\pm$ 0.53 & streaming & 1.95 $\pm$ 0.51 & streaming  \\
          M51 low resolution & & 4.1 & 1.74 $\pm$ 0.40 & streaming & 1.86 $\pm$ 0.49 & streaming   \\
             NGC4321 & & 1.0 & 1.24 $\pm$ 0.38 & diffusion & 1.24 $\pm$ 0.38 & diffusion   \\
      {\bf NGC4321 disk} & & 2.0 & 2.84 $\pm$ 0.94 & streaming & 2.00 $\pm$ 0.74 & streaming  \\
             {\bf NGC4303} & POL & 2.3 & 1.22 $\pm$ 0.36 & diffusion & 2.11 $\pm$ 0.60 & streaming   \\
             {\bf NGC4535} & POL & 2.5 & 2.60 $\pm$ 0.56 & streaming & 2.67 $\pm$ 0.87 & streaming   \\
        NGC4535 VIVA & POL & 1.5 & 1.57 $\pm$ 0.31 &           & 1.44 $\pm$ 0.31 & diffusion   \\
             {\bf NGC4254} & POL & 3.0 & 1.74 $\pm$ 0.48 & streaming & 2.41 $\pm$ 0.87 & streaming  \\
             NGC4501 & POL & 1.5 & 1.45 $\pm$ 0.85 & diffusion & 1.25 $\pm$ 0.81 & diffusion  \\
             NGC4654  & POL & 1.0 & 1.45 $\pm$ 0.79 & diffusion & 1.42 $\pm$ 0.78 & diffusion  \\
\hline
\end{tabular}
\begin{tablenotes}
  \item $^{\rm a}$ Increased source term proportional to the degree of polarization.
    \end{tablenotes}
\end{center}
\end{table*}

\subsection{The diffusion coefficient \label{sec:diffcoeff}}

For the two diffusion-dominated galaxies, NGC~6946 and M~51, we can determine the diffusion coefficient $D$.
The Syrovatsky (1959) solution of the diffusion equation is
\begin{equation}
N \propto \exp(-r^2/(4\,D\,t_{\rm syn}))\ .
\end{equation}
With the Gaussian convolution of Eq.~\ref{eq:kernel} this yields $l_{\rm diff}^2=4\, D\,t_{\rm syn}$.
We adopt a total magnetic field strength of $B_{\rm tot} = 16 \pm 3$~$\mu$G for $\dot{\Sigma}_*=8 \times 10^{-3}$~M$_{\odot}$kpc$^{-2}$yr$^{-1}$
derived by Tabatabaei et al. (2013a). 
This value is somewhat lower than the magnetic field strength at the same star formation $B_{\rm tot} \sim 20$~$\mu$G 
in NGC~4254 (Chy\.zy et al. 2008). Moreover, we use the average diffusion length at $6$~cm $l_{\rm diff}=1.6 \pm 0.2$~kpc (Table~\ref{tab:table_6cm_ccn}, upper part).
With a frequency dependence of $l_{\rm diff} \propto \nu^{-0.25}$ this corresponds to a diffusion length scale of $l_{\rm diff}=2.2$~kpc
at $20$~cm (Table~\ref{tab:table_20cm_ccn}). With these numbers we derive a diffusion coefficient of $D=(1.8 \pm 0.6) \times 10^{28}$~cm$^2$s$^{-1}$
for $B_{\rm ord}/B_{\rm turb}=0.8$ (Table~\ref{tab:gals}).
For comparison, Mulcahy et al. (2016) found a diffusion coefficient of $D=(6.6 \pm 0.2) \times 10^{28}$~cm$^2$s$^{-1}$ in M~51 with 
$B_{\rm ord}/B_{\rm turb}=0.7$ (Table~\ref{tab:gals}), Heesen et al. (2019) found $D=(0.13-1.5) \times 10^{28}$~cm$^2$s$^{-1}$ at 1~GeV, which
corresponds to a frequency of $\sim 200$~MHz for $B_{\rm tot} = 16$~$\mu$G. Moreover, the Milky Way diffusion coefficient is
$D=3 \times 10^{28}$~cm$^2$s$^{-1}$ (Strong et al. 2007).

\subsection{Summary}

Diffusion or streaming along the large-scale ordered magnetic field line only plays a major role in NGC~4535 and a minor role in NGC~4501 and NGC~4654. 
Based on the ratio $B_{\rm ord}/B_{\rm turb}$ we conclude that diffusion is the main transport mechanism in NGC~6946, M~51, NGC~4535, and the central region of NGC~4321.
Cosmic ray electron streaming dominates in NGC~4303, NGC~4254, NGC~4501, NGC~4654, and the disk of NGC~4321
The classifications on $l_{\rm 6cm}/l_{\rm 20cm}$ and $B_{\rm ord}/B_{\rm turb}$ are well consistent and complementary.
However, the star formation dependence of the smoothing lengthscale cannot be used to determine the cosmic ray transport mechanism, because
of a local dependence of the smoothing length on $B_{\rm ord}/B_{\rm turb}$.
Models based on a star formation dependence of the smoothing lengthscale indicate $(B_{\rm ord}/B_{\rm turb}) \propto \SFR^{-m}$ with $m > 0$.

The following overall picture emerges: whereas the streaming lengthscale is constant, the diffusion lengthscale decreases with $B_{\rm ord}/B_{\rm turb}$.
As long as the diffusion lengthscale is larger than the streaming lengthscale, diffusion is the dominant
cosmic ray electron transport mechanism. Once the diffusion lengthscale is smaller than the streaming lengthscale, the transport is
dominated by streaming. The transition occurs around $\log(B_{\rm ord}/B_{\rm turb}) \sim -0.3$ at 6cm and $\log(B_{\rm ord}/B_{\rm turb}) \sim -0.15$ at 20cm, 
which corresponds to a lengthscale of $l \sim 1$~kpc at 6~cm and $l \sim 1.9$~kpc at 20~cm.

We derive a diffusion coefficient of $D=(1.8 \pm 0.6) \times 10^{28}$~cm$^2$s$^{-1}$ for $B_{\rm ord}/B_{\rm turb}=0.8$.

\section{Discussion \label{sec:discussion}}

Cosmic ray electron diffusion occurs on scales which are much smaller than our spatial resolution of $\sim 1$~kpc.
The results of Sect.~\ref{sec:ratiord} imply that the diffusion lengthscale depends on the ordered magnetic field.
It is not obvious why this should be the case, because diffusion is a small-scale process. 
One possibility is that the ordered magnetic field is dominated by anisotropic turbulent
small-scale magnetic fields. In this case the small scales would have a direct impact at large scales.

Berkhuijsen et al. (2013) found smoothing lengthscales for M~31 and M~33 which are well comparable to our results 
(Tables~\ref{tab:table_bestfitcc} and \ref{tab:table_bestfitc}). The smoothing lengthscales at $1.4$~GHz found by Heesen et al. (2019) for three local
spiral galaxies ($1.0-3.5$~kpc) also agree with our values.
A direct comparison with the work of Murphy et al. (2009) is difficult, because these authors used FIR maps as source maps.
Nevertheless, we can have a look at the ratios between the smoothing lengthscales of the galaxies that we have in common:
NGC~6946, M~51, NGC~4321, and NGC~4254 (Table~\ref{tab:table_bestfitcc_20}). The ratio between our smoothing lengthscale and those of Murphy et al. (2009)
are $1.8,\ 1.3,\ 1.4,\ 0.4$. The mean ratio for the first three galaxies is $1.5 \pm 0.3$. The results seem thus to be consistent. However, the results for
NGC~4254 are vastly different, the ratio being $0.4$. The reason for this discrepancy is that the smoothing lengthscale 
determined for NGC~4254 by Murphy et al. (2009) is about three times higher than those of the other galaxies of their sample.

Can we detect the presence of a radio halo as observed in many local spiral galaxies with high star formation rates (e.g., Krause et al. 2018)?
We think that the decrease of goodness $\phi$ by allowing a loss term might also be interpreted as an indication for the existence of
a radio halo. We expect the radio halo to be a dimmed and smoothed-out version of the underlying disk emission.
It is therefore not surprising that we obtain a better fit if we remove the small-scale regions of high local star formation rates
from the source maps of NGC~6946 and NGC~4303. In this case, the normalization factors $Q$ with and without losses
give a rough idea about the radio continuum emission contained in the radio halo: 
$loss/\SFR=Q_{\rm loss}(1/Q_{\rm loss}-1/Q_{\rm no\ loss}) \sim 30$\,\% in NGC~6946 and $\sim 20$\,\% in NGC~4303.
We interpret this result as evidence for the existence of a radio halo in these two galaxies.

We now go back to the question about the dependence of the magnetic field on the star formation rate $B \propto \dot{\Sigma}_{*}^j$ (Sect.~\ref{sec:integ}).
Whereas Heesen et al. (2014) found $j=0.3$, we found $j=0.44$ or $j=0.13 \pm 0.1$ (Fig.~\ref{fig:sfr_Bfield}) for the integrated values
where the magnetic field is calculated under the assumption of energy equipartition between the magnetic field and the cosmic ray electrons.
On the other hand, the model of Vollmer \& Leroy (2011) 
together with the assumption of equipartition between the kinetic and magnetic energy densities yields $j=0.5$ for the local values.
This model assumes conservation of the energy flux injected by supernova explosions and the turbulent energy flux
$\Sigma v_{\rm turb}^3/l_{\rm driv}=\xi \dot{\Sigma}_*$. The height of the gas disk is then proportional to the turbulent driving lengthscale $l_{\rm driv}$
and the turbulent velocity dispersion $v_{\rm turb}$ is constant. On the other hand, in the presence of a constant scale height of the gas disk,
a constant turbulent velocity, and $\Sigma \propto \dot{\Sigma}_*^{1.4}$, the expected exponent is $j=0.36$.
The exponent of the SFR-dependence of the smoothing lengthscale is a mixture
of two exponents (Sect.~\ref{sec:method}): $j$ with $B \propto \dot{\Sigma}_{*}^j$ and $m$ with $(B_{\rm ord}/B_{\rm turb}) \propto \dot{\Sigma}_{*}^{-m}$ (Eq.~\ref{eq:pdegree}).
The relation between the different exponents is $n=-(0.75\,j+m)$, which yields $j=-(m+n)/0.75$.

\begin{figure}[!ht]
  \centering
  \resizebox{\hsize}{!}{\includegraphics{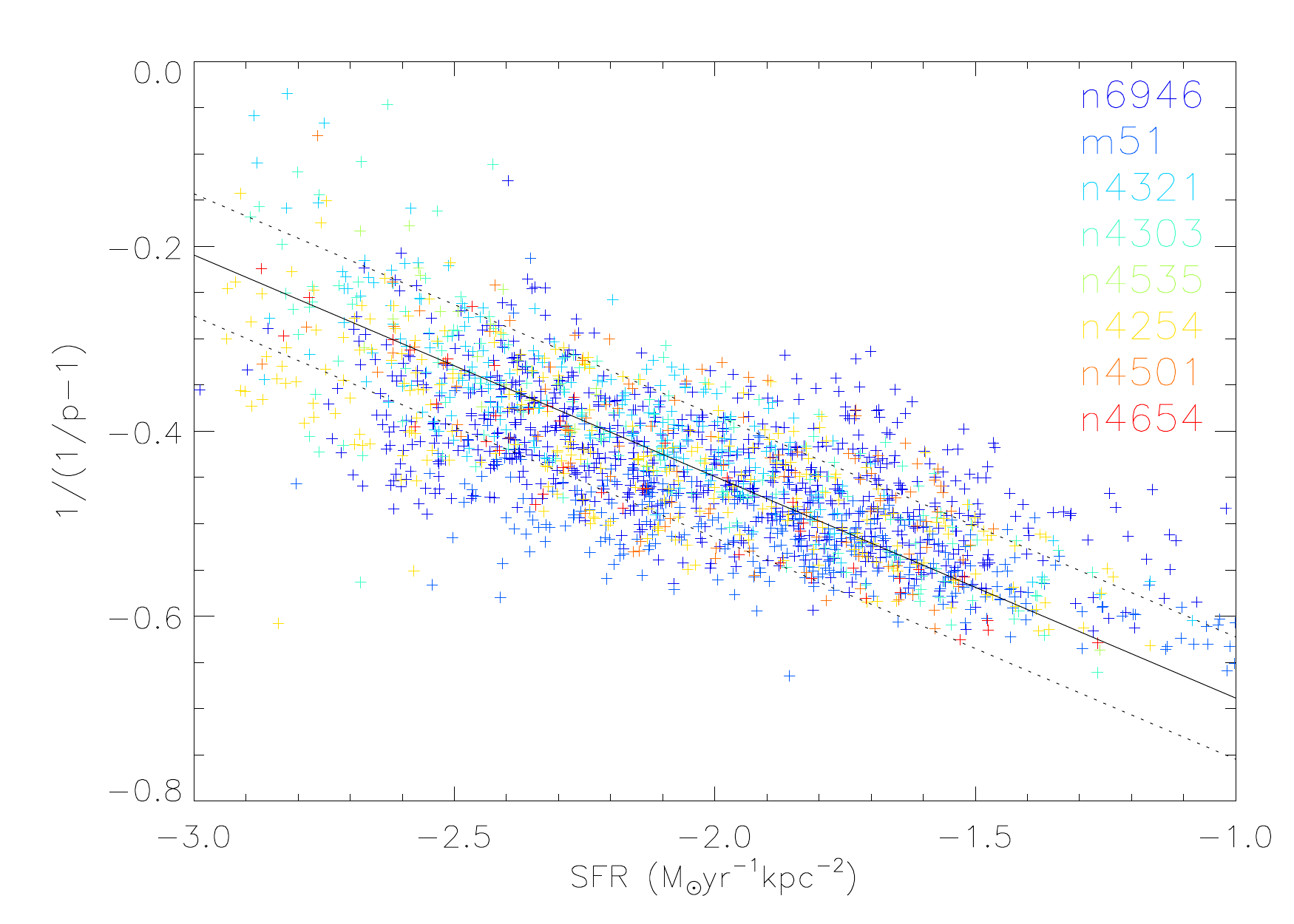}}
  \caption{Ratio between the ordered and turbulent magnetic field strengths $(B_{\rm ord}/B_{\rm turb})=1/(1/p-1)$ as a
    function of the local star formation rate.
  \label{fig:test1color_all_new}}
\end{figure}

To investigate the dependence of $(B_{\rm ord}/B_{\rm turb}) \propto \dot{\Sigma}_{*}^{-m}$ on the star formation rate, we make use of the observed degree of polarization
$p=B_{\rm ord}/B_{\rm tot}=B_{\rm ord}/(B_{\rm turb}+B_{\rm ord})$, where $B_{\rm tot}$ is the total magnetic field strength.
We thus obtain $(B_{\rm ord}/B_{\rm turb})=1/(1/p-1)$. This quantity is shown in Fig.~\ref{fig:test1color_all_new} as a function of
the local star formation rate.
The relation (in log-log scale) is straight and quite tight.
A robust bisector fit yields a slope of $m=0.24$ and $\sigma=0.07$~dex.
With a Bayesian approach to linear regression we obtain $m=0.221 \pm 0.005$.

Table~\ref{tab:expm} shows the exponent $j=-(m+n)/0.75$ for diffusion and $j=-n/1.5$ for streaming for our sample galaxies.
\begin{table}[!ht]
\begin{center}
\caption{Dependence of the magnetic field strength on the star formation rate $B \propto \dot{\Sigma}_{*}^j$.
The relation between the diffusion lengthscale and the star formation rate is $l \propto \dot{\Sigma}_{*}^n$.
In addition, we use $(B_{\rm ord}/B_{\rm turb}) \propto \dot{\Sigma}_{*}^{-m}$ with $m=0.24$.  \label{tab:expm}}
\begin{tabular}{lcccc}
\hline
name  & $n$ (6cm) & $j$ (6cm) & $n$ (20cm) & $j$ (20cm) \\ 
\hline
 & diffusion & & &  \\
\hline
NGC6946 & -0.93 & 0.92 & -0.70 & 0.61 \\
M51 & -0.53 & 0.39 & -0.54 & 0.40 \\
N4535 VIVA & -0.50 & 0.35 & -0.54 & 0.40 \\
\hline
 & streaming & & & \\
\hline
NGC4321 disk & -0.12 & 0.08 & -0.04 & 0.03 \\
NGC4303 & -0.33 & 0.22 & -0.08 & 0.05 \\
NGC4254 & -0.28 & 0.19 & -0.17 & 0.11 \\
NGC4501 & -0.17 & 0.11 & -0.10 & 0.07 \\
NGC4654 & -0.76 & 0.51 & -0.57 & 0.38 \\ 
\hline
\end{tabular}
\end{center}
\end{table}
The mean of all exponents is $j=0.30 \pm 0.24$. We note that the exponents of the galaxies whose cosmic ray
electron transport is dominated by streaming, except that of NGC~4654, are significantly smaller ($j=0.17 \pm 0.16$)
than those of the galaxies whose transport is dominated by diffusion ($j=0.51 \pm 0.22$).

If the magnetic field within the galactic disk is enhanced by an external interaction and if the enhancement of the total magnetic field by the 
additional magnetic field component is highest in regions of low local star formation, i.e. in interarm regions, the dependence of the magnetic field 
on star formation is expected to be weakened. 
In this case the exponent $j$ ($B \propto \SFR^j$) is expected to be smaller than $0.5$.
In Sect.~\ref{sec:integ} we showed that there is no clear dependence of the average magnetic field, which is dominated by the turbulent field, 
on the integrated star formation rate (Fig.~\ref{fig:sfr_Bfield}). On the other hand, there is a clear dependence of $(B_{\rm ord}/B_{\rm turb})$ on the 
local star formation rate $\dot{\Sigma}_{*}$ and the degree of polarization is not exceptionally high in the Virgo cluster galaxies compared to field galaxies
(Fig.~\ref{fig:test1color_all_new}). Therefore, we think that environmental interactions
lead to an enhancement of both, the ordered and turbulent magnetic fields. Presumably, this enhancement with respect to the
already existing magnetic field is more important in the interarm regions than in the arm regions.

We conclude that our results are broadly consistent with energy equipartition 
between the energy density of the magnetic field and the kinetic energy density of the interstellar medium ($j=0.5$) in isolated systems.
In cluster galaxies, the exponent $j$ is smaller presumably due to an enhancement of the magnetic field caused by interactions with the environment.

The mean exponent is steeper than that found in NGC~6946 ($j=0.14$) by Tabatabaei et al. (2013a) and comparable to
that found in NGC~4254 ($j=0.26$) by Chy{\.z}y (2008).
The discrepancy probably lies in the determination of the total magnetic field strength by Tabatabaei et al. (2013a) and Chy{\.z}y (2008), where equipartition 
between the energy densities of the magnetic field and cosmic rays was assumed.
It should also be taken into account that our method is an indirect way to determine the exponent $j$, which is independent of
equipartition hypotheses.
Further investigations are needed to determine which equipartition is relevant in the disks of spiral galaxies, between the
energy densities of the magnetic field and that of (i) the cosmic rays and/or (ii) the kinetic energy of the gas.

\section{Conclusions\label{sec:conclusions}}

The transport mechanism of cosmic ray electrons within galactic disks can be diffusion, as a result of random motions across
tangled magnetic field lines, or streaming, as a result of streaming which is uni-directional down a cosmic ray pressure gradient.
The two transport mechanisms give rise to different dependencies of the kernel lengthscale with respect to
the magnetic field strength and the frequencies of radio continuum observations.
Since the magnetic field strength is related to the star formation rate (e.g., Heesen et al. 2014), we expect that the
smoothing kernel is proportional to the local star formation rate.
Star formation maps were constructed from Spitzer and Herschel infrared and GALEX UV observations.
We convolved the star formation maps of eight rather face-on galaxies with adaptive Gaussian and exponential smoothing kernels to obtain
model radio continuum emission maps (Sect.~\ref{sec:method}). 
The smoothing lengthscales depend on the observation frequency and the star formation 
rate (Eq.~\ref{eq:akernel}). The dependencies are different for cosmic ray electron diffusion and streaming.
The model radio continuum maps were compared to $6$~cm and $20$~cm continuum observations to determine the dominant
cosmic ray electron transport mechanism in the disks of the $8$ spiral galaxies (Sect.~\ref{sec:results}).

The comparison between the model and observed radio continuum maps showed that the residuals are dominated by galaxy-wide large-scale
asymmetries. These cannot be removed by our model. 
The discrimination between the two cosmic ray electron transport mechanisms is based on 
(i) the convolution kernel (Gaussian or exponential),
(ii) the dependence of the smoothing kernel on the local magnetic field and hence on the local star formation rate,
(iii) the ratio between the two smoothing lengthscales via the frequency-dependence of the smoothing kernel, and 
(iv) the dependence of the smoothing kernel on the ratio between the ordered and the turbulent magnetic field.
The results of method (ii) depend on the image resolution, whereas the results of method (iii)
are sensitive to an extended diffuse radio continuum emission beyond the optical radius of the galactic disk (Sect.~\ref{sec:testsres}).
These two effects have to be taken into account for the interpretation of the results of our adaptive kernel smoothing experiments. 
We introduced losses of cosmic ray electrons in regions of high local star formation rates caused by advection into the halo, 
i.e. a galactic wind, in our models.
Two galaxies, in which losses play a role, are tentatively identified: NGC~6946 and NGC~4303 (Sect.~\ref{sec:discussion}).
Methods (i) and (ii) cannot be used to determine the cosmic ray transport mechanism. 
Important asymmetric large-scale residuals and a local dependence of the smoothing length on $B_{\rm ord}/B_{\rm turb}$ are most probably responsible 
for the failure of method (i) and (ii), respectively.
On the other hand, the classifications based on $l_{\rm 6cm}/l_{\rm 20cm}$ (method iii) and $B_{\rm ord}/B_{\rm turb}$ (method iv) are well consistent and complementary.

From the analysis of our adaptive kernel smoothing experiments we draw the following conclusions:
\begin{enumerate}
\item
In asymmetric ridges of polarized radio continuum emission the total power emission is enhanced with respect to the star formation rate 
(Fig.~\ref{fig:rcfir_spixx1c1_nice_pol_smoothing}) due to ISM compression.
\item
At a characteristic star formation rate of $\dot{\Sigma}_*=8 \times 10^{-3}$~M$_{\odot}$yr$^{-1}$kpc$^{-2}$ the typical lengthscale for the
transport of cosmic ray electrons is $l=0.9 \pm 0.3$~kpc at $6$~cm (Table~\ref{tab:table_6cm_ccn}) and  
$l=1.8 \pm 0.5$~kpc at $20$~cm (Table~\ref{tab:table_20cm_ccn}).
\item 
Perturbed spiral galaxies tend to have smaller lengthscales. This is a natural consequence of the enhancement of the magnetic 
field caused by the interaction (see also Otmianowska-Mazur \& Vollmer 2003; Drzazga et al. 2011).
\item
Models with advection losses through galactic winds are viable for NGC~6946 and NGC~4303, the two galaxies with the highest local star 
formation rates. Alternatively, this can be interpreted as the existence of a radio halo around their disks.
\item
The determination of the dominating cosmic ray electron transport mechanism is based on the frequency-dependence of 
the smoothing lengthscale. Diffusion is the main transport mechanism in unperturbed galactic disks (NGC~6946, M~51, and the central region of NGC~4321).
Streaming dominates in the disks of the perturbed Virgo spiral galaxies NGC~4321, NGC~4303, NGC~4254, NGC~4501, and NGC~4654.
\item
Diffusion or streaming along the ordered magnetic field is not a measurable effect in our galaxy sample, except for NGC~4535.
In most of the Virgo galaxies where streaming is the dominant transport mechanism the cosmic ray electrons
travel most probably along the anisotropic component of the turbulent magnetic field.
\item
The streaming lengthscale at a constant star formation rate $\dot{\Sigma}_{*\,0}$ is constant (right panels of Fig.~\ref{fig:tabadiff}).
On the other hand, the diffusion lengthscale decreases with $B_{\rm ord}/B_{\rm turb}$.
As long as the diffusion lengthscale is larger than the streaming lengthscale, diffusion is the dominant
cosmic ray electron transport mechanism. Once the diffusion lengthscale is smaller than the streaming lengthscale, the transport is
dominated by streaming. The transition occurs around $\log(B_{\rm ord}/B_{\rm turb}) \sim -0.3$ at 6cm and $\log(B_{\rm ord}/B_{\rm turb}) \sim -0.15$ at 20cm, 
which corresponds to a lengthscale of $l \sim 1$~kpc at 6~cm and $l \sim 1.9$~kpc at 20~cm (Fig.~\ref{fig:difflengthsketch}).
\item
We derived a diffusion coefficient of $D=(B_{\rm ord}/B_{\rm turb}) \times (9.3 \pm 2.8) \times 10^{28}$~cm$^2$s$^{-1}$ for $(B_{\rm ord}/B_{\rm turb}) > 0.5$.
\item
We determined a streaming velocity of $v_{\rm stream} \sim 50$~km\,s$^{-1}$.
\item
The observed independence of the smoothing lengthscale on the star formation rate (small $n$ in Table~\ref{tab:expm}) indicates  that the 
magnetic field in the interarm regions of perturbed Virgo spiral galaxies might be mostly independent of star formation. 
\item
In our sample galaxies the ratio between the ordered and turbulent magnetic field strengths is $(B_{\rm ord}/B_{\rm turb}) \propto \SFR^{\ -0.24}$
(Fig.~\ref{fig:test1color_all_new}). From this we obtained the relation between the local magnetic field strength and star formation rate
$B \propto \dot{\Sigma}_{*}^{\ 0.51 \pm 0.21}$ for the diffusion-dominated galaxies NGC~6946, M~51, and NGC~4535 and $B \propto \dot{\Sigma}_{*}^{\ 0.30 \pm 0.24}$
for the whole sample. This result is broadly consistent with energy equipartition 
between the energy density of the magnetic field and the kinetic energy density of the interstellar medium ($j=0.5$) in isolated systems.
\item
We argue that in the Virgo spiral galaxies the turbulent magnetic field is globally enhanced in the disk. Therefore, the regions where the magnetic field 
is independent of the star formation rate are more common. In addition, $B_{\rm ord}/B_{\rm turb}$ decreases leading to a diffusion
lengthscale which is smaller than the streaming lengthscale. Therefore, cosmic ray electron streaming dominates in most of the Virgo
spiral galaxies.
\end{enumerate}

\begin{acknowledgements}
The authors would like to thank J.D.P.~Kenney, E.J.~Murphy, and Olaf Wucknitz for useful comments on the article.
\end{acknowledgements}

\clearpage

\appendix

\section{Tests}

\begin{figure*}[!ht]
  \centering
  \resizebox{18cm}{!}{\includegraphics{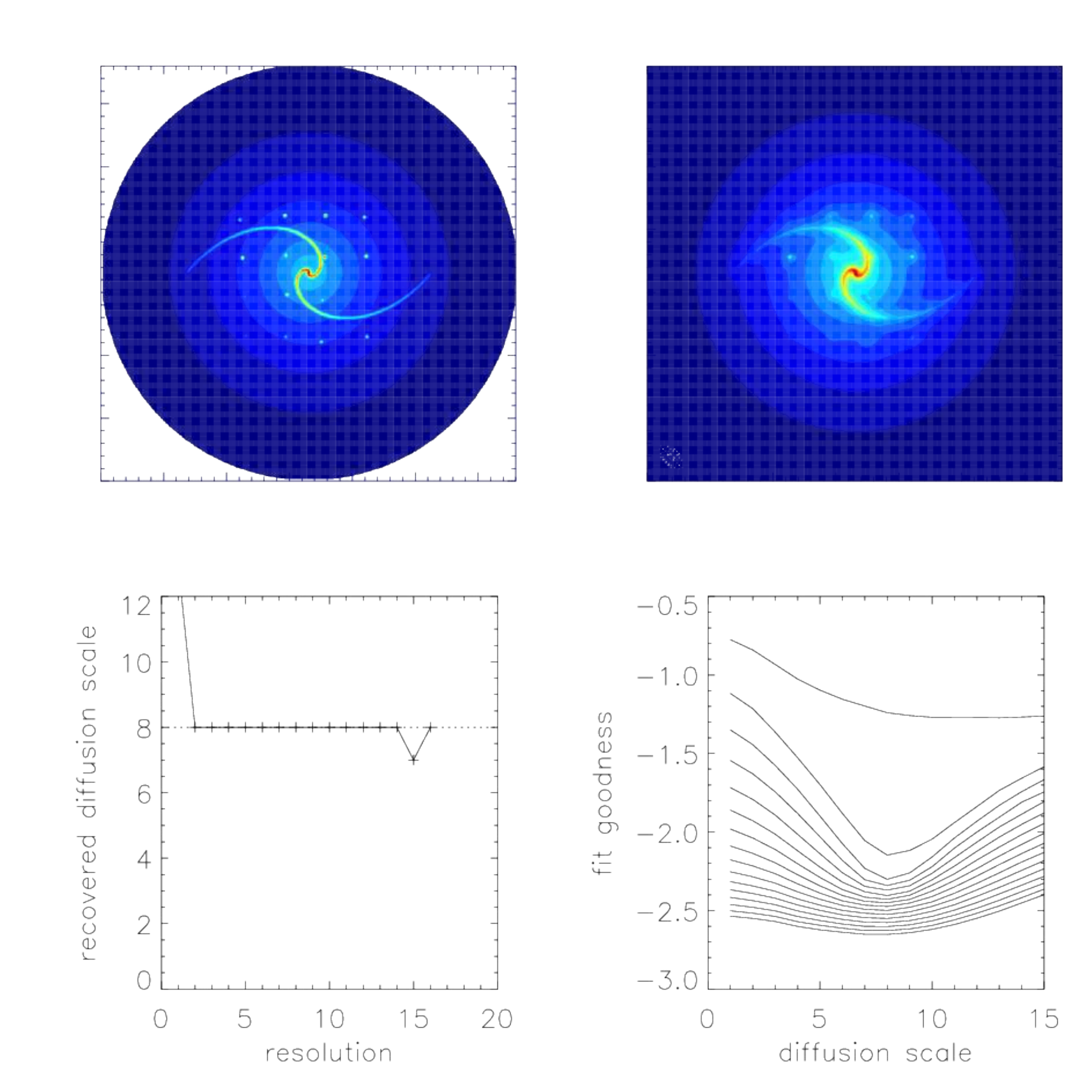}\includegraphics{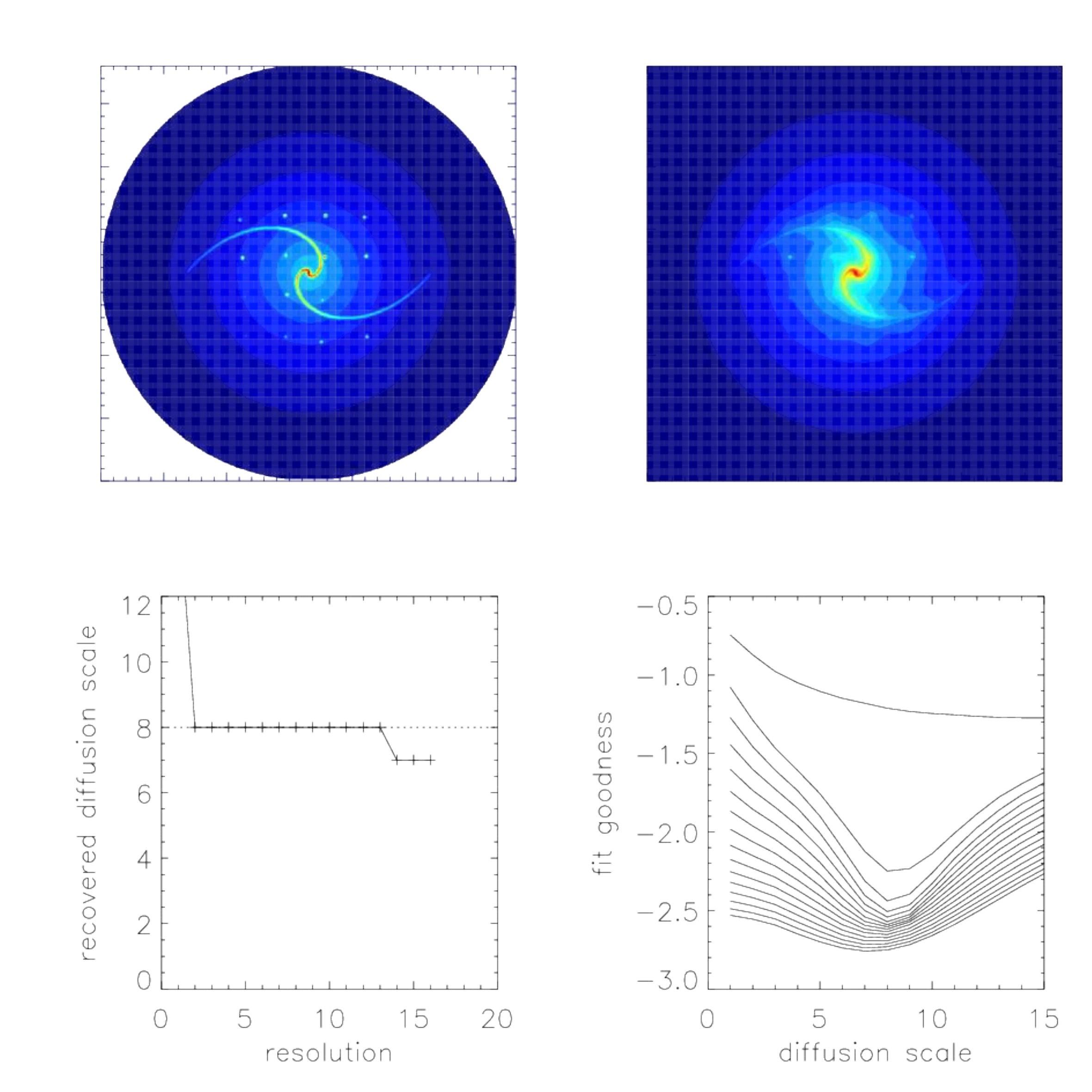}}
  \resizebox{18cm}{!}{\includegraphics{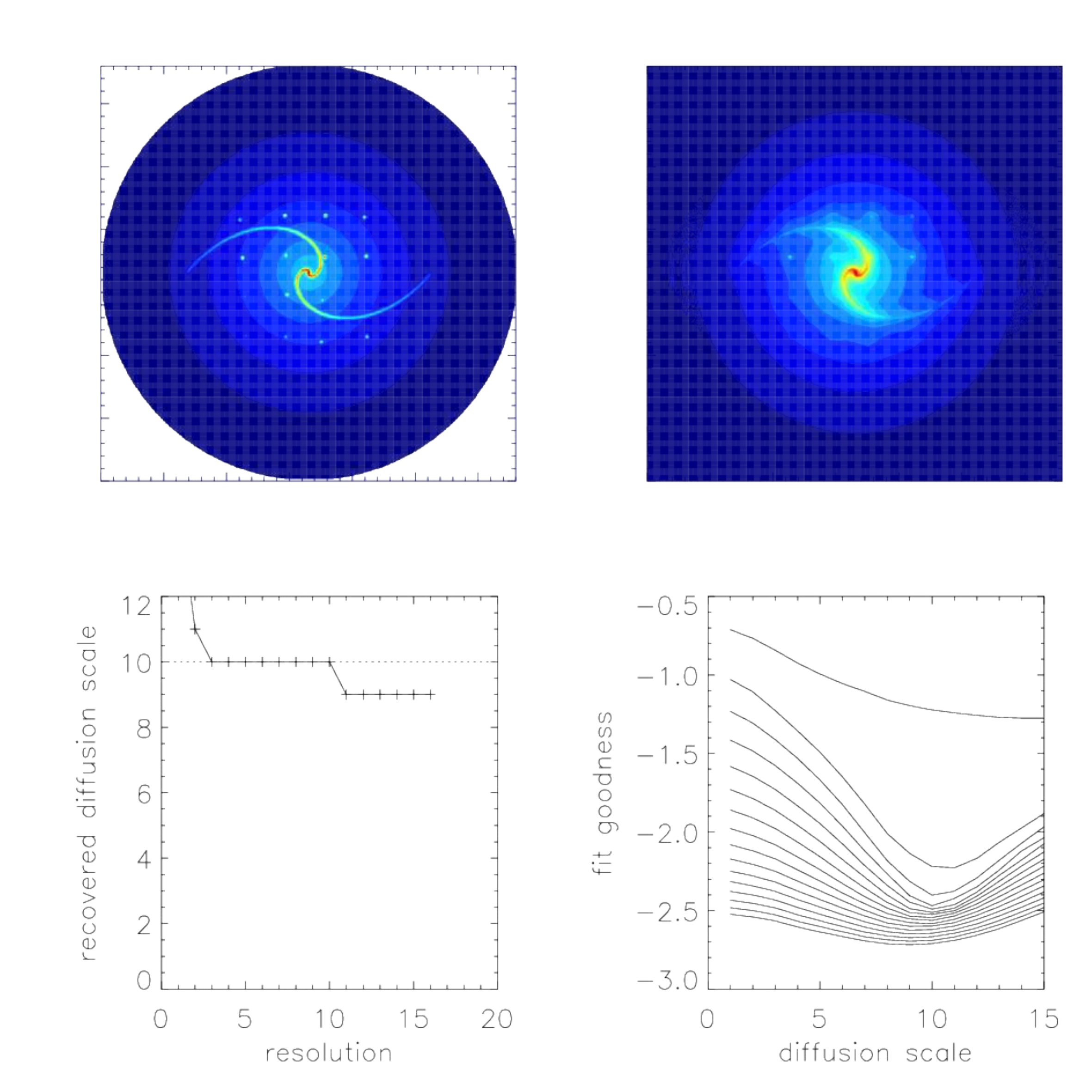}\includegraphics{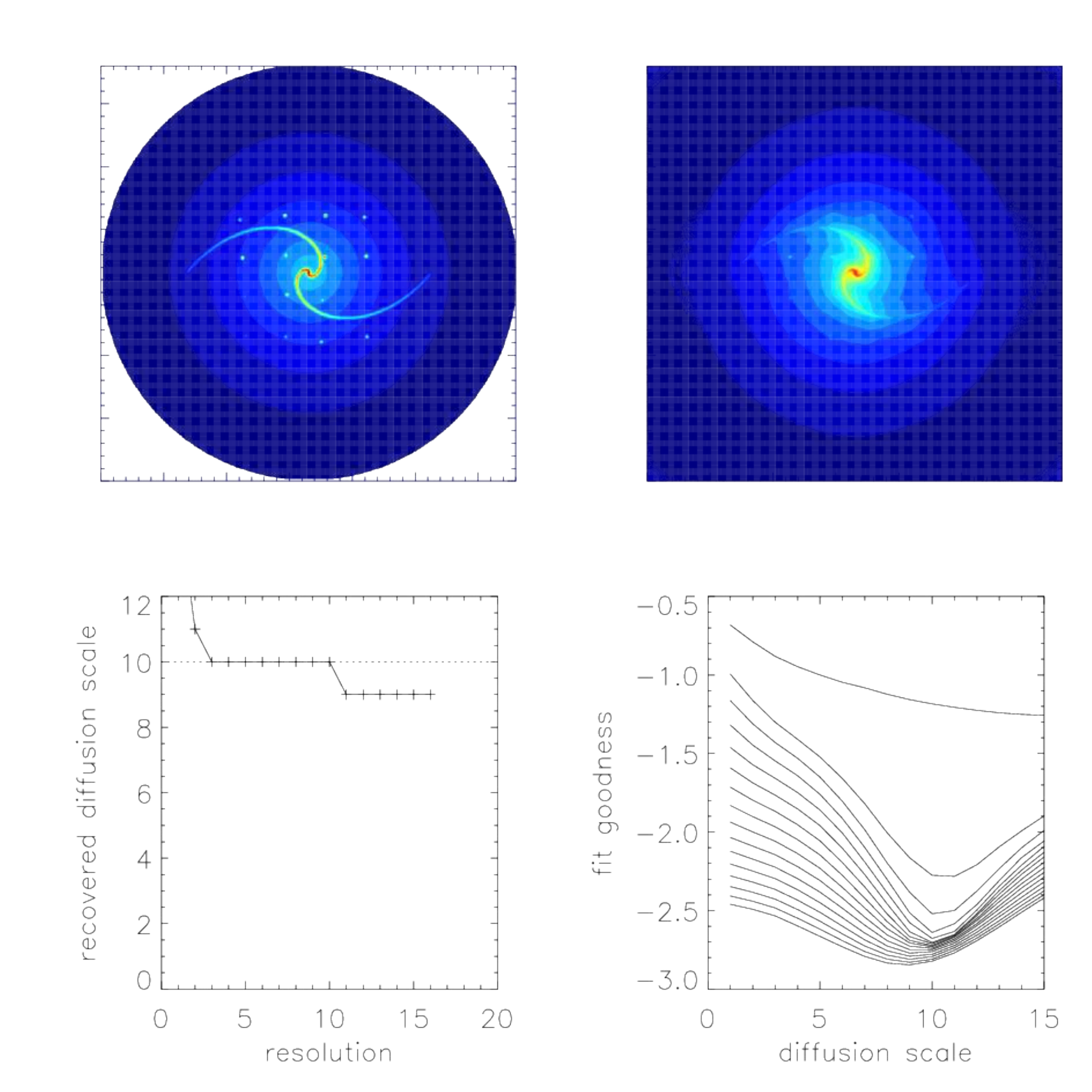}}
  \caption{Same as Fig.~\ref{fig:testgal_adapt_4b}, left row: Gaussian kernels, right row: exponential kernels, 
    but with smoothing lengthscales at a given local star formation rate equal to 8 pixels 
    (upper four panels) and 10 pixels (lower four panels). The resolution and diffusion scale correspond to $1.67 \, FWHM$.
  \label{fig:testgal_adapt_8b}}
\end{figure*}

\section{``Best fit'' model radio continuum maps}

\begin{figure*}[!ht]
  \centering
  \resizebox{16cm}{!}{\includegraphics{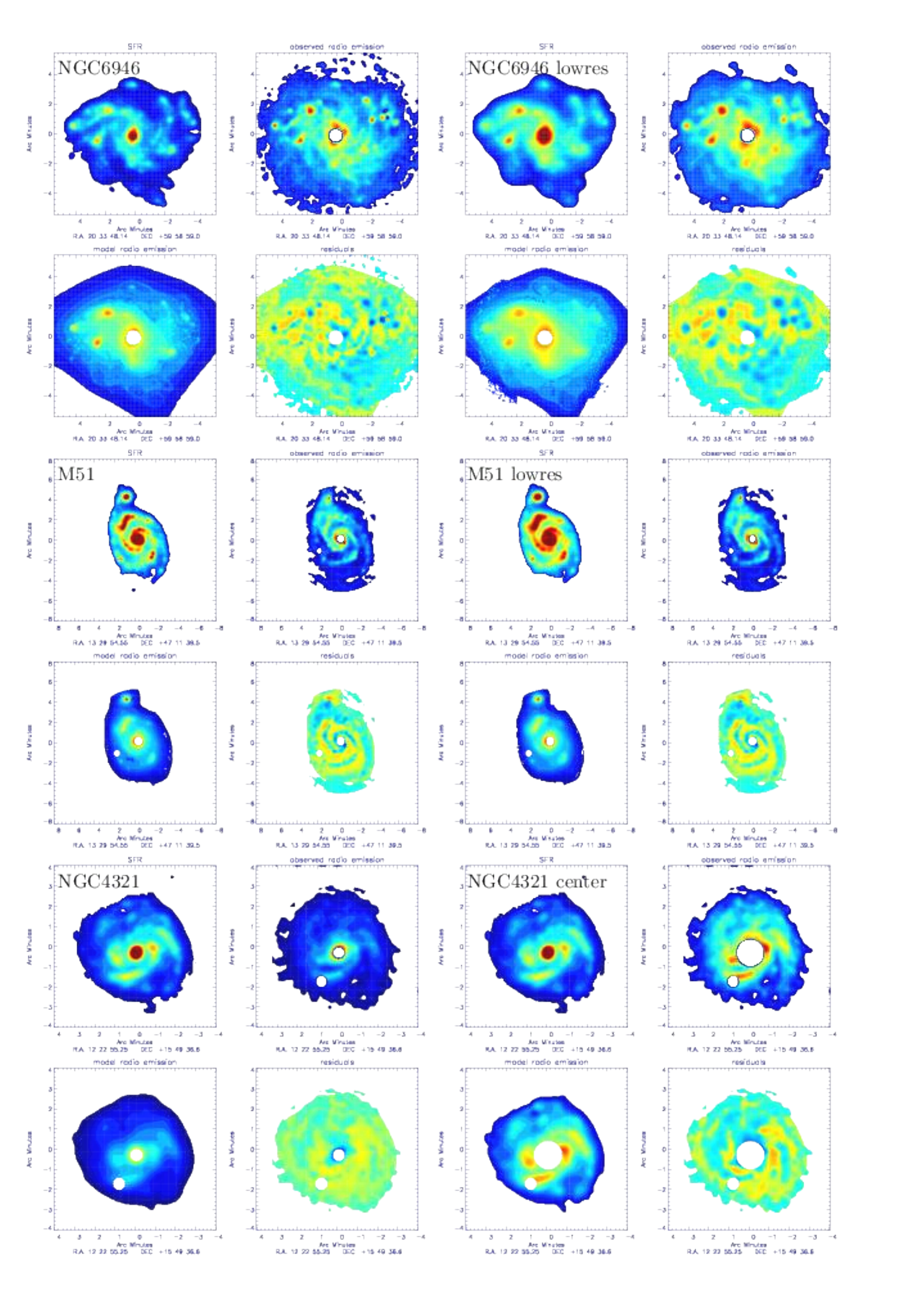}}
  \caption{Exponential convolution: ``best fit'' model radio continuum maps at $6$~cm. Four maps are shown for each galaxy.
    Upper left: observed star formation; upper right: observed radio continuum emission; lower left: model radio continuum;
    lower right: residuals; blue is radio-bright, red radio-dim.
  \label{fig:zusammenplots1exp}}
\end{figure*}

\section{Curves of goodness}

\begin{figure*}[!ht]
  \centering
  \resizebox{\hsize}{!}{\includegraphics{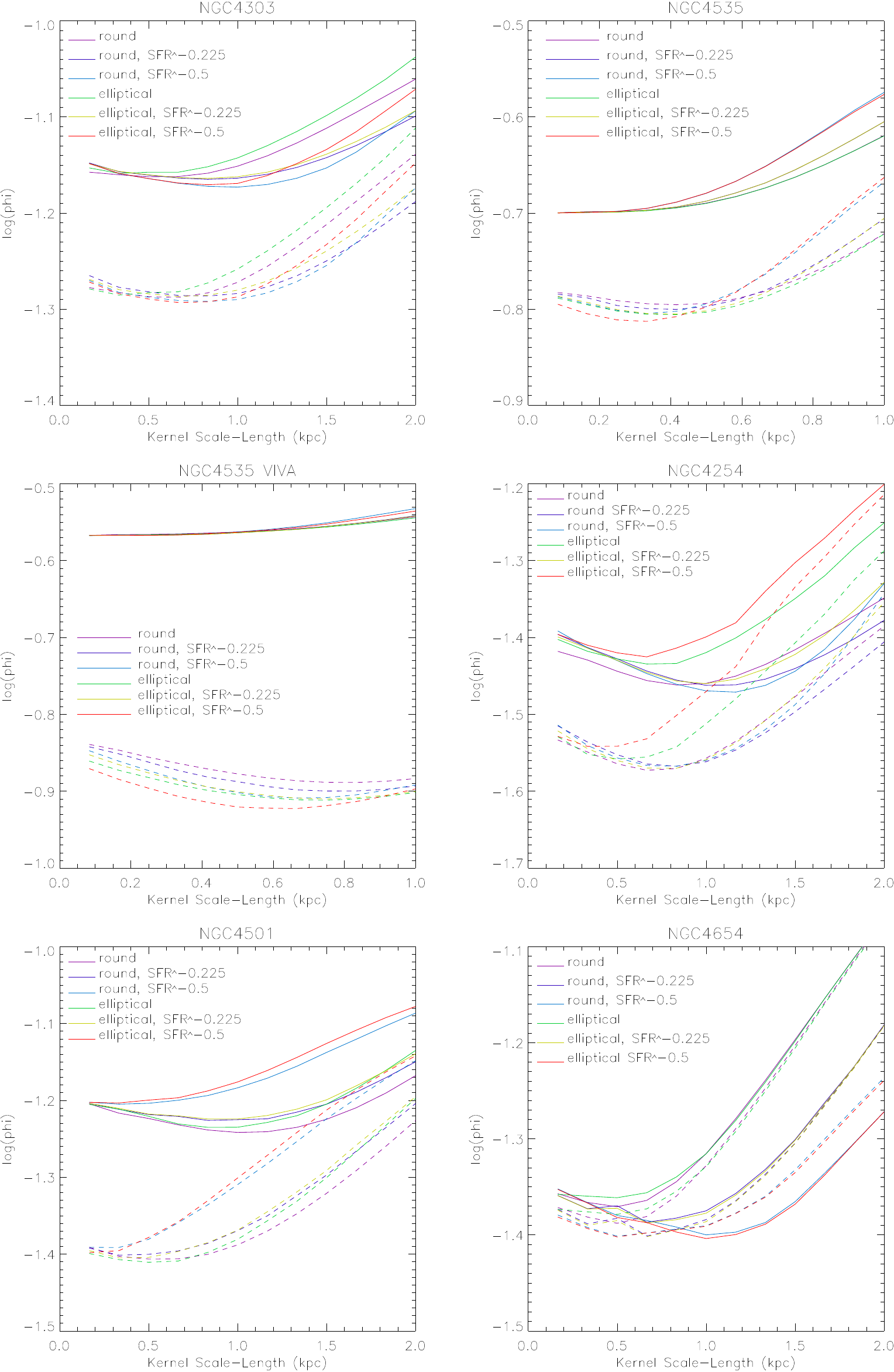}\includegraphics{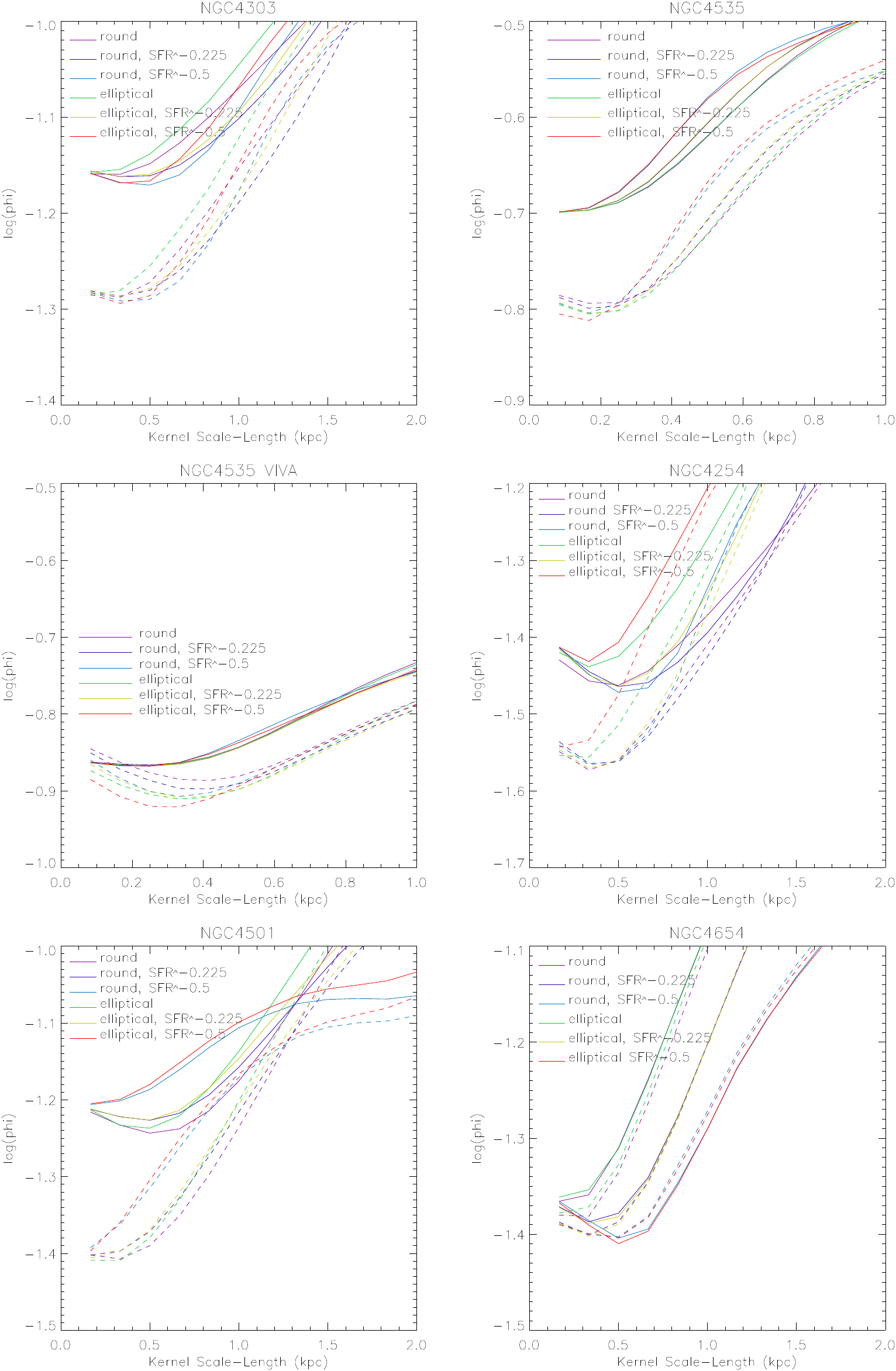}}
    \put(-315,290){Gaussian}
  \put(-445,290){Gaussian}
   \put(-60,290){exponential}
  \put(-190,290){exponential}
  \put(-315,155){Gaussian}
  \put(-445,155){Gaussian}
   \put(-60,155){exponential}
  \put(-190,155){exponential}
  \put(-315,20){Gaussian}
  \put(-445,20){Gaussian}
   \put(-60,20){exponential}
  \put(-190,20){exponential}
  \caption{Same as Fig.~\ref{fig:zusammen1}.
  \label{fig:zusammen2}}
\end{figure*}
\begin{figure*}[!ht]
  \centering
  \resizebox{\hsize}{!}{\includegraphics{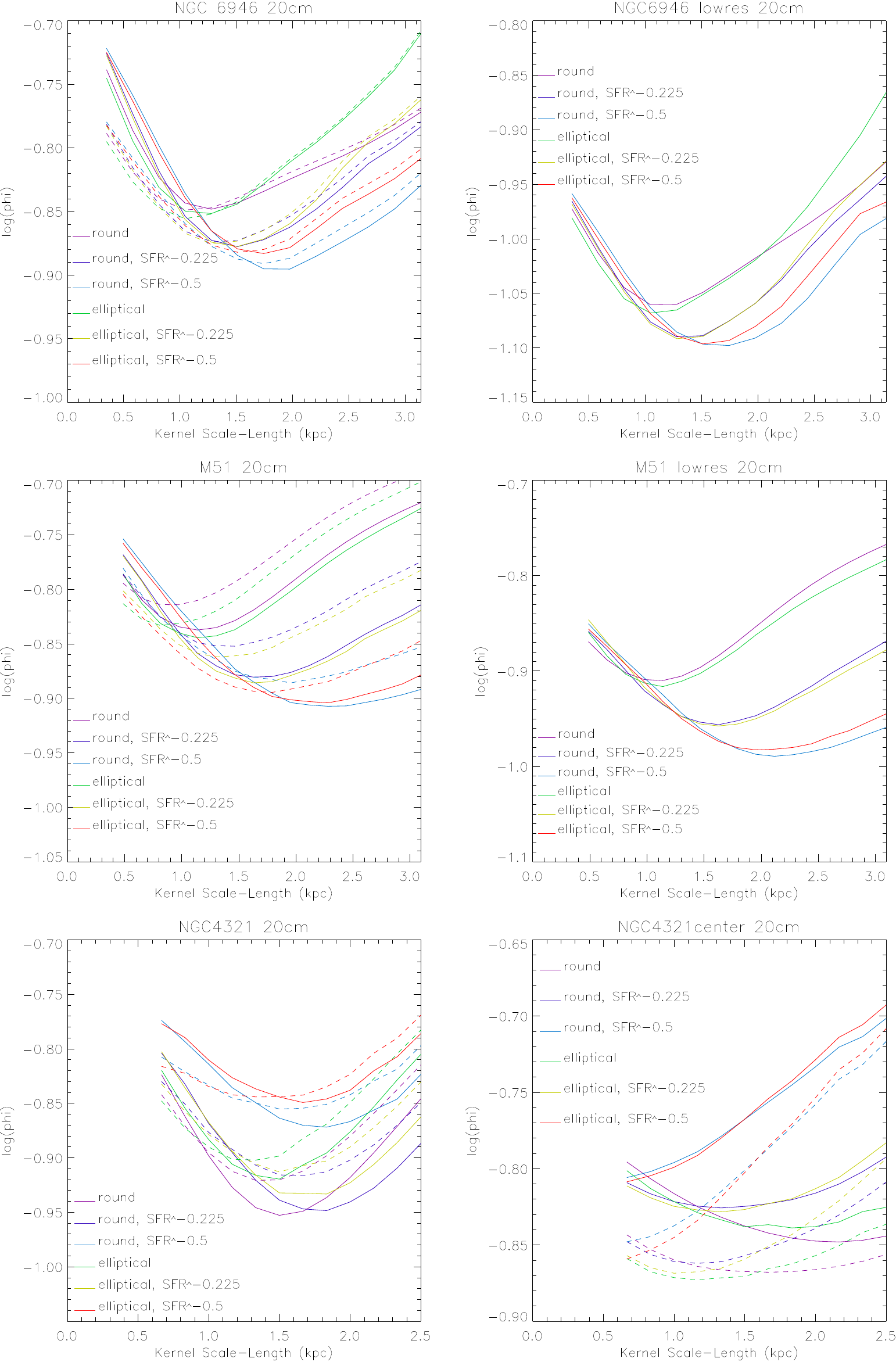}\includegraphics{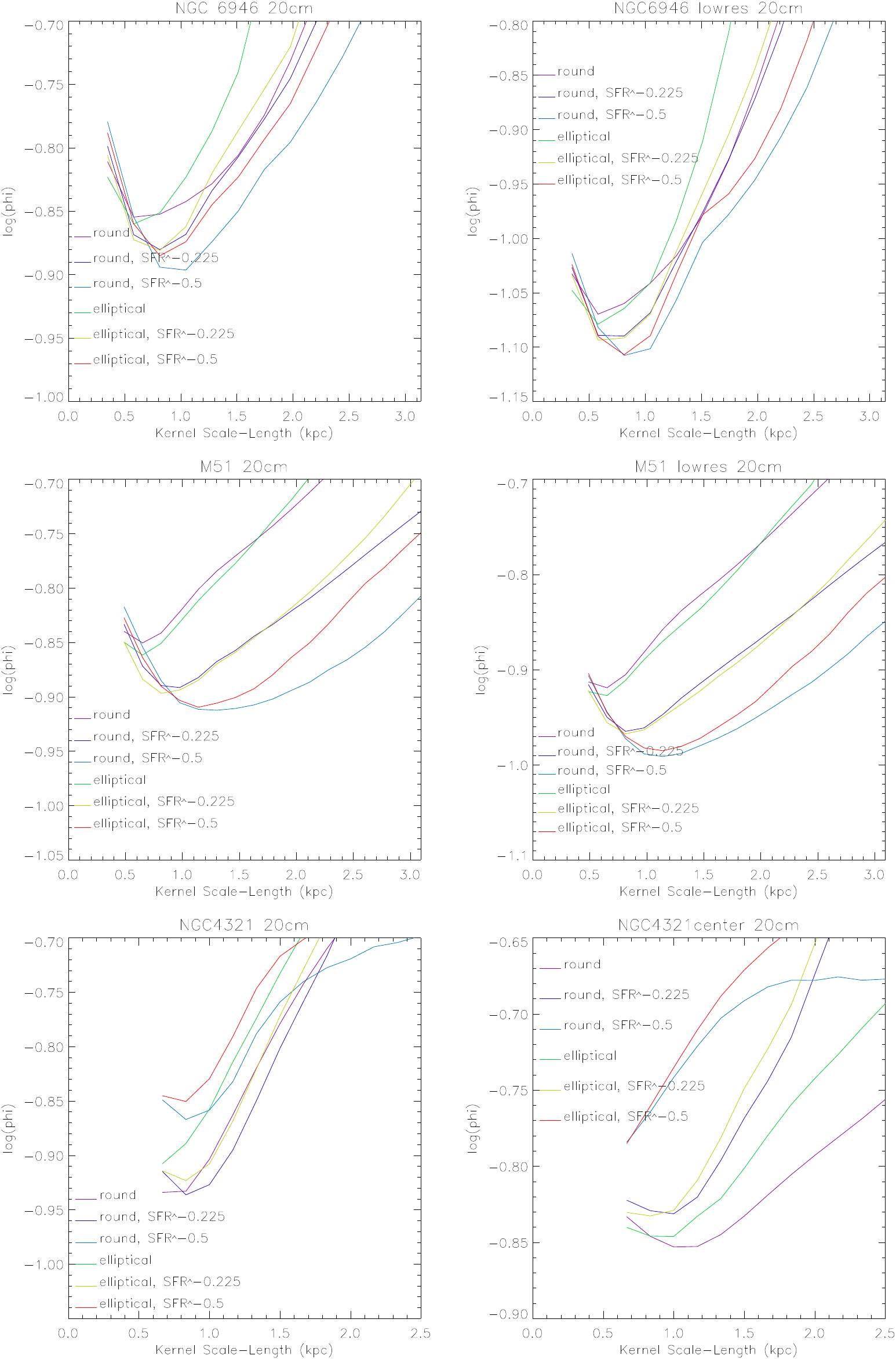}}
    \put(-315,290){Gaussian}
  \put(-445,290){Gaussian}
   \put(-60,290){exponential}
  \put(-190,290){exponential}
  \put(-315,155){Gaussian}
  \put(-445,155){Gaussian}
   \put(-60,155){exponential}
  \put(-190,155){exponential}
  \put(-315,20){Gaussian}
  \put(-445,20){Gaussian}
   \put(-60,20){exponential}
  \put(-190,20){exponential}
  \caption{Same as Fig.~\ref{fig:zusammen1}.
  \label{fig:zusammen3}}
\end{figure*}
\begin{figure*}[!ht]
  \centering
  \resizebox{\hsize}{!}{\includegraphics{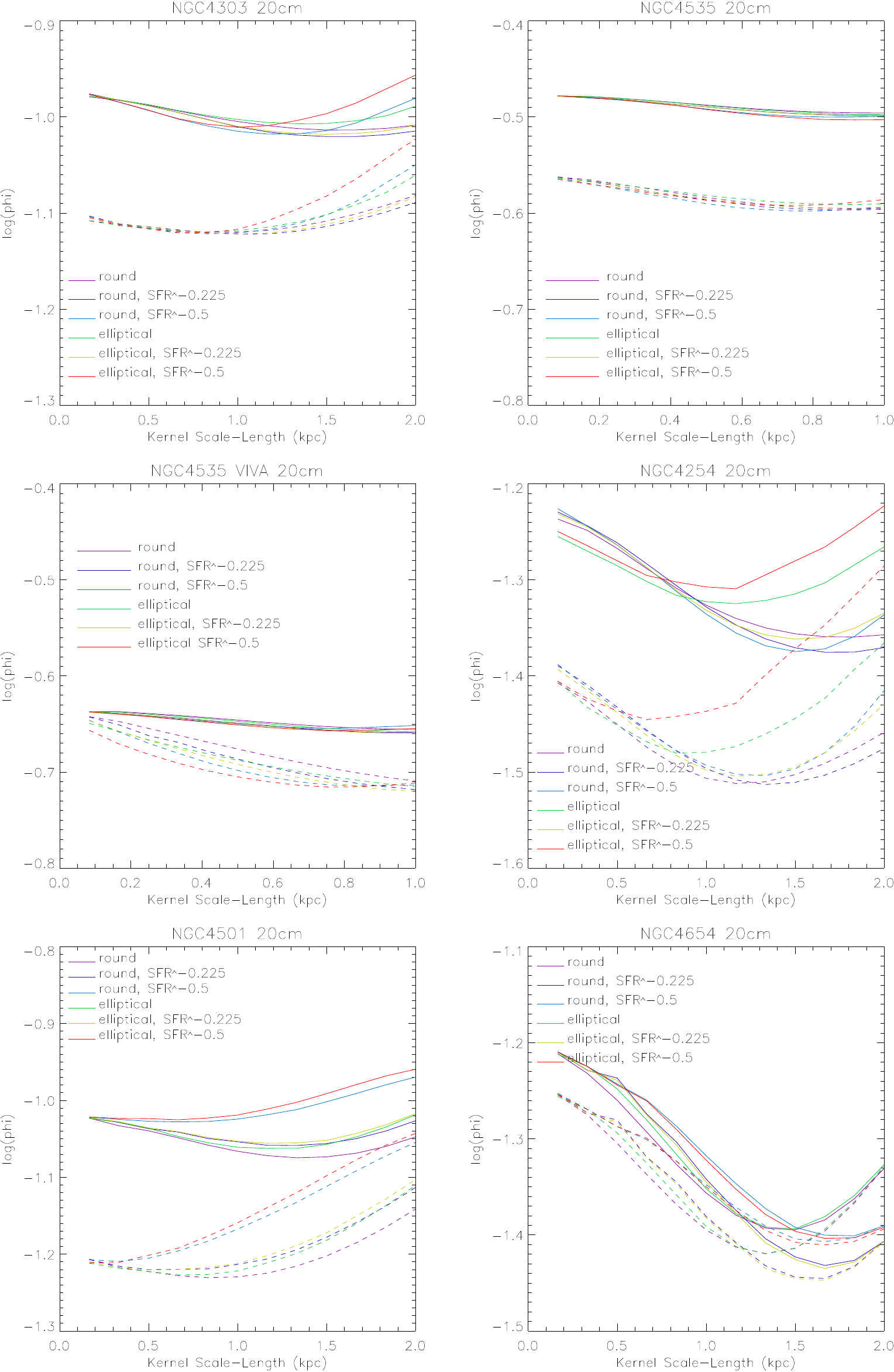}\includegraphics{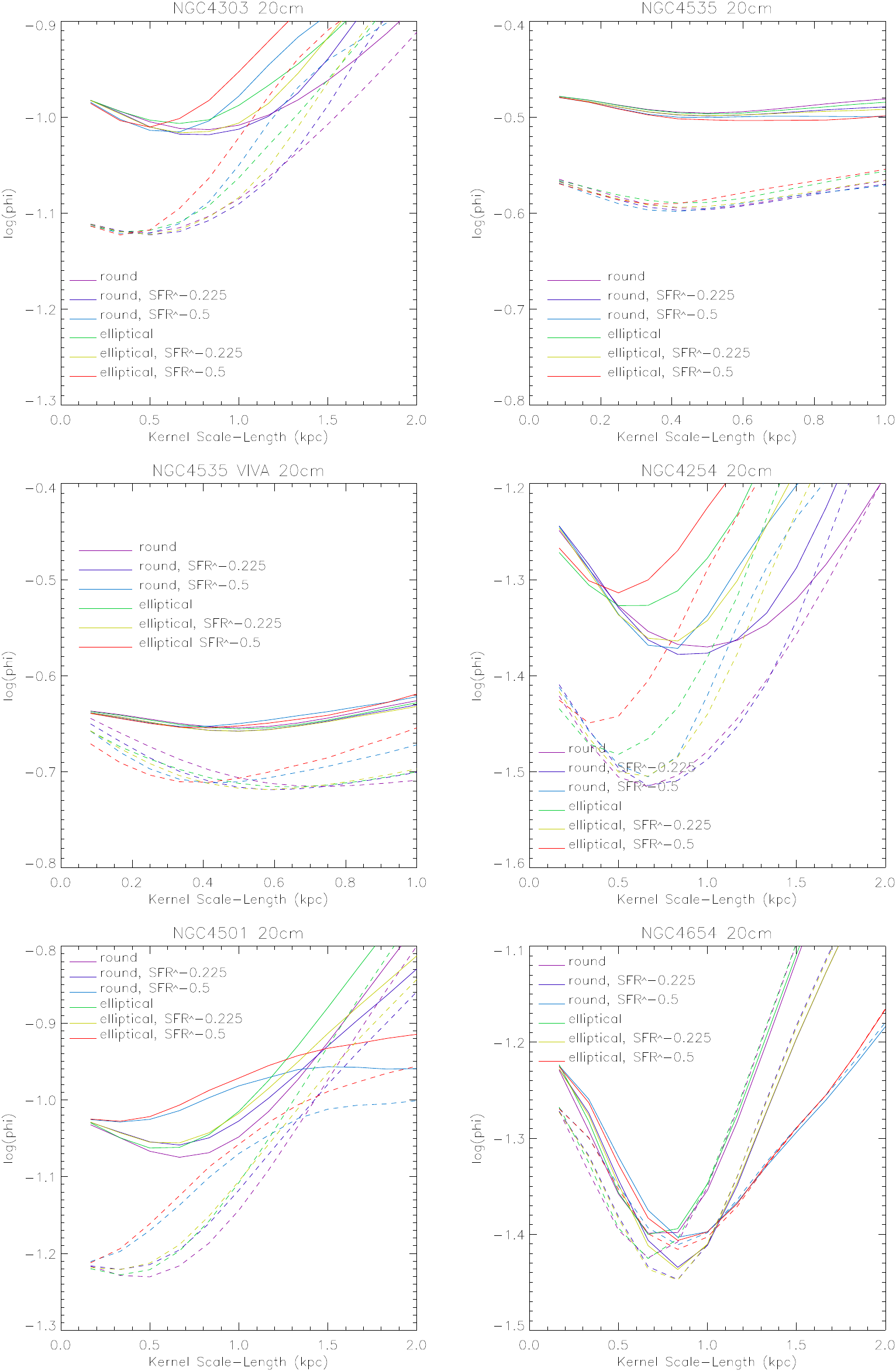}}
    \put(-315,290){Gaussian}
  \put(-445,290){Gaussian}
   \put(-60,290){exponential}
  \put(-190,290){exponential}
  \put(-315,155){Gaussian}
  \put(-445,155){Gaussian}
   \put(-60,155){exponential}
  \put(-190,155){exponential}
  \put(-315,20){Gaussian}
  \put(-445,20){Gaussian}
   \put(-60,20){exponential}
  \put(-190,20){exponential}
  \caption{Same as Fig.~\ref{fig:zusammen1}.
  \label{fig:zusammen4}}
\end{figure*}

\clearpage

\section{Tables}

\begin{table}[!ht]
\begin{center}
\caption{``Best fit'' models at 6~cm.\label{tab:table_bestfitcc}}
\begin{tabular}{lccc}
\hline
   Gaussian convolution & & & \\
\hline
	name &  $l$ &  $Q$ & $\log(\phi)$\\
\hline
             NGC6946 & 0.58 & 19.1 & -0.79 \\ 
      NGC6946 low resolution & 0.12 & 19.9 & -0.91 \\
      NGC6946 clipped & 0.47 & 19.2 & -0.82 \\
                 M51 & 0.65 & 16.8 & -0.81 \\
          M51 low resolution & 0.49 & 16.8 & -0.87 \\
          M51 clipped & 1.30 & 17.9 & -0.77 \\
             NGC4321 & 1.33 & 15.5 & -1.02 \\
      NGC4321 disk & 1.00 & 19.2 & -1.17 \\
             NGC4303 & 0.83 & 14.5 & -1.26 \\
             NGC4535 & 0.42 & 35.1 & -0.80 \\
        NGC4535 VIVA & 0.83 & 36.5 & -0.89 \\
             NGC4254 & 0.83 & 12.7 & -1.55 \\
             NGC4501 & 0.50 & 12.0 & -1.40 \\
             NGC4654 & 0.50 & 17.6 & -1.39 \\
\hline
      exponential convolution & & & \\
\hline
             NGC6946 &  0.35 & 19.1 & -0.80 \\ 
      NGC6946 low resolution &  0.12 & 19.8 & -0.91 \\
      NGC6946 clipped & 0.35 & 19.2 & -0.82 \\
                 M51 &  0.33 & 16.9 & -0.81 \\
          M51 low resolution &  0.16 & 17.1 & -0.87 \\
          M51 clipped & 1.14 & 17.9 & -0.77 \\
             NGC4321 &  0.67 & 15.4 & -1.00 \\
      NGC4321 disk &  0.50 & 19.2 & -1.17 \\
             NGC4303 &  0.50 & 14.8 & -1.26 \\
             NGC4535 &  0.17 & 34.8 & -0.81 \\
        NGC4535 VIVA &  0.33 & 36.5 & -0.92 \\
             NGC4254 &  0.33 & 12.8 & -1.55 \\
             NGC4501 &  0.33 & 11.9 & -1.41 \\
             NGC4654 &  0.83 & 18.2 & -1.41 \\
\hline
\end{tabular}
%\begin{tablenotes}
%  \item $^{\rm a}$ Moorwood et al. (1996), Pier et al. (1994)
%    \end{tablenotes}
\end{center}
\end{table}

\begin{table}[!ht]
\begin{center}
\caption{``Best fit'' models at 6~cm with $\SFR$ dependence.\label{tab:table_bestfitc}}
\begin{tabular}{lccccc}
\hline
        Gaussian & & & & & \\
\hline
	name &  $l$ &  $Q$ & $n$ & $e$ & $\log(\phi)$\\
\hline
             NGC6946 & 1.86 & 20.4 & -0.9 & 1 & -0.89 \\ 
      NGC6946 lr & 1.28 & 20.5 & -1.1 & 1 & -0.98 \\ 
      NGC6946 cl & 1.28 & 20.0 & -0.7 & -1 & -0.88 \\
                 M51 & 1.63 & 17.3 & -0.5 & 1 & -0.82 \\
          M51 lr & 1.47 & 17.4 & -0.7 & -1 & -0.89 \\
          M51 cl & 1.14 & 17.5 & -1.1 & -1 & -0.78 \\
             NGC4321 & 1.33 & 15.5 & 0.0 & -1 & -1.02 \\
      NGC4321 d & 1.00 & 19.2 & 0.0 & -1 & -1.17 \\
             NGC4303 & 0.83 & 14.5 & 0.0 & -1 & -1.26 \\
             NGC4535 & 0.33 & 34.8 & -0.5 & 1 & -0.81 \\
        N4535 V & 0.67 & 36.2 & -0.5 & 1 & -0.92 \\
             NGC4254 & 0.83 & 12.7 & 0.0 & -1 & -1.55 \\
             NGC4501 & 0.50 & 11.9 & 0.0 & 1 & -1.41 \\
             NGC4654 & 0.67 & 17.6 & -0.225 & 1 & -1.40 \\
\hline
        exponential & & & & & \\
\hline
             NGC6946 & 1.51 & 20.5 & -1.1 & -1 & -0.93 \\ 
      NGC6946 lr & 0.81 & 20.5 & -1.1 & -1 & -0.99 \\ 
      NGC6946 cl & 1.05 & 20.2 & -0.7 & 1 & -0.88 \\
                 M51 & 0.81 & 17.3 & -0.5 & 1 & -0.83 \\ 
          M51 lr & 0.65 & 17.1 & -0.5 & 1 & -0.89 \\ 
          M51 cl & 1.14 & 17.5 & -0.5 & 1 & -0.78 \\
             NGC4321 & 0.67 & 15.4 & 0.0 & -1 & -1.00 \\ 
      NGC4321 d & 0.50 & 19.2 & 0.0 & -1 & -1.17 \\ 
             NGC4303 & 0.50 & 14.8 & -0.5 & -1 & -1.26 \\ 
             NGC4535 & 0.17 & 34.8 & -0.5 & 1 & -0.81 \\ 
        NGC4535 V & 0.33 & 36.4 & -0.5 & 1 & -0.92 \\ 
             NGC4254 & 0.33 & 12.8 & 0.0 & -1 & -1.55 \\ 
             NGC4501 & 0.33 & 11.9 & 0.0 & 1 & -1.41 \\ 
             NGC4654 & 0.83 & 18.2 & -1.1 & 1 & -1.41 \\

\hline
\end{tabular}
%\begin{tablenotes}
%  \item $^{\rm a}$ Moorwood et al. (1996), Pier et al. (1994)
%    \end{tablenotes}
\end{center}
\end{table}

\begin{table}[!ht]
\begin{center}
\caption{``Best fit'' models at 6~cm with $\SFR$ dependence and losses.\label{tab:table_bestfit}}
\begin{tabular}{lccccccc}
\hline
       Gaussian & & & & & & & \\
\hline
	name &  $l$ &  $Q$ & $n$ & $e$ & $k$ & $c$ & $\log(\phi)$\\
\hline
             NGC6946 & 1.86 & 10.9 & -1.1 & 1 & 1.2 & 0.24 & -0.91 \\
      (NGC6946 lr & 0.81 & 9.4 & -0.5 & 1 & 1.2 & 0.32 & -1.03) \\
          NGC6946 lr & 0.93 & 14.5 & -0.5 & -1 & 1.6 & 0.04 & -1.01 \\
          NGC6946 cl & 1.16 & 10.7 & -0.7 & -1 & 1.2 & 0.24 & -0.89 \\
                 M51 & 1.63 & 17.3 & -0.5 & 1 & -   & 0.0  & -0.82 \\ 
          M51 lr & 1.47 & 17.4 & -0.5 & -1 & - & 0.0 & -0.89 \\
          M51 cl & 1.14 & 17.5 & -1.1 & -1 & - & 0.0 & -0.78 \\
             NGC4321 & 1.33 & 15.5 & 0.0 & -1 & - & 0.0 & -1.02 \\
      (NGC4321 d & 0.67 & 12.2 & 0.0 & -1 & 1.2 & 0.24 & -1.19) \\
             NGC4321 d & 0.67 & 15.6 & 0.0 & -1 & 1.6 & 0.05 & -1.19 \\
             NGC4303 & 0.67 & 12.3 & -0.5 & 1 & 2.0 & 0.006 & -1.29 \\ 
             NGC4535 & 0.33 & 34.8 & -0.5 & 1 & - & 0.0 & -0.81 \\
        NGC4535 V & 0.67 & 36.1 & -0.5 & 1 & - & 0.0 & -0.92 \\
             NGC4254 & 0.67 & 11.8 & 0.0 & -1 & 2.2 & 0.002 & -1.57 \\
             NGC4501 & 0.50 & 11.9 & 0.0 & 1 & - & 0.0 & -1.41 \\
             NGC4654 & 0.50 & 15.13 & -0.5 & 1 & 1.2 & 0.08 & -1.40 \\
\hline
       exponential & & & & & & & \\
\hline
              NGC6946 & 1.51 & 10.6 & -1.1 & -1 & 1.2 & 0.24 & -0.95 \\
      NGC6946 lr & 1.16 & 8.0 & -1.1 & 1 & 1.2 & 0.32 & -1.07 \\
      NGC6946 cl & 1.05 & 13.9 & -0.7 & 1 & 1.2 & 0.32 & -0.89 \\
                 M51  & 0.81 & 17.3 & -0.5 & 1 & - & 0.0 & -0.83 \\
          M51 lr & 0.65 & 17.1 & -0.5 & 1 & - & 0.0 & -0.89 \\
             NGC4321 & 0.67 & 15.4 & 0.0 & -1 & - & 0.0 & -1.00 \\
      NGC4321 d & 0.33 & 12.3 & 0.0 & -1 & 1.2 & 0.24 & -1.19 \\
             NGC4303  & 0.33 & 12.3 & -0.7 & 1 & 2.0 & 0.01 & -1.29 \\
             NGC4535 & 0.17 & 34.8 & -0.5 & 1 & - & 0.0 & -0.81 \\
        NGC4535 V & 0.33 & 36.5 & -0.5 & 1 & - & 0.0 & -0.92 \\
             NGC4254 & 0.33 & 11.8 & 0.0 & -1 & 2.2 & 0.002 & -1.57 \\
	     NGC451 & 0.33 & 11.9 & 0.0 & 1 & - & 0.0 & -1.41 \\
             NGC4654 & 0.83 & 18.2 & -1.1 & 1 & - & 0.0 & -1.41 \\
\hline
\end{tabular}
%\begin{tablenotes}
%  \item $^{\rm a}$ Moorwood et al. (1996), Pier et al. (1994)
%    \end{tablenotes}
\end{center}
\end{table}

%===================================================================================

\begin{table}[!ht]
\begin{center}
\caption{``Best fit'' models at 20~cm.\label{tab:table_bestfitcc_20}}
\begin{tabular}{lccc}
\hline
      Gaussian & & & \\
\hline
	name &  $l$ &  $Q$ & $\log(\phi)$\\
\hline
             NGC6946 & 1.28 & 6.7 & -0.81 \\
      NGC6946 low resolution & 1.28 & 6.7 & -1.00 \\
                 M51 & 1.14 & 5.4 & -0.84 \\
          M51 low resolution & 1.14 & 5.3 & -0.91 \\
             NGC4321 & 1.50 & 4.1 & -0.95 \\
      NGC4321 disk & 2.17 & 4.8 & -0.85 \\
             NGC4303 & 2.00 & 4.8 & -1.02 \\
             NGC4535 & 0.92 & 8.6 & -0.60 \\
        NGC4535 VIVA & 1.00 & 12.0 & -0.71 \\
             NGC4254 & 1.67 & 4.2 & -1.44 \\
             NGC4501 & 0.83 & 2.6 & -1.23 \\
             NGC4654 & 1.33 & 5.0 & -1.42 \\
\hline
exponential & & & \\
\hline
                 NGC6946 & 1.05 & 6.2 & -0.83 \\
      NGC6946 low resolution & 1.05 & 6.3 & -1.02 \\
                 M51 & 0.65 & 5.3 & -0.85 \\
          M51 low resolution & 0.65 & 5.2 & -0.92 \\
             NGC4321 & 0.67 & 4.0 & -0.93 \\
      NGC4321 disk & 1.00 & 4.9 & -0.85 \\
             NGC4303 & 1.17 & 4.5 & -1.03 \\
             NGC4535 & 0.42 & 8.7 & -0.60 \\
        NGC4535 VIVA & 0.58 & 12.0 & -0.76 \\
             NGC4254 & 1.00 & 4.1 & -1.46 \\
             NGC4501 & 0.50 & 2.6 & -1.23 \\
             NGC4654 & 0.83 & 5.0 & -1.45 \\
\hline
\end{tabular}
%\begin{tablenotes}
%  \item $^{\rm a}$ Moorwood et al. (1996), Pier et al. (1994)
%    \end{tablenotes}
\end{center}
\end{table}

\begin{table}[!ht]
\begin{center}
\caption{``Best fit'' models at 20~cm with $\SFR$ dependence.\label{tab:table_bestfitc_20}}
\begin{tabular}{lccccc}
\hline
        Gaussian & & & & & \\
\hline
	name &  $l$ &  $Q$ & $n$ & $e$ & $\log(\phi)$\\
\hline
             NGC6946 & 1.74 & 7.0 & -0.5 & -1 & -0.87 \\
      N6946 lr & 1.74 & 7.0 & -0.7 & 1 & -1.05 \\
                 M51 & 2.44 & 5.6 & -0.5 & -1 & -0.91 \\
          M51 lr & 2.12 & 5.6 & -0.5 & -1 & -0.99 \\
             NGC4321 & 1.50 & 4.1 & 0.0 & -1 & -0.95 \\
      NGC4321 d & 2.16 & 4.8 & 0.0 & -1 & -0.85 \\
             NGC4303 & 2.00 & 5.0 & -0.5 & -1 & -1.02 \\
             NGC4535 & 0.75 & 8.7 & -0.5 & -1 & -0.60 \\
        N4535 V & 1.00 & 11.5 & -0.225 & 1 & -0.77 \\
             NGC4254 & 1.67 & 4.2 & 0.0 & -1 & -1.44 \\
             NGC4501 & 0.83 & 2.6 & 0.0 & -1 & -1.23 \\
             NGC4654 & 1.67 & 5.0 & -0.225 & 1 & -1.45 \\
\hline
         exponential & & & & & \\
\hline
                  NGC6946 & 1.05 & 6.9 & -0.5 & -1 & -0.87 \\
      NGC6946 lr & 0.81 & 7.0 & -0.7 & 1 & -1.06 \\
                 M51 & 1.47 & 5.4 & -0.5 & -1 & -0.91 \\
          M51 lr & 1.14 & 5.5 & -0.5 & -1 & -0.99 \\
             NGC4321 & 0.83 & 4.1 & -0.225 & -1 & -0.94 \\
      NGC4321 d & 1.00 & 4.9 & 0.0 & -1 & -0.85 \\
             NGC4303 & 1.17 & 4.5 & 0.0 & -1 & -1.03 \\
             NGC4535 & 0.42 & 8.73 & -0.5 & -1 & -0.60 \\
        NGC4535 V & 0.50 & 12.0 & -0.225 & 1 & -0.77 \\
             NGC4254 & 1.00 & 4.1 & 0.0 & -1 & -1.46 \\
             NGC4501 & 0.50 & 2.6 & 0.0 & -1 & -1.23 \\
             NGC4654 & 0.83 & 5.0 & -0.225 & 1 & -1.45 \\
\hline
\end{tabular}
%\begin{tablenotes}
%  \item $^{\rm a}$ Moorwood et al. (1996), Pier et al. (1994)
%    \end{tablenotes}
\end{center}
\end{table}

\begin{table}[!ht]
\begin{center}
\caption{``Best fit'' models at 20~cm with $\SFR$ dependence and losses.\label{tab:table_bestfit_20}}
\begin{tabular}{lccccccc}
\hline
        Gaussian & & & & & & & \\
\hline
	name &  $l$ &  $Q$ & $n$ & $e$ & $k$ & $c$ & $\log(\phi)$\\
\hline
             NGC6946 & 1.98 & 3.6 & -0.5 & -1 & 1.2 & 0.24 & -0.90 \\
      NGC6946 lr & 1.16 & 3.7 & -0.7 & -1 & 1.2 & 0.24 & -1.10 \\
                 M51 & 2.28 & 4.7 & -0.5 & -1 & 1.2 & 0.08 & -0.91 \\
          M51 lr & 2.11 & 4.6 & -0.5 & -1 & 1.2 & 0.08 & -0.99 \\
             NGC4321 & 1.50 & 4.1 & 0.0 & -1 & - & 0.0 & -0.95 \\
      NGC4321 d & 2.17 & 4.8 & 0.0 & -1 & - & 0.0 & -0.85 \\
             (NGC4303 & 1.00 & 2.7 & -0.225 & -1 & 1.4 & 0.13 & -1.12) \\
      NGC4303 & 1.20 & 3.3 & -0.225 & -1 & 1.8 & 0.03 & -1.12 \\
             NGC4535 & 0.75 & 7.7 & -0.5 & -1 & 1.2 & 0.08 & -0.60 \\
        NGC4535 V & 1.00 & 11.3 & -0.225 & 1 & - & 0.0 & -0.77 \\
             NGC4254 & 1.33 & 3.6 & -0.225 & -1 & 2.0 & 0.006 & -1.51 \\
             NGC4501 & 0.83 & 2.6 & 0.0 & -1 & - & 0.0 & -1.23 \\
             NGC4654 & 1.67 & 5.0 & -0.225 & 1 & - & 0.0 & -1.45 \\
\hline
       exponential & & & & & & & \\
\hline
             NGC6946 & 1.05 & 3.6 & -1.1 & -1 & 1.2 & 0.24 & -0.90 \\
      NGC6946 lr & 0.81 & 3.7 & -0.5 & -1 & 1.2 & 0.24 & -1.11 \\
                 M51 & 1.30 & 4.6 & -0.5 & -1 & 1.2 & 0.08 & -0.91 \\
          M51 lr & 1.14 & 4.6 & -0.5 & -1 & 1.2 & 0.08 & -0.99 \\
             NGC4321 & 0.83 & 4.1 & -0.225 & -1 & - & 0.0 & -0.94 \\
      NGC4321 d  & 1.00 & 4.9 & 0.0 & -1 & - & 0.0 & -0.85 \\
             NGC4303 & 0.33 & 4.5 & -0.5 & 1 & 1.4 & 0.13 & -1.12 \\
             NGC4535 & 0.42 & 8.3 & -0.5 & -1 & 1.6 & 0.01 & -0.60 \\
        NGC4535 V & 0.50 & 12.0 & -0.225 & 1 & - & 0.0 & -0.77 \\
             NGC4254 & 0.67 & 3.6 & 0.0 & -1 & 2.0 & 0.006 & -1.52 \\
             NGC4501 & 0.50 & 2.6 & 0.0 & -1 & - & 0.0 & -1.23 \\
             NGC4654 & 0.83 & 5.0 & -0.225 & 1 & - & 0.0 & -1.45 \\
\hline
\end{tabular}
%\begin{tablenotes}
%  \item $^{\rm a}$ Moorwood et al. (1996), Pier et al. (1994)
%    \end{tablenotes}
\end{center}
\end{table}

\clearpage

%===================================================================================================================

\begin{table}[!ht]
\begin{center}
\caption{Most probable parameters for the model with $\SFR$ dependence at 6~cm.\label{tab:table_6cm_ccn}}
\begin{tabular}{lccccccc}
\hline
       Gaussian & & & & \\
\hline
	name &  $l$ &  $Q$ & $n$ & $e$ \\
\hline
             NGC6946 &   1.69 $\pm$   0.12 &   20.4 $\pm$    0.2 &  -0.93 &   0.12 \\ 
      NGC6946 lowres &   1.22 $\pm$   0.43 &   19.6 $\pm$    4.2 &  -0.93 &  -0.04 \\ 
      NGC6946 clipped & 1.29 $\pm$ 0.17 & 20.1 $\pm$ 0.19 & -0.71 & -0.50 \\
                 M51 &   1.63 $\pm$   0.24 &   17.6 $\pm$    0.3 &  -0.53 &   0.17 \\ 
          M51 lowres &   1.36 $\pm$   0.18 &   17.2 $\pm$    0.3 &  -0.58 &   0.39 \\
          M51 clip   &  1.02  $\pm$   0.28 &   17.4 $\pm$    0.4 &  -0.94 &   0.00 \\
             NGC4321 &   1.46 $\pm$   0.23 &   15.6 $\pm$    0.2 &  -0.15 &  -0.50 \\ 
      NGC4321 disk &   0.94 $\pm$   0.13 &   19.3 $\pm$    0.2 &  -0.08 &   0.00 \\ 
             NGC4303 &   0.97 $\pm$   0.18 &   14.6 $\pm$    0.1 &  -0.25 &  -1.00 \\ 
             NGC4535 &   0.35 $\pm$   0.08 &   34.9 $\pm$    0.1 &  -0.36 &   1.00 \\ 
        NGC4535 VIVA &   0.62 $\pm$   0.14 &   36.2 $\pm$    0.1 &  -0.50 &   1.00 \\ 
             NGC4254 &   0.78 $\pm$   0.08 &   12.8 $\pm$    0.1 &  -0.21 &  -0.33 \\ 
             NGC4501 &   0.50 $\pm$   0.14 &   11.9 $\pm$    0.1 &  -0.04 &   0.33 \\ 
             NGC4654 &   0.69 $\pm$   0.18 &   17.7 $\pm$    0.1 &  -0.42 &   0.33 \\ 
\hline
      exponential & & & & \\
\hline
      NGC6946 &   1.47 $\pm$   0.25 &   20.4 $\pm$    0.1 &  -1.09 &   0.12 \\ 
      NGC6946 lowres &   0.82 $\pm$   0.22 &   20.7 $\pm$    0.0 &  -1.06 &   0.38 \\ 
                 M51 &   0.79 $\pm$   0.20 &   17.3 $\pm$    0.5 &  -0.47 &   0.00 \\ 
          M51 lowres &   0.73 $\pm$   0.13 &   17.2 $\pm$    0.1 &  -0.68 &   0.17 \\ 
             NGC4321 &   0.68 $\pm$   0.16 &   15.5 $\pm$    0.2 &  -0.15 &  -0.17 \\ 
      NGC4321 disk &   0.50 $\pm$   0.14 &   19.4 $\pm$    0.3 &  -0.12 &   0.00 \\ 
             NGC4303 &   0.49 $\pm$   0.11 &   14.7 $\pm$    0.1 &  -0.33 &  -0.17 \\ 
             NGC4535 &   0.17 $\pm$   0.05 &   34.9 $\pm$    0.2 &  -0.29 &   0.67 \\ 
        NGC4535 VIVA &   0.35 $\pm$   0.06 &   36.5 $\pm$    0.1 &  -0.29 &   1.00 \\ 
             NGC4254 &   0.37 $\pm$   0.10 &   12.8 $\pm$    0.1 &  -0.28 &  -0.17 \\ 
             NGC4501 &   0.28 $\pm$   0.13 &   11.9 $\pm$    0.1 &  -0.17 &   0.00 \\ 
             NGC4654 &   0.69 $\pm$   0.33 &   17.9 $\pm$    0.2 &  -0.76 &   0.00 \\ 
\hline
\end{tabular}
%\begin{tablenotes}
%  \item $^{\rm a}$ Moorwood et al. (1996), Pier et al. (1994)
%    \end{tablenotes}
\end{center}
\end{table}

\begin{table*}[!ht]
\begin{center}
\caption{Most probable parameters for the model with $\SFR$ dependence and advection losses at 6~cm.\label{tab:table_6cm}}
\begin{tabular}{lccccccc}
\hline
        Gaussian & & & & & & \\
\hline
name &  $l$ &  $Q$ & $n$ & $e$ & $k$ & $c$ \\
\hline
             NGC6946 &   1.72 $\pm$   0.14 &   13.5 $\pm$    2.7 &  -0.96 &   0.18 &   1.36 $\pm$   0.19 & 1.3E-01 $\pm$ 9.3E-02 \\ 
      (NGC6946 lowres &   0.71 $\pm$   0.36 &    9.0 $\pm$    2.8 &  -0.45 &  -0.08 &   1.23 $\pm$   0.27 & 2.3E-01 $\pm$ 1.1E-01) \\ 
       NGC6946 lowres &   0.82 $\pm$   0.23 &   14.5 $\pm$    1.0 &  -0.50 &   0.08 &   1.67 $\pm$   0.10 & 3.5E-02 $\pm$ 1.4E-02 \\ 
       NGC6946 clipped & 1.12 $\pm$ 0.19 & 13.5 $\pm$ 2.4 & -0.78 & -0.51 & 1.30 $\pm$ 0.14 & 1.4E-01 $\pm$ 8.4E-03 \\
                 M51 &   1.50 $\pm$   0.23 &   16.9 $\pm$    0.9 &  -0.54 &   0.49 &   1.71 $\pm$   0.37 & 9.2E-03 $\pm$ 2.3E-02 \\ 
          M51 lowres &   1.36 $\pm$   0.18 &   17.2 $\pm$    0.3 &  -0.58 &   0.39 &   - & 0.00 \\ 
          M51 clip & 1.00 $\pm$ 0.27 & 17.2 $\pm$ 0.7 & -1.0 & 0.01 & 1.69 $\pm$ 0.35 & 2.9E-3 $\pm$ 1.3E-02 \\
             NGC4321 &   1.46 $\pm$   0.23 &   15.6 $\pm$    0.2 &  -0.15 &  -0.50 &   - & 0.00 \\ 
      (NGC4321 disk &   0.68 $\pm$   0.13 &   13.4 $\pm$    2.1 &  -0.11 &  -0.39 &   1.37 $\pm$   0.17 & 1.6E-01 $\pm$ 1.0E-01) \\
             NGC4321 disk &   0.72 $\pm$   0.11 &   16.3 $\pm$    0.8 &  -0.11 &  -0.33 &   1.73 $\pm$   0.15 & 3.4E-02 $\pm$ 1.7E-02 \\
             NGC4303 &   0.78 $\pm$   0.14 &   11.6 $\pm$    1.5 &  -0.50 &   0.06 &   1.74 $\pm$   0.31 & 4.6E-02 $\pm$ 7.3E-02 \\ 
             NGC4535 &   0.31 $\pm$   0.07 &   34.4 $\pm$    0.8 &  -0.55 &   1.00 &   1.69 $\pm$   0.38 & 4.4E-03 $\pm$ 1.4E-02 \\ 
        NGC4535 VIVA &   0.59 $\pm$   0.11 &   36.1 $\pm$    0.2 &  -0.50 &   1.00 &   1.76 $\pm$   0.36 & 1.8E-04 $\pm$ 5.1E-04 \\ 
             NGC4254 &   0.76 $\pm$   0.09 &   11.9 $\pm$    0.3 &  -0.18 &  -0.39 &   1.99 $\pm$   0.24 & 4.9E-03 $\pm$ 7.3E-03 \\ 
             NGC4501 &   0.50 $\pm$   0.14 &   11.9 $\pm$    0.1 &  -0.04 &   0.33 &   - & 0.00 \\ 
             NGC4654 &   0.68 $\pm$   0.18 &   17.4 $\pm$    0.8 &  -0.41 &   0.32 &   1.64 $\pm$   0.36 & 8.5E-03 $\pm$ 2.4E-02 \\ 
\hline
        exponential & & & & & & \\
\hline
           NGC6946 &   1.52 $\pm$   0.21 &   15.3 $\pm$    2.6 &  -1.10 &  -0.53 &   1.52 $\pm$   0.31 & 7.8E-02 $\pm$ 8.2E-02 \\ 
      NGC6946 lowres &   0.91 $\pm$   0.23 &    12.1 $\pm$    2.8 &  -1.08 &   0.35 &   1.48 $\pm$   0.27 & 1.4E-01 $\pm$ 1.2E-01 \\ 
                 M51 &   0.80 $\pm$   0.18 &   16.7 $\pm$    0.8 &  -0.48 &   0.44 &   1.70 $\pm$   0.35 & 7.9E-03 $\pm$ 2.1E-02 \\ 
          M51_lowres &   0.73 $\pm$   0.13 &   17.2 $\pm$    0.3 &  -0.68 &   0.17 &   1.70 $\pm$   0.34 & 0.0E+00 $\pm$ 0.0E+00 \\ 
             NGC4321 &   0.68 $\pm$   0.16 &   15.4 $\pm$    0.4 &  -0.15 &  -0.20 &   1.68 $\pm$   0.35 & 3.2E-03 $\pm$ 1.6E-02 \\ 
      NGC4321 disk &   0.35 $\pm$   0.11 &   14.4 $\pm$    2.4 &  -0.14 &  -0.09 &   1.49 $\pm$   0.25 & 1.2E-01 $\pm$ 1.0E-01 \\ 
             NGC4303 &   0.36 $\pm$   0.06 &   11.5 $\pm$    1.5 &  -0.54 &  -0.06 &   1.75 $\pm$   0.32 & 4.5E-02 $\pm$ 7.0E-02 \\ 
             NGC4535 &   0.15 $\pm$   0.03 &   34.6 $\pm$    0.8 &  -0.38 &   0.69 &   1.73 $\pm$   0.35 & 3.6E-03 $\pm$ 1.4E-02 \\ 
        NGC4535 VIVA &   0.33 $\pm$   0.05 &   36.3 $\pm$    0.7 &  -0.38 &   1.00 &   1.74 $\pm$   0.35 & 2.6E-03 $\pm$ 1.3E-02 \\ 
             NGC4254 &   0.39 $\pm$   0.09 &   11.8 $\pm$    0.5 &  -0.26 &  -0.49 &   1.89 $\pm$   0.29 & 1.1E-02 $\pm$ 2.3E-02 \\ 
             NGC4501 &   0.23 $\pm$   0.08 &   11.9 $\pm$    0.4 &  -0.18 &   0.11 &   1.72 $\pm$   0.36 & 3.6E-03 $\pm$ 1.6E-02 \\ 
             NGC4654 &   0.68 $\pm$   0.33 &   17.9 $\pm$    0.5 &  -0.74 &   0.00 &   1.69 $\pm$   0.35 & 2.2E-03 $\pm$ 1.3E-02 \\ 
\hline
\end{tabular}
%\begin{tablenotes}
%  \item $^{\rm a}$ Moorwood et al. (1996), Pier et al. (1994)
%    \end{tablenotes}
\end{center}
\end{table*}

%++++++++++++++++++++++++++++++++++++++++++++++++++++++++++++++++++++++++++++++++++++++++++++++++++++++++++++++++++++++++++++++++++++++++=

\begin{table}[!ht]
\begin{center}
\caption{Most probable parameters for the model with $\SFR$ dependence at 20~cm.\label{tab:table_20cm_ccn}}
\begin{tabular}{lccccccc}
\hline
       Gaussian & & & & \\
\hline
	name &  $l$ &  $Q$ & $n$ & $e$ \\
\hline
             NGC6946 &   2.14 $\pm$   0.41 &    7.2 $\pm$    0.1 &  -0.70 &  -0.38 \\ 
      NGC6946 lowres &   1.78 $\pm$   0.22 &    7.0 $\pm$    0.3 &  -0.68 &  -0.08 \\ 
                 M51 &   2.48 $\pm$   0.40 &    5.6 $\pm$    0.1 &  -0.54 &  -0.29 \\ 
          M51 lowres &   2.38 $\pm$   0.27 &    5.6 $\pm$    0.2 &  -0.60 &  -0.25 \\ 
             NGC4321 &   1.71 $\pm$   0.23 &    4.1 $\pm$    0.1 &  -0.17 &  -0.50 \\ 
      NGC4321 disk &   2.08 $\pm$   0.29 &    4.9 $\pm$    0.1 &   0.00 &  -1.00 \\ 
             NGC4303 &   1.83 $\pm$   0.14 &    4.9 $\pm$    0.1 &  -0.12 &  -1.00 \\ 
             NGC4535 &   0.79 $\pm$   0.08 &    8.7 $\pm$    0.1 &  -0.38 &  -1.00 \\ 
        NGC4535 VIVA &   0.94 $\pm$   0.06 &   11.5 $\pm$    0.1 &  -0.54 &   0.33 \\ 
             NGC4254 &   1.64 $\pm$   0.23 &    4.2 $\pm$    0.1 &  -0.04 &  -1.00 \\ 
             NGC4501 &   0.86 $\pm$   0.18 &    2.6 $\pm$    0.0 &   0.00 &  -0.33 \\ 
             NGC4654 &   1.58 $\pm$   0.16 &    5.1 $\pm$    0.0 &  -0.25 &   0.33 \\ 
\hline
       exponential & & & & \\
\hline
       NGC6946 &   1.23 $\pm$   0.32 &    7.1 $\pm$    0.2 &  -0.74 &  -0.38 \\ 
      NGC6946 lowres &   0.94 $\pm$   0.12 &    7.2 $\pm$    0.1 &  -0.62 &   0.23 \\ 
                 M51 &   1.53 $\pm$   0.30 &    5.5 $\pm$    0.2 &  -0.59 &  -0.29 \\ 
          M51 lowres &   1.36 $\pm$   0.27 &    5.6 $\pm$    0.1 &  -0.62 &  -0.18 \\ 
             NGC4321 &   0.85 $\pm$   0.16 &    4.1 $\pm$    0.0 &  -0.15 &  -0.17 \\ 
      NGC4321 disk &   1.00 $\pm$   0.25 &    4.9 $\pm$    0.0 &  -0.04 &  -0.17 \\ 
             NGC4303 &   1.03 $\pm$   0.18 &    4.6 $\pm$    0.1 &  -0.08 &  -1.00 \\ 
             NGC4535 &   0.44 $\pm$   0.06 &    8.7 $\pm$    0.1 &  -0.33 &  -1.00 \\ 
        NGC4535 VIVA &   0.50 $\pm$   0.07 &   12.0 $\pm$    0.1 &  -0.29 &   0.00 \\ 
             NGC4254 &   0.90 $\pm$   0.22 &    4.2 $\pm$    0.1 &  -0.17 &  -0.67 \\ 
             NGC4501 &   0.35 $\pm$   0.16 &    2.6 $\pm$    0.0 &  -0.10 &  -0.17 \\ 
             NGC4654 &   0.99 $\pm$   0.28 &    5.1 $\pm$    0.1 &  -0.57 &  -0.17 \\
\hline
\end{tabular}
%\begin{tablenotes}
%  \item $^{\rm a}$ Moorwood et al. (1996), Pier et al. (1994)
%    \end{tablenotes}
\end{center}
\end{table}

\begin{table*}[!ht]
\begin{center}
\caption{Most probable parameters for the model with $\SFR$ dependence and advection losses at 20~cm.\label{tab:table_20cm}}
\begin{tabular}{lccccccc}
\hline
       Gaussian & & & & & & \\
\hline
name &  $l$ &  $Q$ & $n$ & $e$ & $k$ & $c$ \\
\hline
              NGC6946 &   2.04 $\pm$   0.33 &    4.9 $\pm$    1.0 &  -0.62 &  -0.57 &   1.38 $\pm$   0.25 & 1.2E-01 $\pm$ 9.3E-02 \\ 
      NGC6946 lowres &   1.66 $\pm$   0.24 &    4.6 $\pm$    0.8 &  -0.55 &  -0.09 &   1.38 $\pm$   0.18 & 1.3E-01 $\pm$ 9.2E-02 \\ 
                 M51 &   2.33 $\pm$   0.25 &    5.0 $\pm$    0.6 &  -0.53 &  -0.55 &   1.57 $\pm$   0.34 & 4.1E-02 $\pm$ 6.0E-02 \\ 
          M51 lowres &   2.35 $\pm$   0.26 &    5.0 $\pm$    0.4 &  -0.61 &  -0.80 &   1.58 $\pm$   0.35 & 3.4E-02 $\pm$ 5.3E-02 \\ 
             NGC4321 &   1.71 $\pm$   0.23 &    4.1 $\pm$    0.1 &  -0.17 &  -0.50 &   - & 0.00 \\ 
      NGC4321 disk &   2.08 $\pm$   0.17 &    4.7 $\pm$    0.2 &   0.00 &  -1.00 &   1.67 $\pm$   0.36 & 1.1E-02 $\pm$ 2.5E-02 \\ 
             (NGC4303 &   0.92 $\pm$   0.19 &    2.4 $\pm$    0.3 &  -0.22 &  -0.09 &   1.31 $\pm$   0.10 & 2.1E-01 $\pm$ 9.6E-02) \\
      NGC4303 &   0.94 $\pm$   0.19 &    3.1 $\pm$    0.2 &  -0.29 &  -0.09 &   1.68 $\pm$   0.10 & 4.5E-02 $\pm$ 1.7E-02 \\
             NGC4535 &   0.80 $\pm$   0.04 &    8.4 $\pm$    0.4 &  -0.50 &  -1.00 &   1.77 $\pm$   0.35 & 1.4E-02 $\pm$ 3.2E-02 \\ 
             NGC4535 VIVA &   1.00 $\pm$   0.02 &   11.3 $\pm$    1.3 &  -0.57 &   0.77 &   1.66 $\pm$   0.32 & 6.7E-02 $\pm$ 8.0E-02 \\ 
             NGC4254 &   1.28 $\pm$   0.14 &    3.3 $\pm$    0.5 &  -0.12 &  -1.00 &   1.74 $\pm$   0.35 & 5.8E-02 $\pm$ 8.5E-02 \\ 
             NGC4501 &   0.86 $\pm$   0.18 &    2.6 $\pm$    0.0 &   0.00 &  -0.33 &   - & 0.00 \\ 
             NGC4654 &   1.53 $\pm$   0.11 &    5.0 $\pm$    0.2 &  -0.25 &   0.15 &   1.74 $\pm$   0.36 & 4.4E-03 $\pm$ 1.8E-02 \\ 
\hline
        exponential & & & & & & \\
\hline
        NGC6946 &   1.11 $\pm$   0.23 &    5.1 $\pm$    1.0 &  -0.67 &  -0.63 &   1.45 $\pm$   0.26 & 9.6E-02 $\pm$ 8.9E-02 \\ 
      NGC6946 lowres &   0.89 $\pm$   0.14 &    4.8 $\pm$    0.8 &  -0.62 &  -0.09 &   1.43 $\pm$   0.24 & 1.2E-01 $\pm$ 9.9E-02 \\ 
                 M51 &   1.43 $\pm$   0.28 &    5.0 $\pm$    0.6 &  -0.58 &  -0.59 &   1.56 $\pm$   0.33 & 3.4E-02 $\pm$ 5.3E-02 \\ 
          M51 lowres &   1.27 $\pm$   0.19 &    5.2 $\pm$    0.4 &  -0.62 &  -0.55 &   1.61 $\pm$   0.35 & 2.5E-02 $\pm$ 4.5E-02 \\ 
             NGC4321 &   0.85 $\pm$   0.16 &    4.0 $\pm$    0.1 &  -0.16 &  -0.31 &   1.67 $\pm$   0.35 & 4.5E-03 $\pm$ 1.9E-02 \\ 
      NGC4321 disk &   0.99 $\pm$   0.13 &    4.6 $\pm$    0.3 &   0.00 &  -0.40 &   1.66 $\pm$   0.34 & 2.1E-02 $\pm$ 4.1E-02 \\ 
             NGC4303 &   0.44 $\pm$   0.11 &    2.5 $\pm$    0.4 &  -0.27 &  -0.09 &   1.41 $\pm$   0.20 & 1.7E-01 $\pm$ 1.1E-01 \\ 
             NGC4535 &   0.40 $\pm$   0.03 &    8.4 $\pm$    0.4 &  -0.46 &  -1.00 &   1.75 $\pm$   0.35 & 1.2E-02 $\pm$ 3.1E-02 \\ 
             NGC4535 VIVA &   0.51 $\pm$   0.06 &    11.8 $\pm$    0.2 &  -0.25 &   0.26 &   1.73 $\pm$   0.36 & 2.5E-03 $\pm$ 1.3E-02 \\ 
             NGC4254 &   0.68 $\pm$   0.11 &    3.3 $\pm$    0.4 &  -0.16 &  -0.71 &   1.76 $\pm$   0.34 & 5.0E-02 $\pm$ 7.6E-02 \\ 
             NGC4501 &   0.34 $\pm$   0.16 &    2.6 $\pm$    0.0 &  -0.09 &  -0.11 &   1.73 $\pm$   0.35 & 6.3E-05 $\pm$ 2.9E-04 \\ 
             NGC4654 &   1.00 $\pm$   0.25 &    5.1 $\pm$    0.2 &  -0.61 &  -0.03 &   1.74 $\pm$   0.36 & 4.8E-03 $\pm$ 1.9E-02 \\ 
\hline
\end{tabular}
%\begin{tablenotes}
%  \item $^{\rm a}$ Moorwood et al. (1996), Pier et al. (1994)
%    \end{tablenotes}
\end{center}
\end{table*}

\clearpage

\section{Goodness distributions}

\begin{figure*}[!ht]
  \centering
  \resizebox{\hsize}{!}{\includegraphics{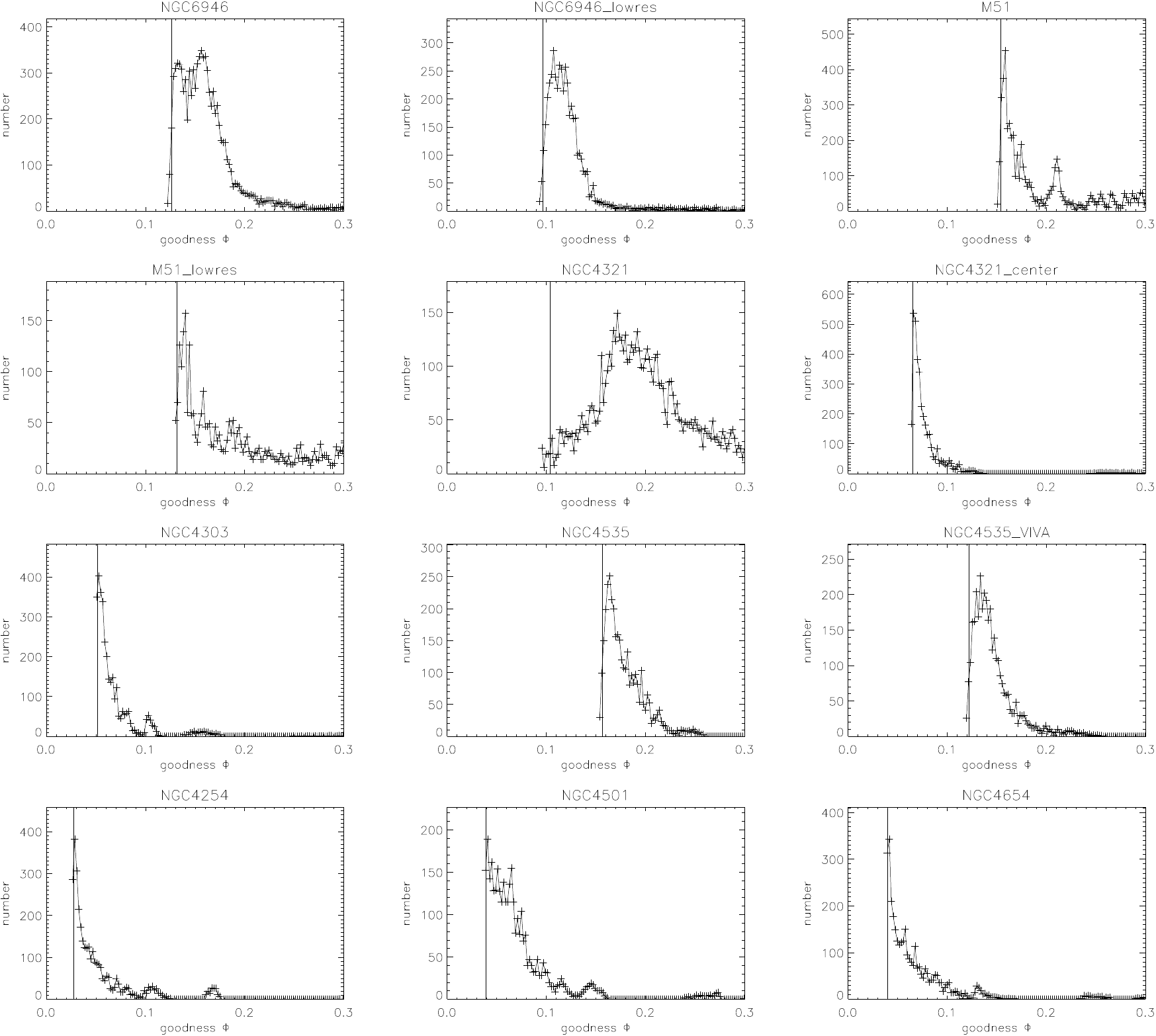}\includegraphics{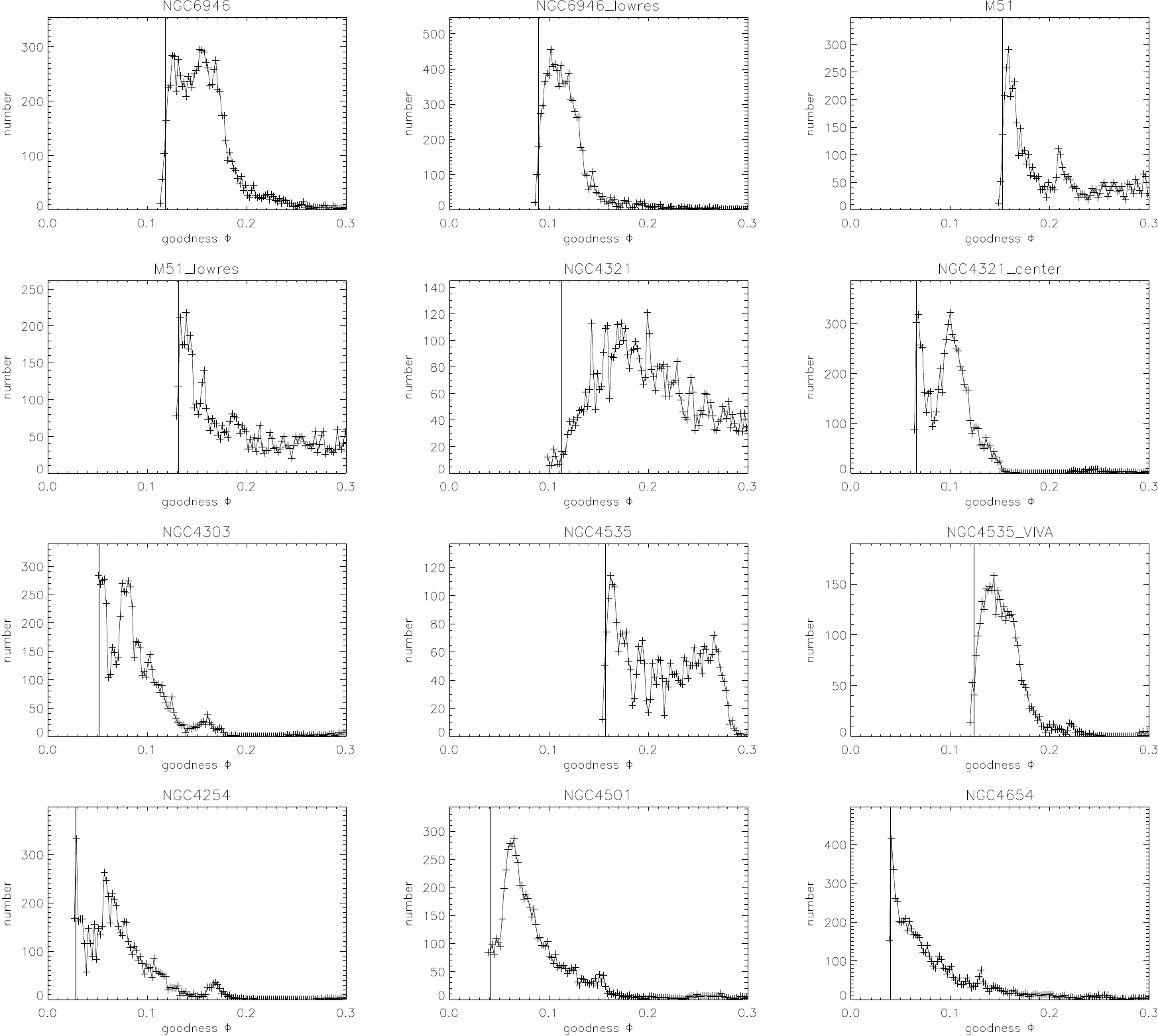}}
  \put(-475,220){Gauss}
  \put(-380,220){Gauss}
  \put(-290,220){Gauss}
  \put(-200,220){exp}
  \put(-110,220){exp}
  \put(-20,220){exp}
  \put(-475,162){Gauss}
  \put(-380,162){Gauss}
  \put(-290,162){Gauss}
  \put(-200,162){exp}
  \put(-110,162){exp}
  \put(-20,162){exp}
  `\put(-475,102){Gauss}
  \put(-380,102){Gauss}
  \put(-290,102){Gauss}
  \put(-200,102){exp}
  \put(-110,102){exp}
  \put(-20,102){exp}
    `\put(-475,42){Gauss}
  \put(-380,42){Gauss}
  \put(-290,42){Gauss}
  \put(-200,42){exp}
  \put(-110,42){exp}
  \put(-20,42){exp}
  \caption{Distribution of the goodness parameter $\phi$ based on the $6$~cm radio continuum data. 
    The cut for the accepted models is shown as a vertical line.
    No losses are included in these models.
  \label{fig:zusammenphi1}}
\end{figure*}

\begin{figure*}[!ht]
  \centering
  \resizebox{\hsize}{!}{\includegraphics{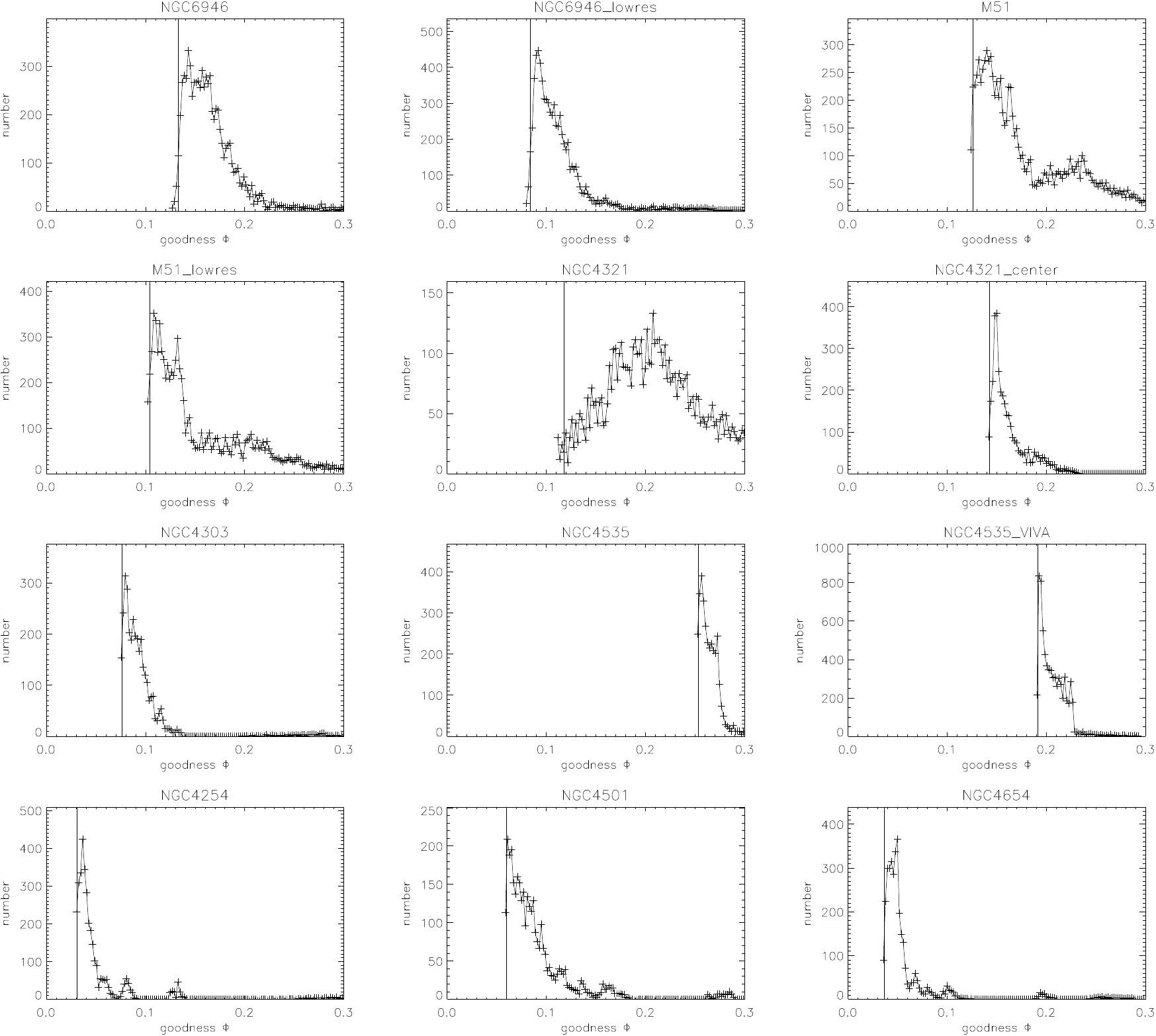}\includegraphics{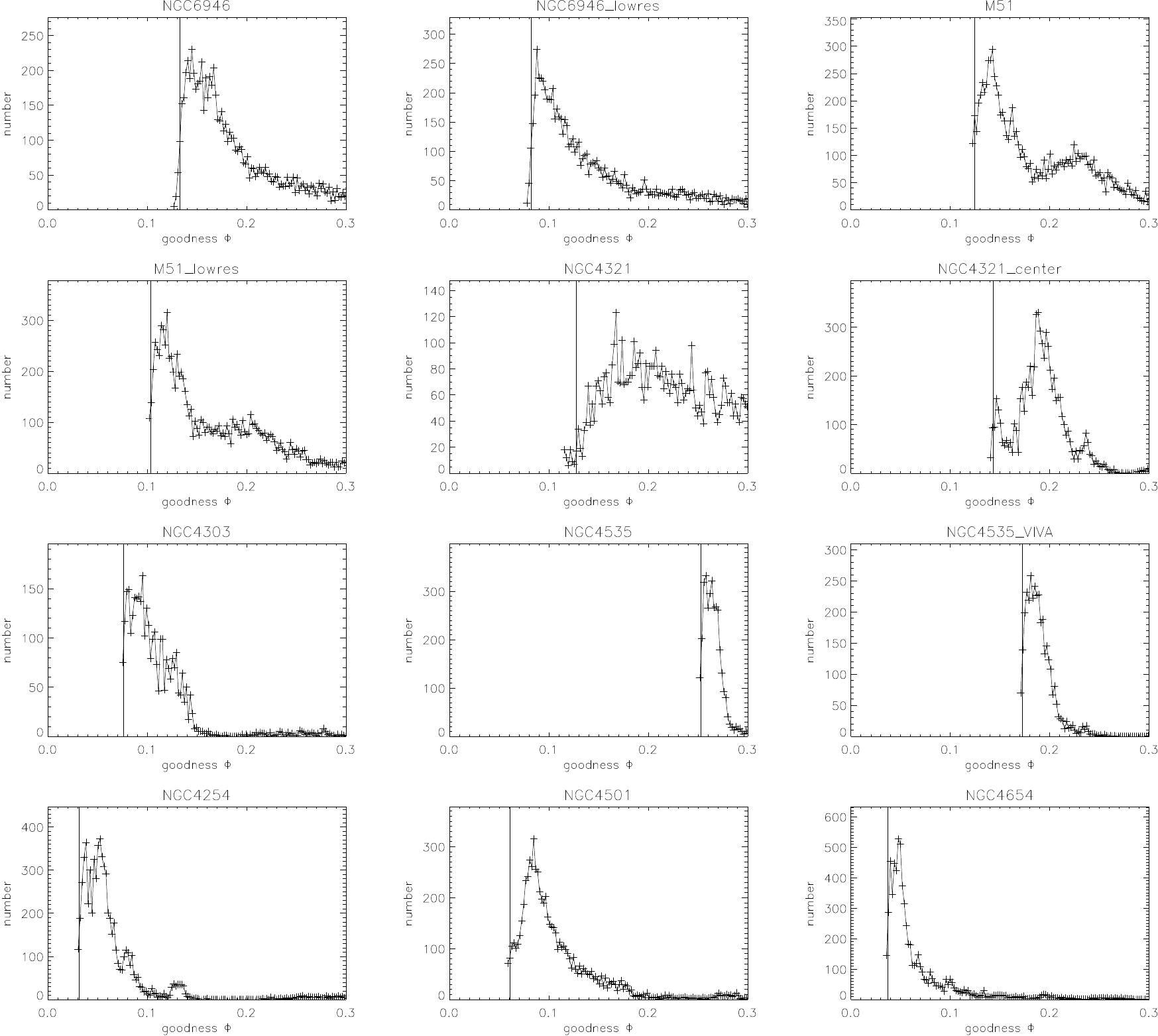}}
  \put(-475,220){Gauss}
  \put(-380,220){Gauss}
  \put(-290,220){Gauss}
  \put(-200,220){exp}
  \put(-110,220){exp}
  \put(-20,220){exp}
  \put(-475,162){Gauss}
  \put(-380,162){Gauss}
  \put(-290,162){Gauss}
  \put(-200,162){exp}
  \put(-110,162){exp}
  \put(-20,162){exp}
  `\put(-475,102){Gauss}
  \put(-380,102){Gauss}
  \put(-290,102){Gauss}
  \put(-200,102){exp}
  \put(-110,102){exp}
  \put(-20,102){exp}
    `\put(-475,42){Gauss}
  \put(-380,42){Gauss}
  \put(-290,42){Gauss}
  \put(-200,42){exp}
  \put(-110,42){exp}
  \put(-20,42){exp}
  \caption{Distribution of the goodness parameter $\phi$ based on the $20$~cm radio continuum data. 
    The cut for the accepted models is shown as a vertical line.
    No losses are included in these models.
  \label{fig:zusammenphi2}}
\end{figure*}
\begin{figure*}[!ht]
  \centering
  \resizebox{\hsize}{!}{\includegraphics{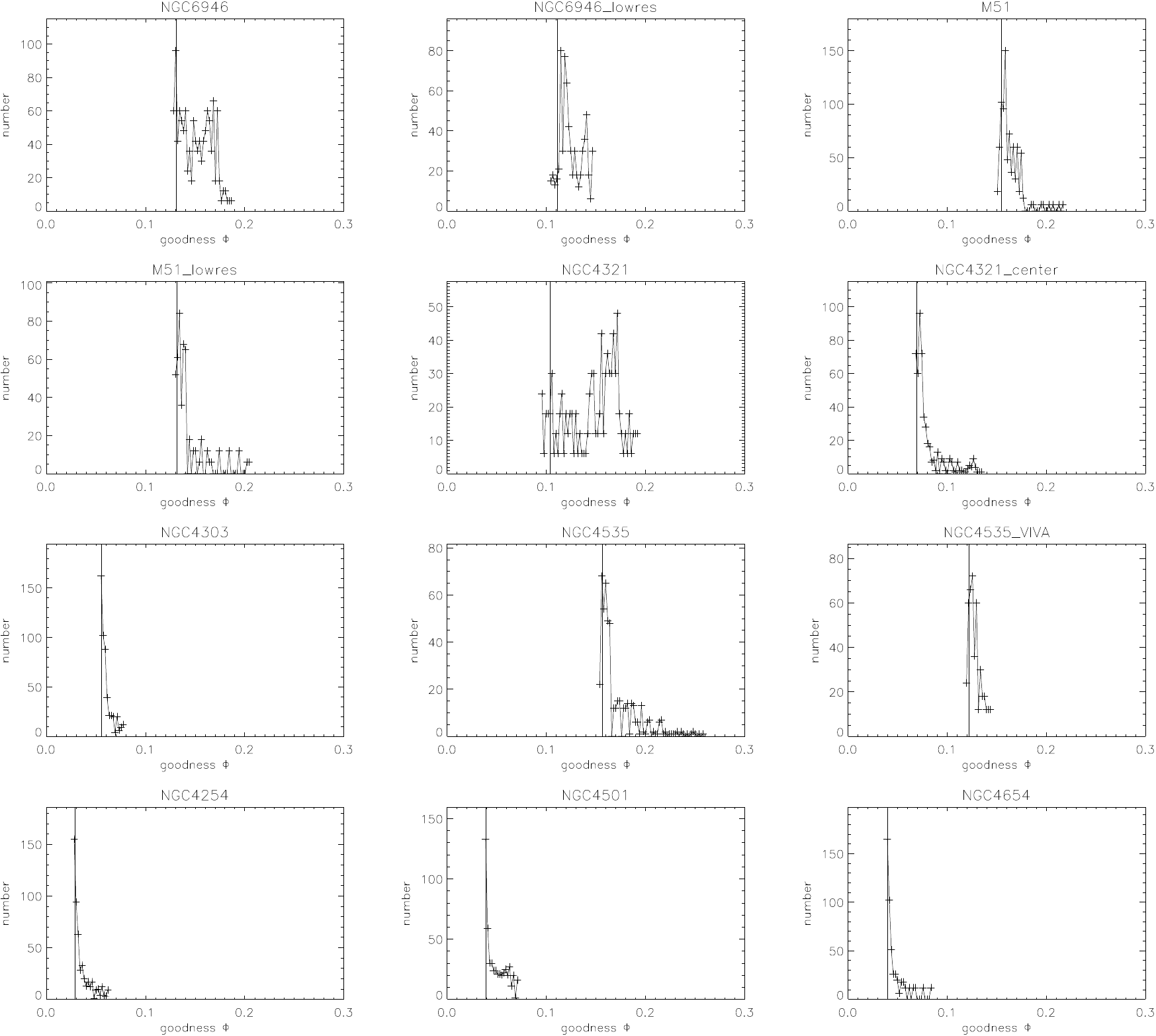}\includegraphics{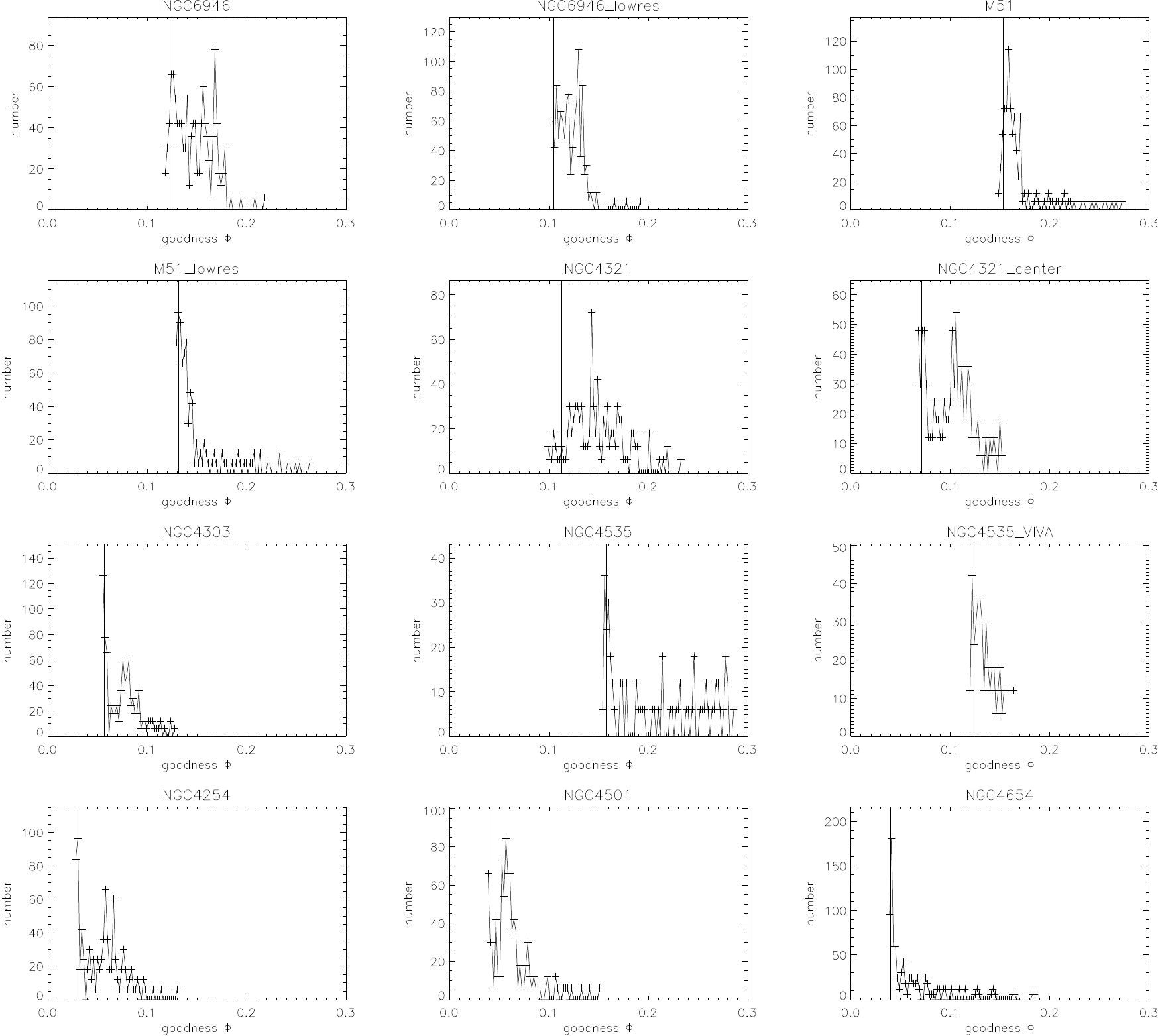}}
  \put(-475,220){Gauss}
  \put(-380,220){Gauss}
  \put(-290,220){Gauss}
  \put(-200,220){exp}
  \put(-110,220){exp}
  \put(-20,220){exp}
  \put(-475,162){Gauss}
  \put(-380,162){Gauss}
  \put(-290,162){Gauss}
  \put(-200,162){exp}
  \put(-110,162){exp}
  \put(-20,162){exp}
  `\put(-475,102){Gauss}
  \put(-380,102){Gauss}
  \put(-290,102){Gauss}
  \put(-200,102){exp}
  \put(-110,102){exp}
  \put(-20,102){exp}
    `\put(-475,42){Gauss}
  \put(-380,42){Gauss}
  \put(-290,42){Gauss}
  \put(-200,42){exp}
  \put(-110,42){exp}
  \put(-20,42){exp}
  \caption{Distribution of the goodness parameter $\phi$ based on the $6$~cm radio continuum data. 
    The cut for the accepted models is shown as a vertical line.
    Losses are included in these models.
  \label{fig:zusammenphi3}}
\end{figure*}
\begin{figure*}[!ht]
  \centering
  \resizebox{\hsize}{!}{\includegraphics{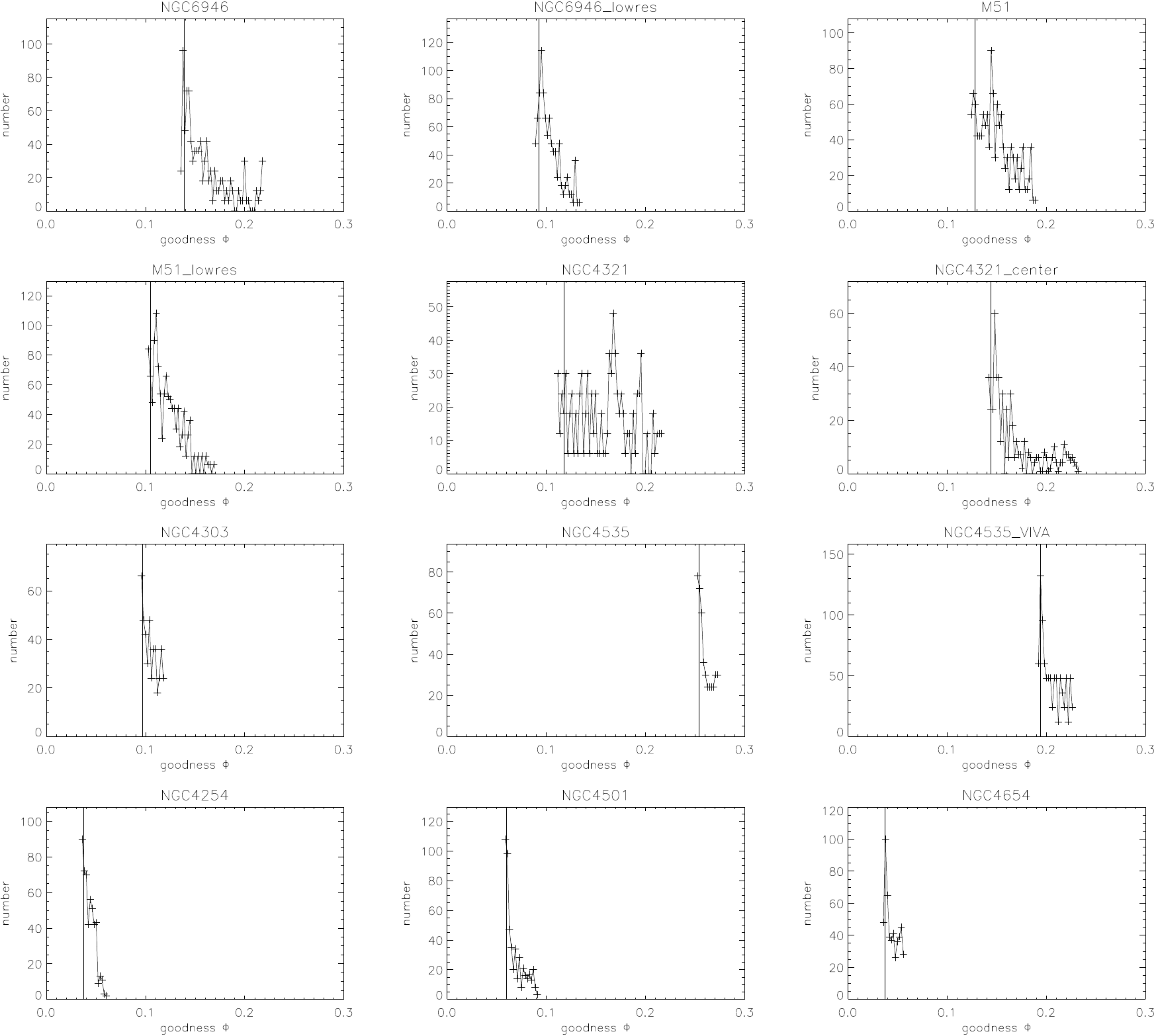}\includegraphics{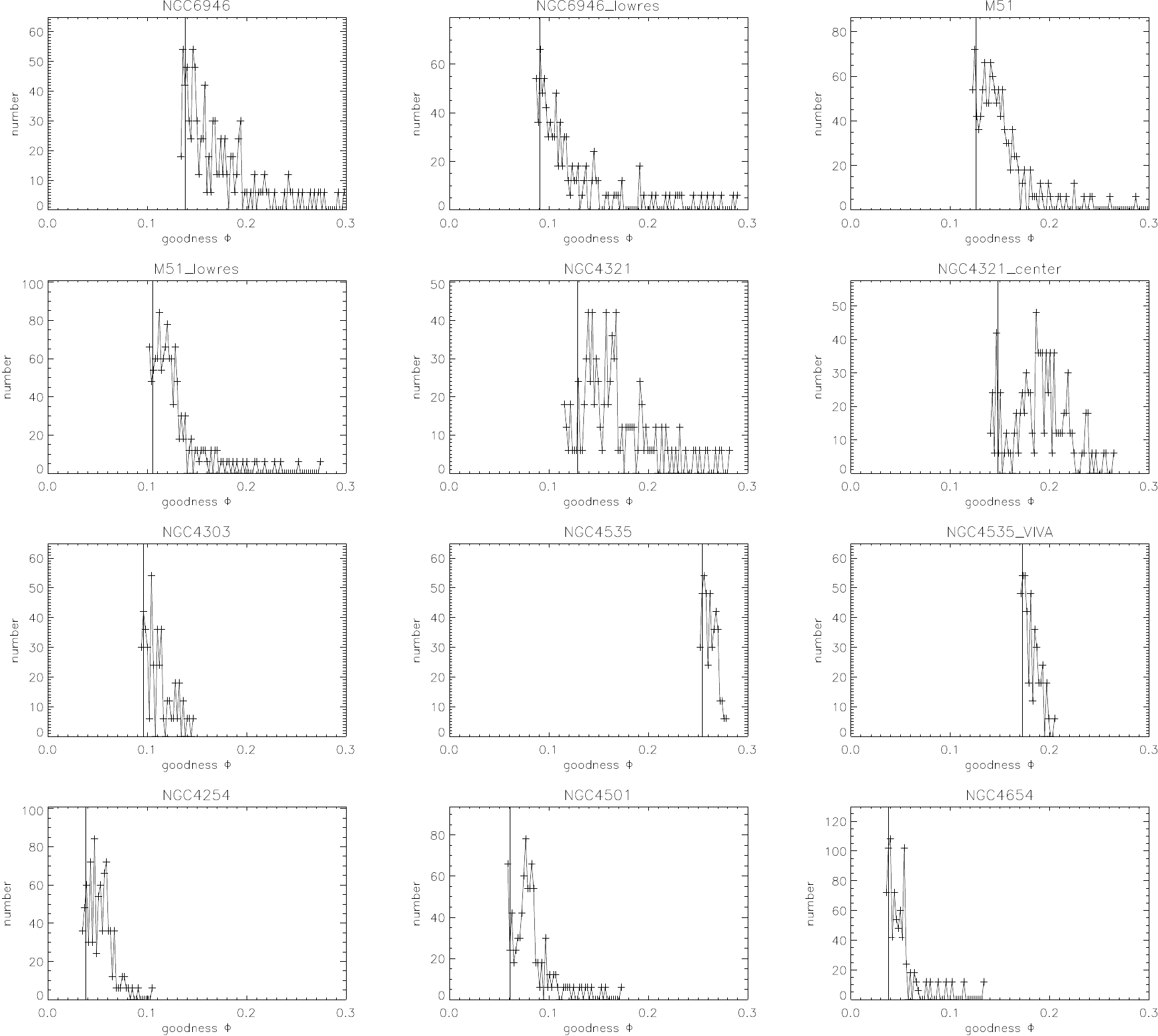}}
  \put(-475,220){Gauss}
  \put(-380,220){Gauss}
  \put(-290,220){Gauss}
  \put(-200,220){exp}
  \put(-110,220){exp}
  \put(-20,220){exp}
  \put(-475,162){Gauss}
  \put(-380,162){Gauss}
  \put(-290,162){Gauss}
  \put(-200,162){exp}
  \put(-110,162){exp}
  \put(-20,162){exp}
  `\put(-475,102){Gauss}
  \put(-380,102){Gauss}
  \put(-290,102){Gauss}
  \put(-200,102){exp}
  \put(-110,102){exp}
  \put(-20,102){exp}
    `\put(-475,42){Gauss}
  \put(-380,42){Gauss}
  \put(-290,42){Gauss}
  \put(-200,42){exp}
  \put(-110,42){exp}
  \put(-20,42){exp}
  \caption{Distribution of the goodness parameter $\phi$ based on the $20$~cm radio continuum data. 
    The cut for the accepted models is shown as a vertical line.
    Losses are included in these models.
  \label{fig:zusammenphi4}}
\end{figure*}

\clearpage

\section{``Best fit'' model radio continuum maps}

\begin{figure*}[!ht]
  \centering
  \resizebox{16cm}{!}{\includegraphics{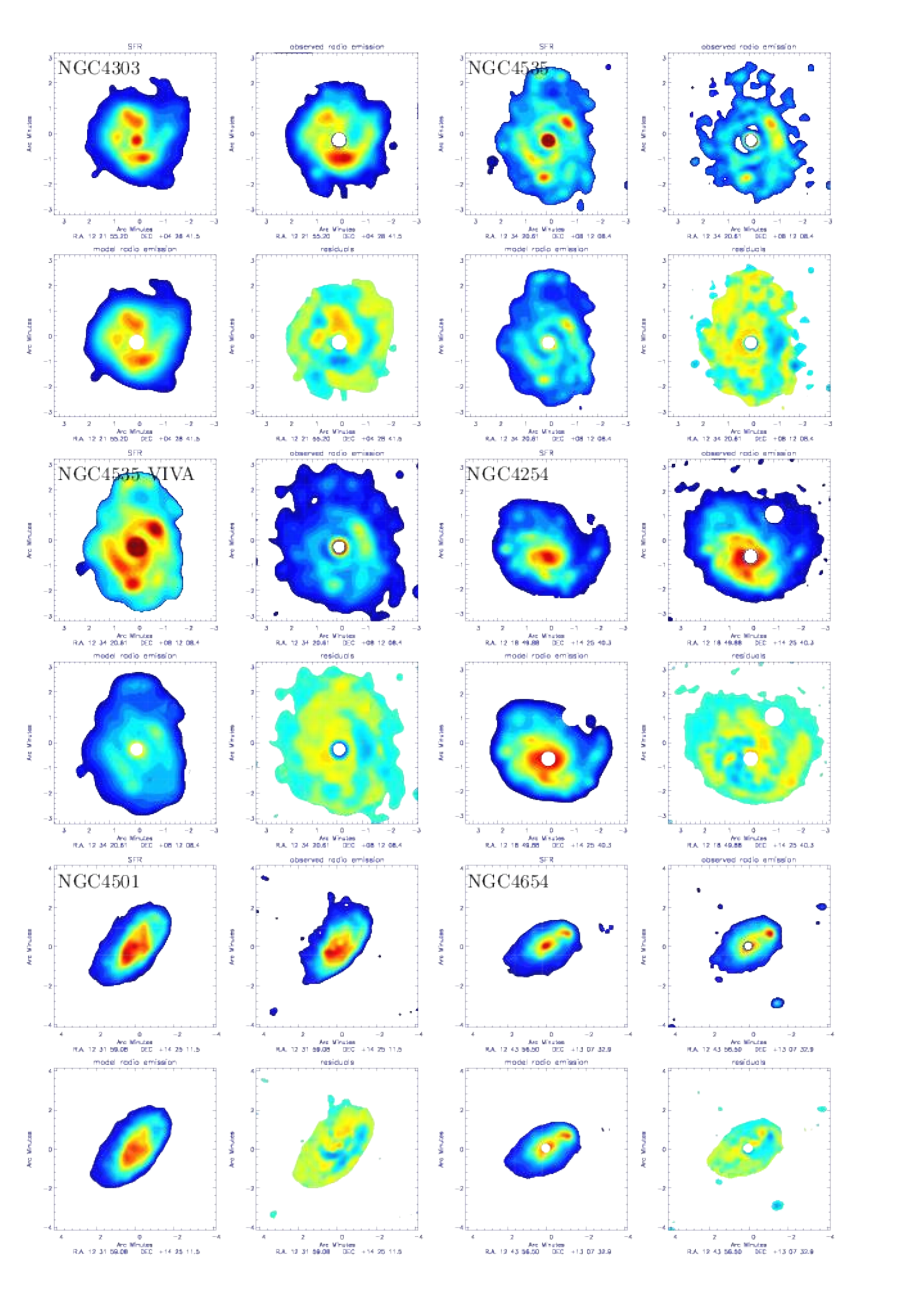}}
  \caption{ Gaussian convolution: ``best fit'' model radio continuum maps at $6$~cm. Four maps are shown for each galaxy.
    Upper left: observed star formation; upper right: observed radio continuum emission; lower left: model radio continuum;
    lower right: residuals; blue is radio bright, red radio dim.
  \label{fig:zusammenplots2}}
\end{figure*}
\begin{figure*}[!ht]
  \centering
  \resizebox{16cm}{!}{\includegraphics{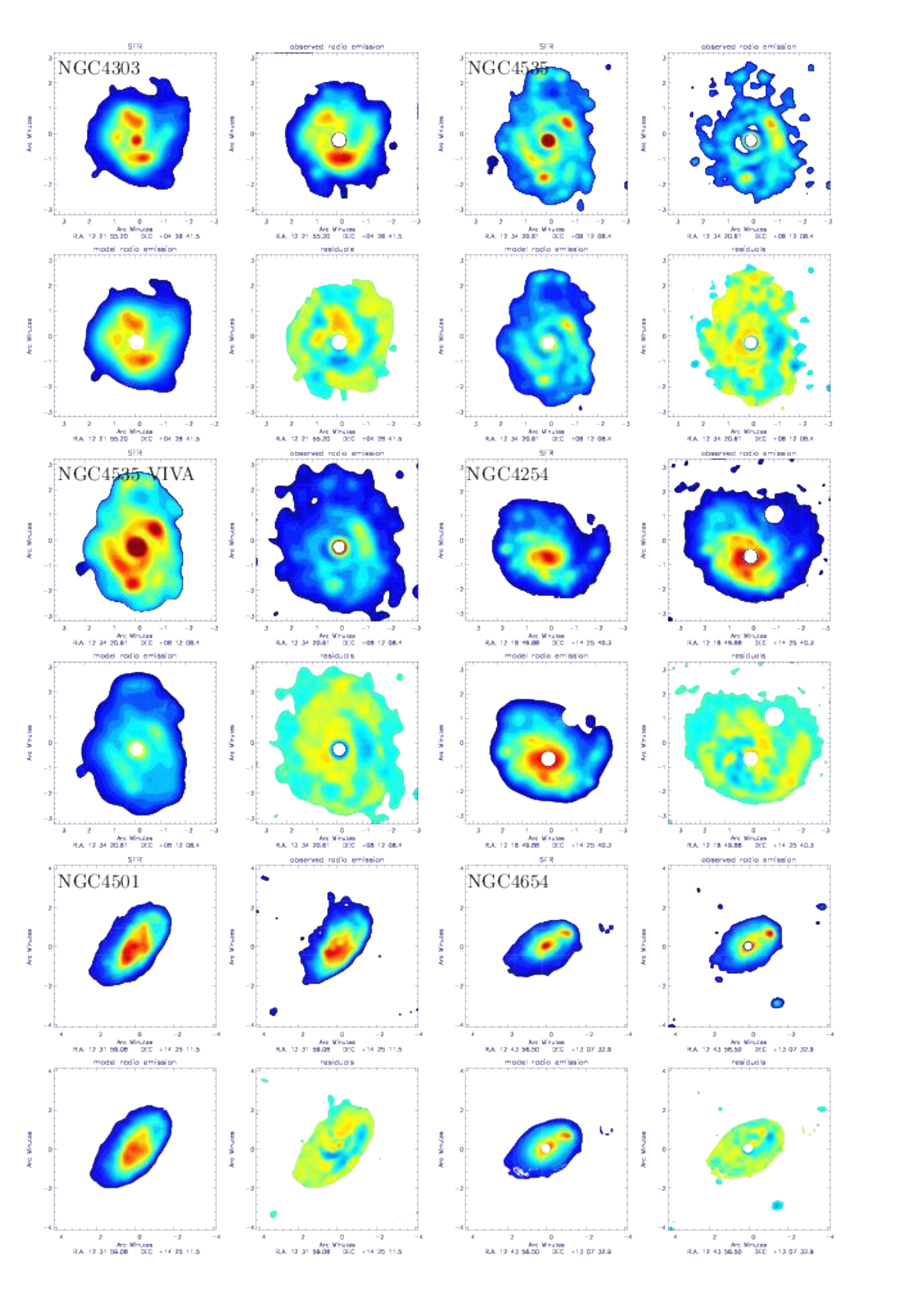}}
  \caption{ Exponential convolution: ``best fit'' model radio continuum maps at $6$~cm. Four maps are shown for each galaxy.
    Upper left: observed star formation; upper right: observed radio continuum emission; lower left: model radio continuum;
    lower right: residuals; blue is radio bright, red radio dim.
  \label{fig:zusammenplots2a}}
\end{figure*}

\begin{figure*}[!ht]
  \centering
  \resizebox{16cm}{!}{\includegraphics{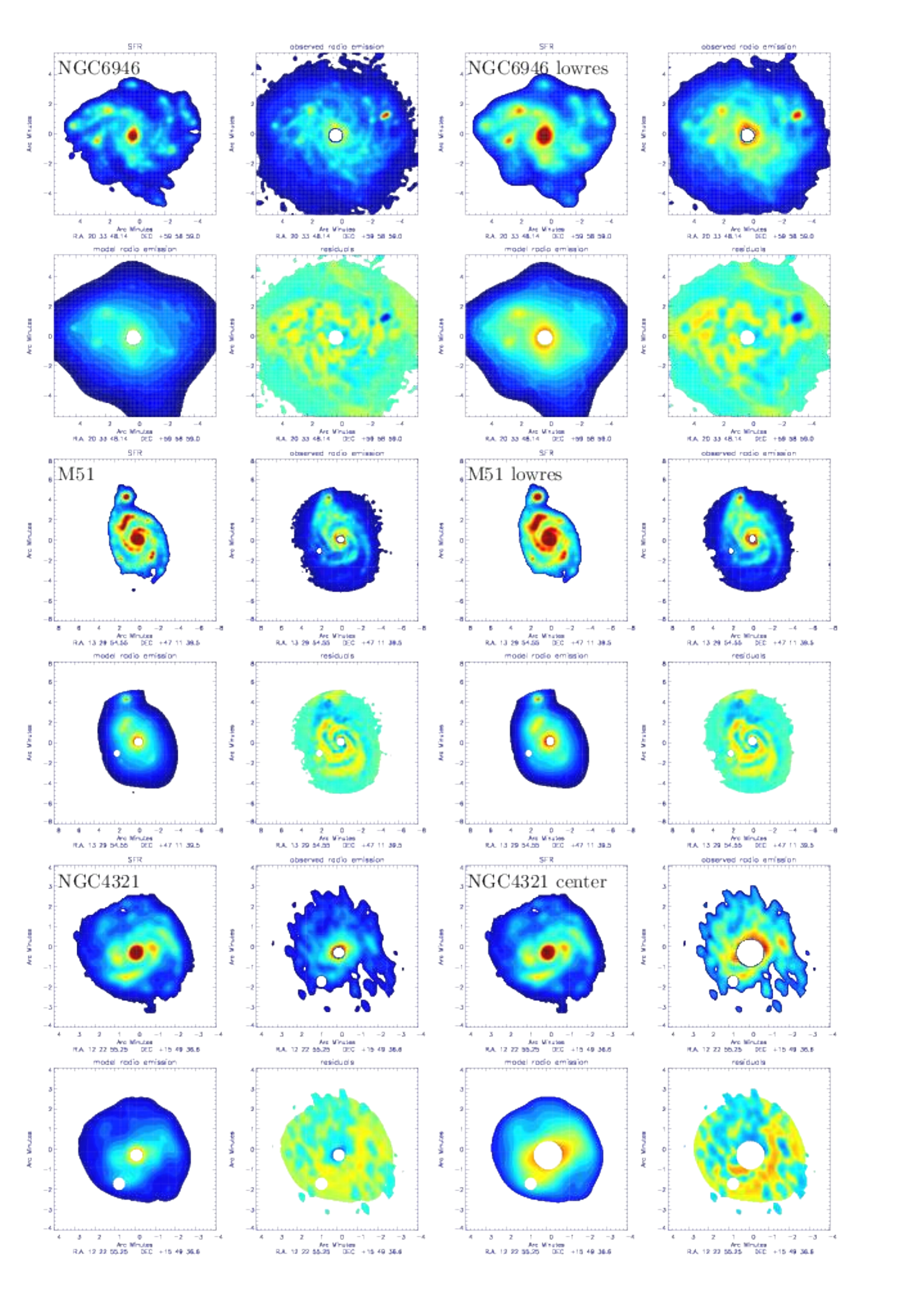}}
  \caption{Same as Fig.~\ref{fig:zusammenplots2} for the $20$~cm data.
  \label{fig:zusammenplots3}}
\end{figure*}
\begin{figure*}[!ht]
  \centering
  \resizebox{16cm}{!}{\includegraphics{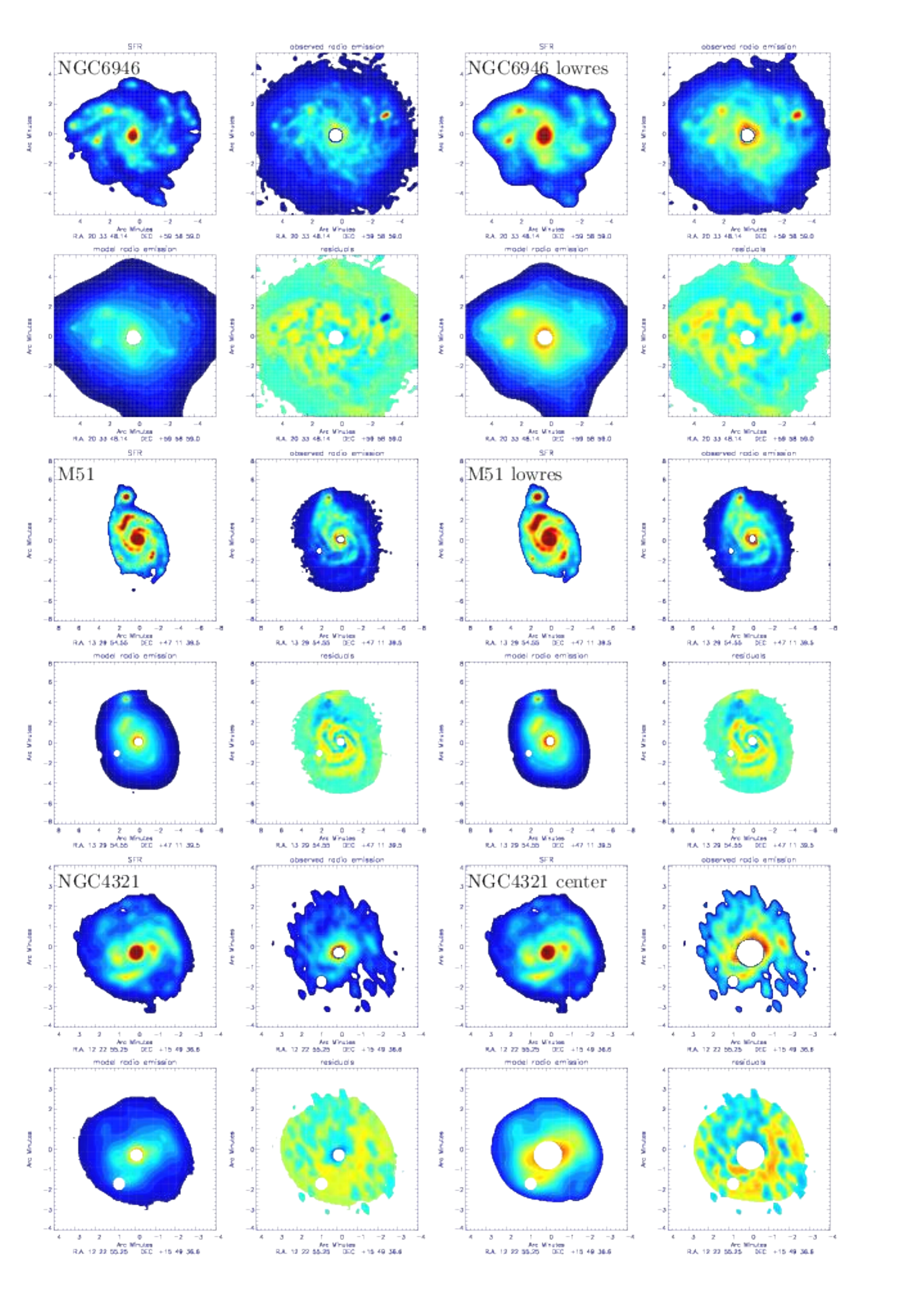}}
  \caption{Same as Fig.~\ref{fig:zusammenplots2} for the $20$~cm data.
  \label{fig:zusammenplots3a}}
\end{figure*}

\begin{figure*}[!ht]
  \centering
  \resizebox{16cm}{!}{\includegraphics{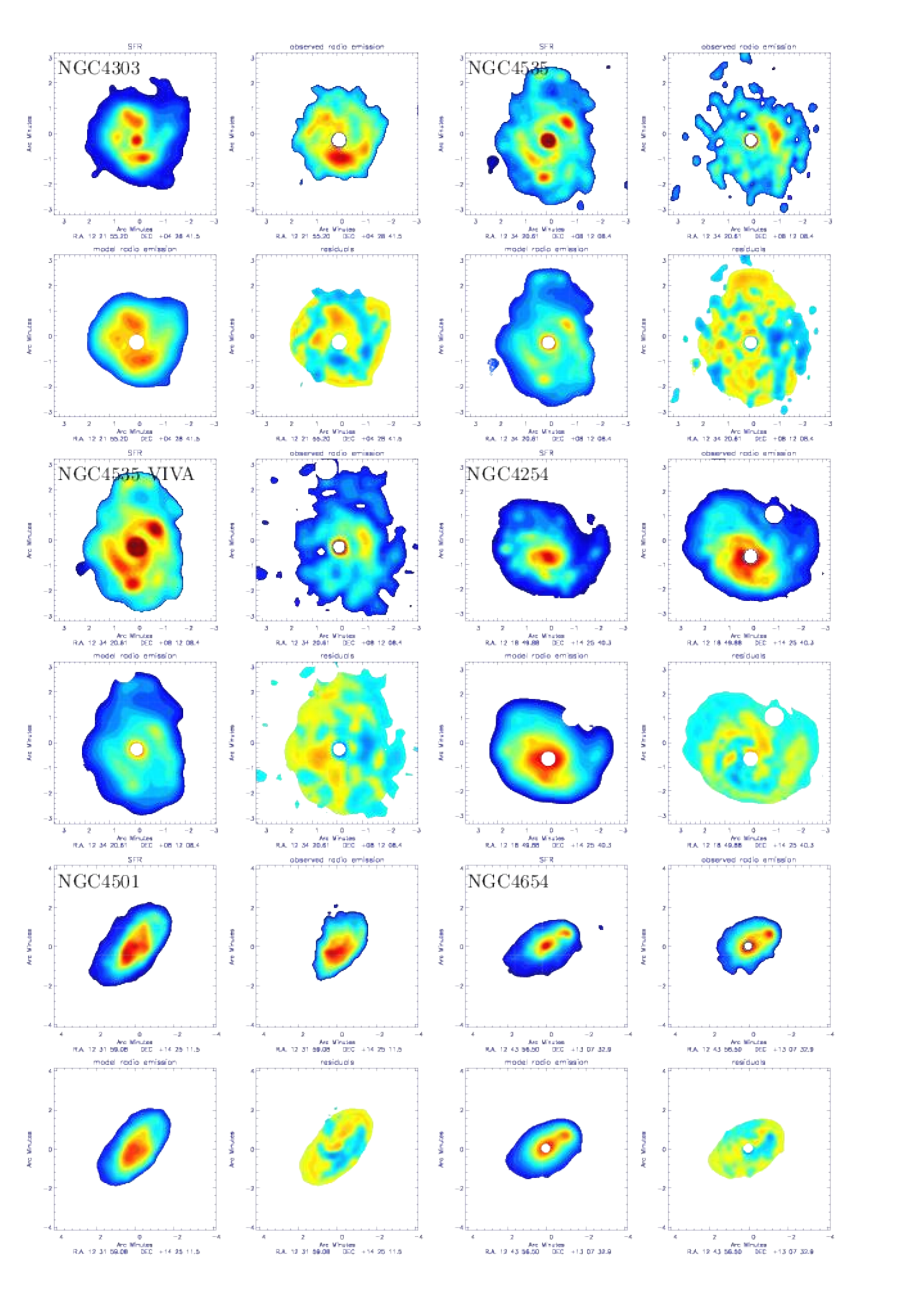}}
  \caption{Fig.~\ref{fig:zusammenplots3} -- continued.
  \label{fig:zusammenplots4}}
\end{figure*}
\begin{figure*}[!ht]
  \centering
  \resizebox{16cm}{!}{\includegraphics{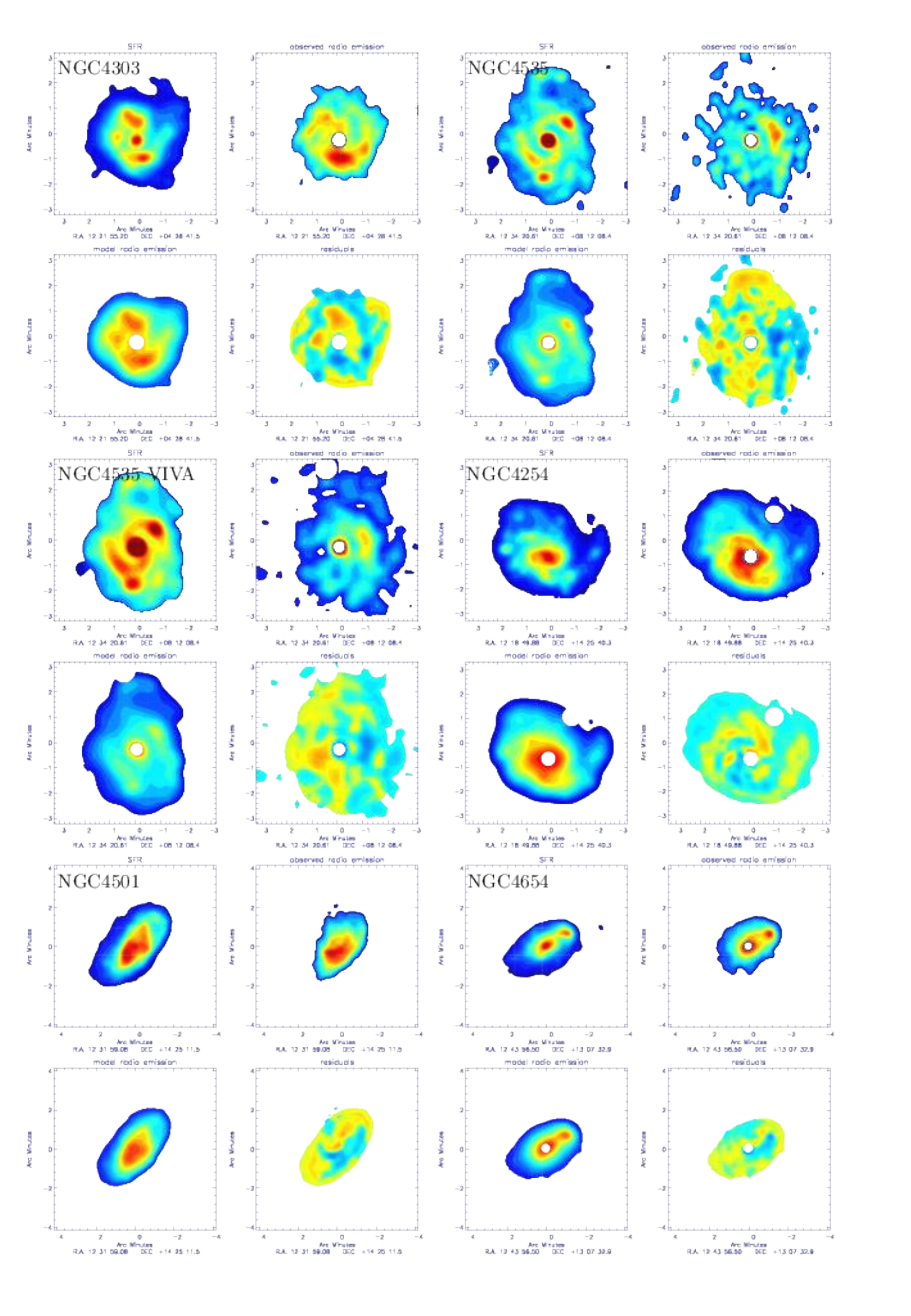}}
  \caption{Fig.~\ref{fig:zusammenplots3} -- continued.
  \label{fig:zusammenplots4a}}
\end{figure*}

\begin{figure*}[!ht]
  \centering
  \resizebox{16cm}{!}{\includegraphics{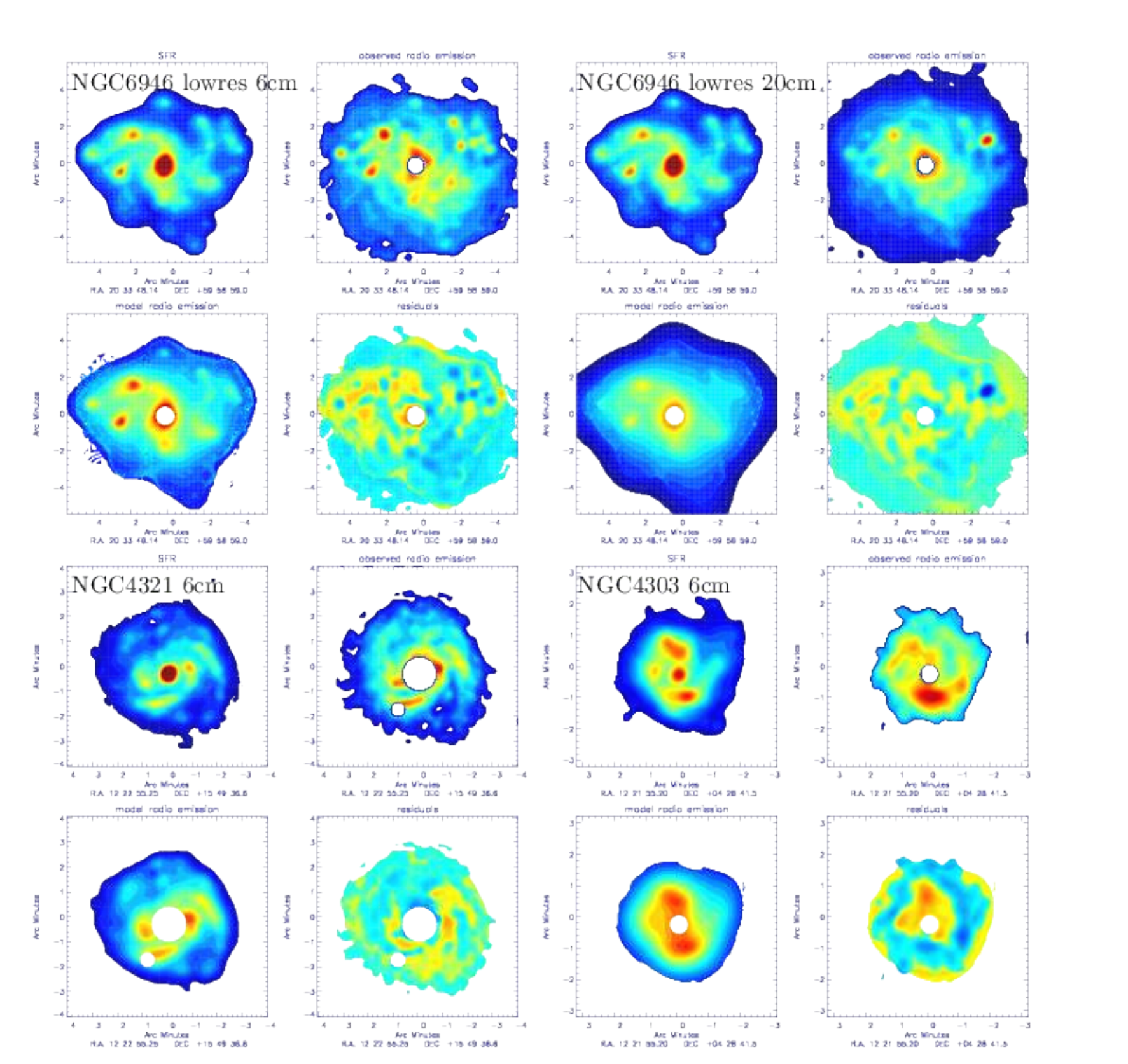}}
  \caption{``Best fit'' model radio continuum maps for additional studies (low resolution or removed central emission).
  \label{fig:zusammenplots5}}
\end{figure*}

\end{document}